%% file: pebmc.tex
\documentclass[onefignum,onetabnum]{siamonline250211}

\input{ex_shared}

\begin{document}
\maketitle

\begin{abstract}
Calibrating a simulation model involves estimating its parameters by comparing model outputs with experimental data, 
so that simulation results faithfully reproduce the experimental observations. When the outputs are functions of time, 
there are multiple ways to quantify the discrepancy between experimental and simulated curves. A recent approach based on 
elastic functional data analysis decomposes a functional output into two components: a function temporally aligned to a template, 
and the corresponding warping function. This decomposition splits the problem into two independent calibration tasks, thereby addressing
functional misalignment. However, it assumes that experimental and simulated curves share the same temporal support, an assumption often violated
in practice when initial or end times are themselves uncertain or depend on the calibration parameters. In this work, we reinterpret the decomposition 
step as an approximation to a more general Bayesian calibration problem that incorporates an error term on the time axis. This perspective allows us 
to naturally extend the framework to a broader family of time warpings with varying initial or end times, using partial elastic alignment. We illustrate 
the method on a synthetic test case, comparing it with existing Bayesian calibration methods and demonstrating improved surrogate performance and error 
modeling. We then apply the proposed approach to the calibration of an equation of state
 (a thermodynamic equation relating the state variables of a material).
\end{abstract}



\clearpage

\tableofcontents

\clearpage

\section{Introduction}

Computer models are widely used to study complex phenomena in numerous fields such as biology \cite{hwang2025bayesian}, earth sciences \cite{watson2021model}, and solid mechanics \cite{perrin2018calibration}.
These models typically depend on unknown parameters that must be selected from experimental data before the model can be used for predictions.
This task is known as model calibration and consists of comparing the outputs of the model with experimental data. A natural approach is to determine
the model parameters that minimize a goodness-of-fit criterion between the model and the experiments. Various such criteria are discussed in \cite{walter1997identification}, including least squares, least modulus, and maximum likelihood. 
However, this approach provides only point estimates and does not quantify the associated uncertainties.

The seminal paper \cite{kennedy2001bayesian} introduced a Bayesian framework to address this issue. This framework relates experimental observations to simulation outputs through an additive decomposition of 
the various error sources, typically measurement noise and a model error, also called the discrepancy. This Bayesian approach is particularly well suited for jointly estimating and quantifying the uncertainty 
of both the calibration parameters and the simulation outputs.

The Bayesian calibration method has been adapted and improved for numerous settings \cite{sung2024review}. 
In this work, we focus on calibration problems whose outputs are time series. Surrogate modeling and Bayesian calibration for time-dependent (and more generally high-dimensional) outputs raise several specific challenges.
First, one must rely on dimension-reduction strategies or implementation techniques to alleviate the computational burden \cite{higdon2008, bayarri2007computer, perrin, brown}.
Second, since time-dependent outputs can be compared in multiple ways, several authors have proposed alternative error models that account for the shape of the 
functions \cite{francom, guan2019computer, kleiber2014model, cheng2024, xie2024functional}. These approaches, referred to as elastic in this paper, rely on the idea that comparing functions of time requires comparing both their 
common shapes and the time deformations between them.

Elastic calibration frameworks offer two main advantages. First, an elastic transformation can simplify the representation of simulation outputs, thereby improving surrogate model 
performance \cite{REST, guo2022, xie2024functional}. Second, elastic calibration makes it possible to introduce a non-additive discrepancy term, providing a more faithful representation of the error structure 
when the simulator and the experiments differ by time warpings.

\begin{figure}[h]
\begin{minipage}{0.32\linewidth}
    \centering
    Pointwise matching
\includegraphics[width=0.98\textwidth]{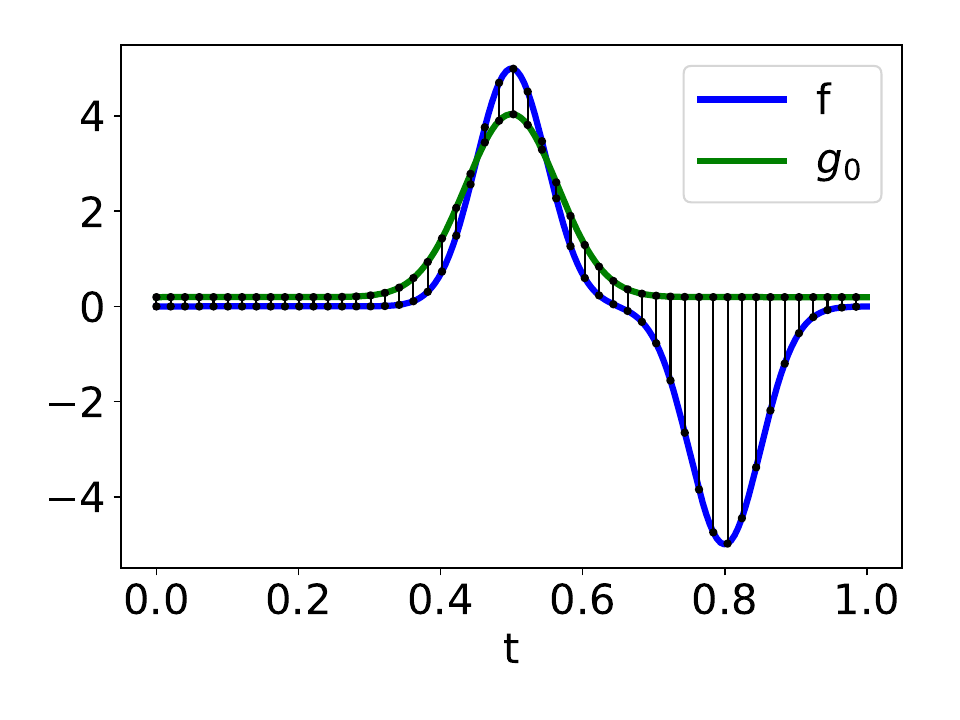}
\end{minipage}
\begin{minipage}{0.32\linewidth}
    \centering
Elastic matching

\includegraphics[width=0.98\textwidth]{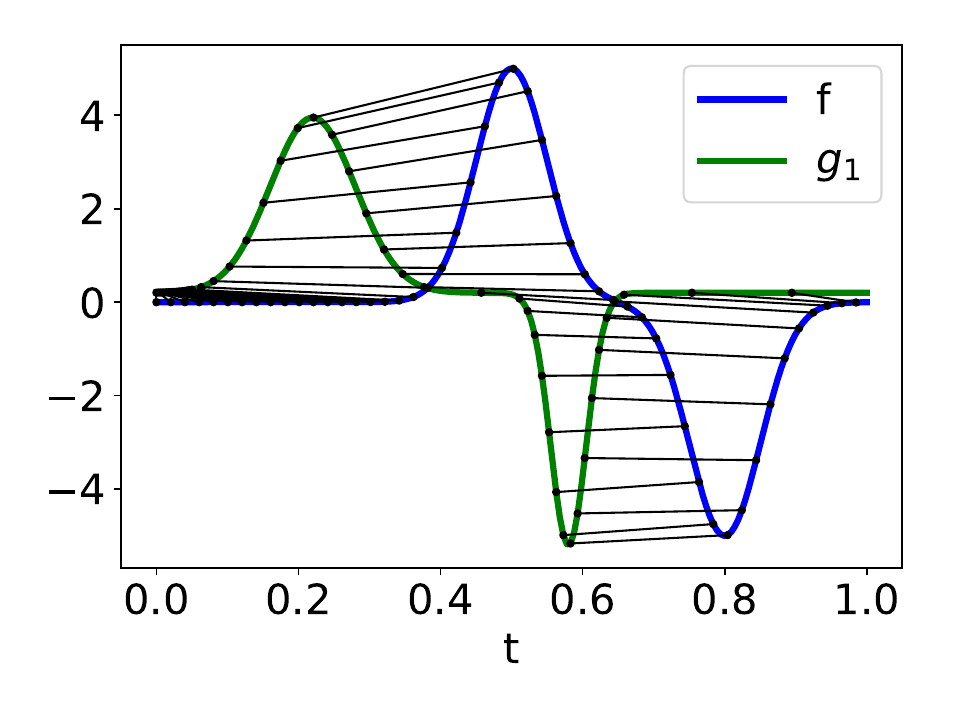}

\end{minipage}
\begin{minipage}{0.32\linewidth}
    \centering
Partial elastic matching
\includegraphics[width=0.98\textwidth]{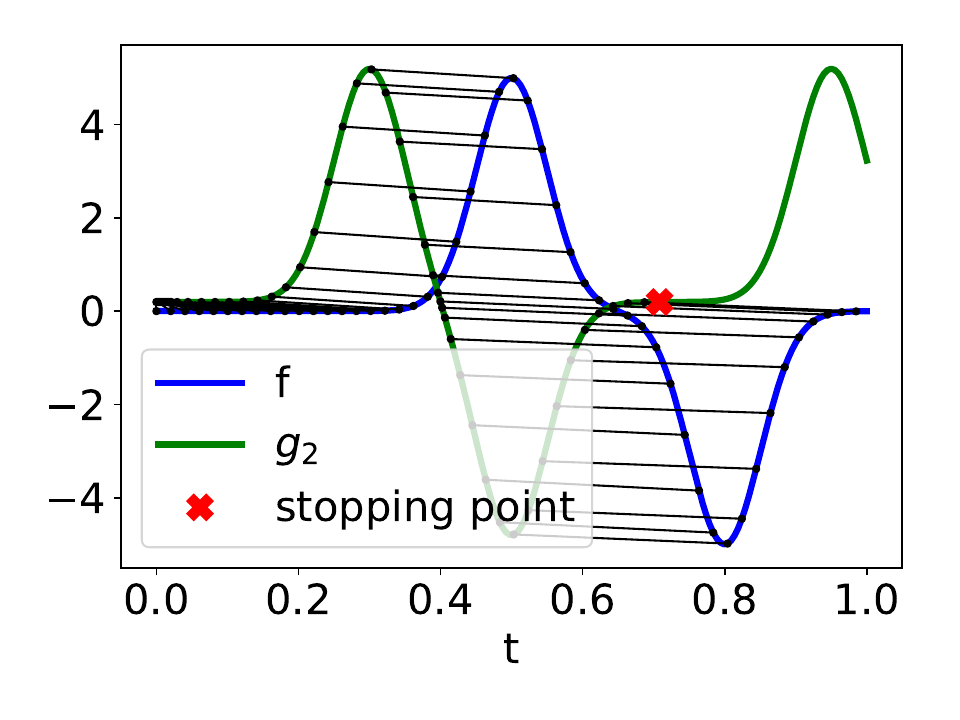}
\end{minipage}
\caption{Comparison of a reference function
$f$ (blue) with three different candidate functions (green) for three different matching strategies. Each candidate is closest to $f$
under one of the three matchings: pointwise (left), elastic (middle), and partial elastic (right). The black lines indicate the time warping correspondences associated with each matching.}
\label{compare_dist}
\end{figure}

For example, Figure \ref{compare_dist} shows a reference function $f$ in blue together with three candidate functions $g_0$, $g_1$ and $g_2$ in green. 
Depending on the comparison method used, any of these candidate functions can be considered the best match for the reference function.
The function $g_0$ has one peak that matches the first peak of $f$. Since the peaks of the other functions are
misaligned, this first function is closest to the reference function with respect to the $L^2$ distance. 
However, the second candidate exhibits the same number of peaks as the 
reference, with comparable amplitudes but shifted in time; after elastic 
matching, it becomes closest to the reference.  Finally,
the first two peaks of the function $g_2$ exactly match the peaks of the reference function. 
The reference could be interpreted as having one additional peak if observed over a longer 
time interval.
A framework accounting for this possibility, which we call partial elastic matching, makes $g_2$ closest to the reference.
Such configurations arise in many experimental conditions when the recording time window is not fully controlled.

In the related field of functional curve alignment, Srivastava \textit{et al.}~\cite{srivastava2011} introduced a method for elastic matching, further developed in the book \cite{fdabook}. This method decomposes a function into two components: 
an amplitude component, which encodes the shape of the function, and a phase component, which encodes the time warping required to match the function to a template. It was generalized by
Bryner and Srivastava~\cite{cut} 
to partial elastic matching, where the time domains of the functions to be compared need not share the same endpoints. 
This generalization enables the matching of a broader class of functions and allows the calibration of phenomena that are only partially observed.

Several articles use an amplitude/phase decomposition for the calibration of simulation codes. Kleiber \textit{et al.}~\cite{kleiber2014model} introduced a method combining model calibration with landmark-based image warping.
Guo \textit{et al.}~\cite{guo2022} propose to perform calibration on the distance to the experiments rather than directly in the amplitude/phase space, which simplifies the calibration problem; however, their procedure does not account 
 for amplitude and phase discrepancies separately. Xie \textit{et al.}~\cite{xie2024functional} use an elastic surrogate model for calibration but ignore phase discrepancies. Finally, Francom \textit{et al.}~\cite{francom} propose to perform 
 the amplitude/phase decomposition as a preprocessing step and then to address two separate calibration problems, one in the amplitude space and one in the phase space. This two-step approach greatly reduces the computational 
 burden compared to the fully Bayesian methods of Cheng~\cite{cheng2024}, since it does not require sampling the phase discrepancy.

In this work, we present an insightful reinterpretation of the preprocessing step of Francom \textit{et al.}~\cite{francom} through the lens of a more general Bayesian model. 
This perspective not only naturally introduces an alignment framework as an approximation of the full posterior, but also clearly highlights the sources of uncertainty that are typically overlooked.
 Building on this foundation, we further enrich the phase space by integrating the partial elastic matching approach of Bryner and Srivastava~\cite{cut} into the elastic calibration framework of Francom \textit{et al.}~\cite{francom}, 
 thereby introducing valuable new degrees of freedom. This enhanced framework achieves more accurate alignment and significantly broadens the applicability of elastic calibration to a diverse 
 range of time-dependent calibration problems. In the examples studied, these advancements lead to improved surrogate model performance and a more refined treatment of non-additive discrepancies, 
 ultimately yielding more precise and robust posterior distributions for the calibration parameters.

In Section~\ref{section2}, we recall the main tools of Bayesian calibration. We then characterize, in Section~\ref{section3}, the approximation underlying the elastic calibration method of Francom \textit{et al.}~\cite{francom}.
 In Section~\ref{section4}, we generalize this approach to handle partial phases. 
Finally, in Section~\ref{section5}, we illustrate the proposed method on two calibration problems: one involving synthetic data, the other based on real data.

\section{Bayesian calibration with additive discrepancy}
\label{section2}
\subsection{Calibration with scalar outputs}

Kennedy and O'Hagan~\cite{kennedy2001bayesian} propose a Bayesian method for calibrating computer codes. 
To calibrate the parameters, we seek a relation between the experimental outputs $y_{\mathrm{exp}}$ associated with the experimental settings $x$,
 and the outputs of a simulation code $y_{\mathrm{sim}}(x,\beta)$
 which depend on the experimental settings $x$ and unknown parameters $\beta$ that must be calibrated. 
Kennedy and O'Hagan~\cite{kennedy2001bayesian} assume that the outputs of the experiment and the simulation are scalar, $y_{\mathrm{exp}}, y_{\mathrm{sim}}(x,\beta)\in\mathbb{R}$, 
 and adopt the following relation between an experiment observed at the experimental setting $x$ and the simulation code:

\begin{equation}
y_{\mathrm{exp}} = y_{\mathrm{sim}}(x,\beta) + \delta(x) + \varepsilon.
\label{calibration}
\end{equation}
Here, $\varepsilon\sim \mathcal{N}(0,\sigma^2)$ is a measurement error on the experiment, and 
$\delta(x)$ represents a model error of the simulation code, commonly referred to as the discrepancy,
 whose prior distribution is a Gaussian process with hyperparameters $\theta_{\delta}$.

Let $(x_i)_{i=1,\ldots,m}$ denote the $m$ known experimental settings. At each experimental setting $x_i$, an experimental observation $y_{\mathrm{exp},i}$ is available. 
A fully Bayesian framework requires specifying prior distributions on $\beta$, $\sigma^2$, and $\theta_{\delta}$,
which are assumed here independent, in order to compute the posterior distribution
\begin{equation}
p(\beta,\sigma^2,\theta_{\delta} \mid (y_{\mathrm{exp},i})_i) \propto p((y_{\mathrm{exp},i})_i \mid \beta,\sigma^2,\theta_{\delta}) \, p_{\mathrm{prior}}^{\beta}(\beta) \, p_{\mathrm{prior}}^{\varepsilon}(\sigma^2) \, p_{\mathrm{prior}}^{\delta}(\theta_{\delta}).
\label{posterior_scalar}
\end{equation}

Higdon \textit{et al.}~\cite{higdon2004combining} sample the posterior distribution \eqref{posterior_scalar} in a fully Bayesian framework,
 which is computationally demanding. Kennedy and O'Hagan~\cite{kennedy2001bayesian} propose a more efficient plug-in approach, in which the hyperparameters $(\sigma^2, \theta_{\delta})$ 
 are first estimated and the calibration parameters $\beta$ are then sampled. In this article, we adopt the plug-in strategy, neglecting the uncertainties 
 in the hyperparameters and fixing them at the maximum of the posterior distribution.

\subsection{Calibration with time-series outputs}
\label{subcalib}

In this article, we study calibration problems with time-dependent outputs \cite{perrin, higdon2008}. 
Observations are available at different experimental settings $(x_i)_i$ and cover a dense, fixed temporal grid. 
The time variable $t$ can be incorporated into Equation~\eqref{calibration}:
\begin{equation}
y_{\mathrm{exp}}(t) = y_{\mathrm{sim}}(x,\beta,t) + \delta(x,t) + \varepsilon(t).
\label{calibration_t}
\end{equation}
The discrepancy $\delta(x,t)$ depends on both the time $t$ and the experimental settings $x$~\cite{higdon2008}.
In our examples, since we have access to only a few experimental settings, the dependence of the discrepancy on $x$ cannot be reliably estimated.
We therefore model the experiments as independent.
For each experimental setting $(x_i)_{i=1,\ldots,m}$, the discrepancy is modeled as a stationary Gaussian process (GP) in time:
$\delta(x_i,\cdot)\sim \mathcal{GP}(0,k_{\delta})$.
After discretizing time on a grid $\bm{t}=(t_k)_{k=1,\ldots,N_t}$, the relation may be written in vector form for each experimental setting $x_i$:
\begin{equation}
y_{\mathrm{exp},i}(\bm{t}) = y_{\mathrm{sim}}(x_i,\beta,\bm{t}) + \delta(x_i,\bm{t}) + \varepsilon_i(\bm{t}),
\label{calibration_f}
\end{equation}
where
 $y_{\mathrm{exp},i}(\bm{t}) = (y_{\mathrm{exp},i}(t_k))_{k=1,\ldots,N_t}$, $y_{\mathrm{sim}}(x_i,\beta,\bm{t}) = (y_{\mathrm{sim}}(x_i,\beta,t_k))_{k=1,\ldots,N_t}$,
 $\delta(x_i,\bm{t})= $\\ $(\delta(x_i,t_k))_{k=1,\ldots,N_t} \sim \mathcal{N}(\mathbf{0}, \Sigma_{\delta})$, with $\Sigma_{\delta} = (k_{\delta}(t_k,t_l))_{k,l=1,\ldots,N_t}$
 and $\varepsilon_i(\bm{t}) \sim \mathcal{N}(\mathbf{0}, \sigma^2 \mathbf{I}_{N_t})$.

\begin{remark}
 Other modeling choices are possible for the discrepancy. For instance, Francom \textit{et al.}~\cite{francom} work with discrepancies expressed as a sum of a small number of basis functions,
whereas Brown \textit{et al.}~\cite{brown} use a white noise discrepancy normalized by the effective sample size so that the likelihood does not decrease with the time resolution.
\end{remark}

For computationally expensive computer codes, direct sampling from the posterior is infeasible, as it would require too many simulator evaluations. 
A surrogate model is therefore constructed as an inexpensive approximation of the simulator. For each experiment, we build a surrogate model of $y_{\mathrm{sim}}(x_i,\cdot,\bm{t})$, 
denoted $\tilde{y}_{\mathrm{sim}}(x_i,\cdot,\bm{t})$, using Gaussian process regression (GPR)~\cite{williams2006gaussian,gramacy2020surrogates}
with a training dataset of simulations.
Perrin~\cite{perrin} distinguishes two families of GP surrogates for time-dependent outputs: 
those based on separable covariance functions, which exploit the Kronecker structure of the covariance matrix, and
 those relying on a dimension-reduction step followed by independent GP regression in a low-dimensional latent space. In this article, 
 we adopt the second approach, using principal component analysis (PCA) for the dimension reduction. 
 This choice is especially relevant for problems where only a few components are needed to obtain a good approximation of the data \cite{higdon2008}.

Since the measurement error $\varepsilon$, the discrepancy $\delta$ and the surrogate model $\tilde{y}_{\mathrm{sim}}$ are assumed Gaussian and independent,
they can be combined into a single multivariate Gaussian distribution.
Let $\tilde{\mu}^y$ and $\tilde{\Sigma}^y$ denote the mean and covariance of the surrogate model given the training data. Under the previous assumptions, 
the likelihood for one experimental setting $x_i$ can be written as
\begin{equation}
y_{\mathrm{exp},i}(\bm{t}) \mid \beta,\sigma^2,\theta_{\delta} \sim \mathcal{N}\bigl(\tilde{\mu}^y(x_i,\beta,\bm{t}),\, \tilde{\Sigma}^y(x_i,\beta) + \Sigma_{\delta} + \sigma^2 \mathbf{I}_{N_t}\bigr).
\end{equation}
One can then apply Bayes' rule to compute the posterior distribution with Equation~\eqref{posterior_scalar}.

\section{Elastic Bayesian calibration}
\label{section3}

In this section, we first introduce and motivate a Bayes\-ian calibration method accounting for a non-additive discrepancy. We
then propose an efficient approximation to this method, and show that this approximation admits a likelihood similar to that of Francom \textit{et al.}~\cite{francom}.

\subsection{Bayesian calibration with phase discrepancy}
\label{sec:bayes-warp}

In time-dependent calibration tasks, errors may arise along the vertical axis (additive discrepancy) but also along the horizontal axis. 
We refer to this second type of error as the phase discrepancy. Its definition relies on tools originally developed for time-series alignment 
\cite{srivastava2011, fdabook}, which are summarized in Appendix~\ref{align}.

 For notational simplicity, we rescale the time domain to the unit interval $[0,1]$. 
 The extension to an arbitrary interval $[T_0, T_f]$ is straightforward. Following Srivastava \textit{et al.}\cite{srivastava2011}, we define the time warping group
\begin{equation*}
\Gamma = \{\gamma : [0,1] \to [0,1] \mid \gamma(0)=0,\, \gamma(1)=1,\, \gamma \text{ is a }\mathcal{C}^1\text{-diffeomorphism}\}.
\end{equation*}

We extend the calibration relation~\eqref{calibration_f} by introducing a time warping $\gamma$ that models a discrepancy on the time axis in addition to the additive discrepancy $\delta$:
\begin{equation}
y_{\mathrm{exp},i}(\bm{t}) = y_{\mathrm{sim}}(x_i,\beta,\gamma(x_i,\bm{t})) + \delta(x_i,\bm{t}) + \varepsilon_i(\bm{t}), \quad \varepsilon_i(\bm{t}) \sim \mathcal{N}(\mathbf{0},\sigma^2 \mathbf{I}_{N_t}),
\label{calibration_phasediscr}
\end{equation}
where $\gamma(x_i,\cdot) \in \Gamma$ warps the simulator time axis for experiment $i$.

We now specify a prior on $\gamma(x_i,\cdot)$. As argued by Francom \textit{et al.}~\cite{francom}, 
the complex geometry of the phase space $\Gamma$ makes it convenient to first transform it into a vector space of so-called shooting vectors $v$, defined in
$
V= \left\{ v \in L^2([0,1]) \,\bigg|\, \int_{0}^{1} v(t)\,\mathrm{d}t = 0 \right\}
$
(see Appendix~\ref{shooting} for details). Following previously proposed Bayesian alignment frameworks \cite{lu2017,tucker2021}, we adopt a Gaussian process prior in the space of shooting vectors, $v(x_i,t) \sim \mathcal{GP}(0,k^v)$,
with hyperparameters $\theta_{v}$, 
with discretized counterpart $v(x_i,\bm{t}) \sim \mathcal{N}(\mathbf{0},\Sigma_v)$, where $\Sigma_v = (k^v(t_k,t_l))_{k,l=1,\ldots,N_t}$. 

In this section, the error hyperparameters $(\sigma^2,\theta_{\delta},\theta_v)$ are assumed to be known.
We can apply Bayes' rule to obtain the posterior distribution:
\begin{equation}
p\bigl(\beta,(v_i(\bm{t}))_i \mid (y_{\mathrm{exp},i}(\bm{t}))_i\bigr) \propto p_{\mathrm{prior}}^{\beta}(\beta) \prod_i p\bigl(y_{\mathrm{exp},i}(\bm{t}) \mid \beta,v_i(\bm{t})\bigr)\, p_{\mathrm{prior}}^{v}(v_i(\bm{t})).
\label{posterior_phasediscr}
\end{equation}

Since, unlike the additive discrepancy, the phase discrepancy cannot be integrated out, it must be sampled jointly with the calibration parameters, 
resulting in a higher-dimensional (and therefore more challenging) sampling problem than the one without phase discrepancy.
 Moreover, the two discrepancies $\delta$ and $\delta_v$ can
both capture differences between the simulation code and the experiment, which may exacerbate identifiability issues. In the next subsection, we propose an approximation
of this posterior distribution that enables us to formulate a calibration problem similar to that of Francom \textit{et al.}~\cite{francom} and mitigates both issues.

\subsection{Approximation of the phase discrepancy}
\label{ela_approx}

To avoid jointly sampling the shooting vectors $(v_i(\bm{t}))_i$ together with the
calibration parameters $\beta$, we assume that the conditional distribution of
$v_i(\bm{t}) \mid \beta, y_{\mathrm{exp},i}(\bm{t})$ is Gaussian:
\begin{equation}
    v_i(\bm{t})|\beta,y_{\mathrm{exp},i}(\bm{t}) \sim \mathcal{N}(v_{\mathrm{sim},i}(\beta,\bm{t}),S_{\beta,y_{\mathrm{exp},i}})
    \label{approx_alignMAP}
\end{equation}
where 
\begin{equation}
    v_{\mathrm{sim},i}(\beta,\bm{t})=\argmax_{v_i \in V} p(v_i(\bm{t})|\beta,y_{\mathrm{exp},i}(\bm{t}))
\label{opti_align}
\end{equation} 
is the mode of the distribution 
and $S_{\beta,y_{\mathrm{exp},i}}$ a covariance matrix.
One may notice that setting $S_{\beta,y_{\mathrm{exp},i}}$ to the inverse of the Hessian of
$-\log p(v_i(\bm{t}) \mid \beta, y_{\mathrm{exp},i}(\bm{t}))$,
evaluated at $v_{\mathrm{sim},i}(\beta,\bm{t})$, recovers the classical Laplace
approximation.

The optimization problem \eqref{opti_align} defines, for each experiment, an alignment problem in which a simulation output is aligned to the associated experimental output: 
\begin{align}
v_{\mathrm{sim},i}(\beta,\bm{t}) &= \argmax_{v_i \in V} \, p\bigl(y_{\mathrm{exp},i}(\bm{t}) \mid \beta,v_i(\bm{t})\bigr) \, p_{\mathrm{prior}}^{v}(v_i(\bm{t})) \nonumber \\
&= \argmin_{v_i \in V} \, \lVert y_{\mathrm{exp},i}(\bm{t}) - y_{\mathrm{sim}}(x_i,\beta,\gamma_i(\bm{t})) \rVert^2_{\Sigma_{\delta} + \sigma^2 \mathbf{I}_{N_t}} + \lVert v_i(\bm{t}) \rVert^2_{\Sigma_v},
\label{opti_alignMAP}
\end{align}
where $\gamma_i$ is the phase associated with the shooting vector $v_i$.

\begin{figure}[h]
    \centering
    \begin{subfigure}{0.49\textwidth}
        \centering
        \includegraphics[width=0.9\textwidth]{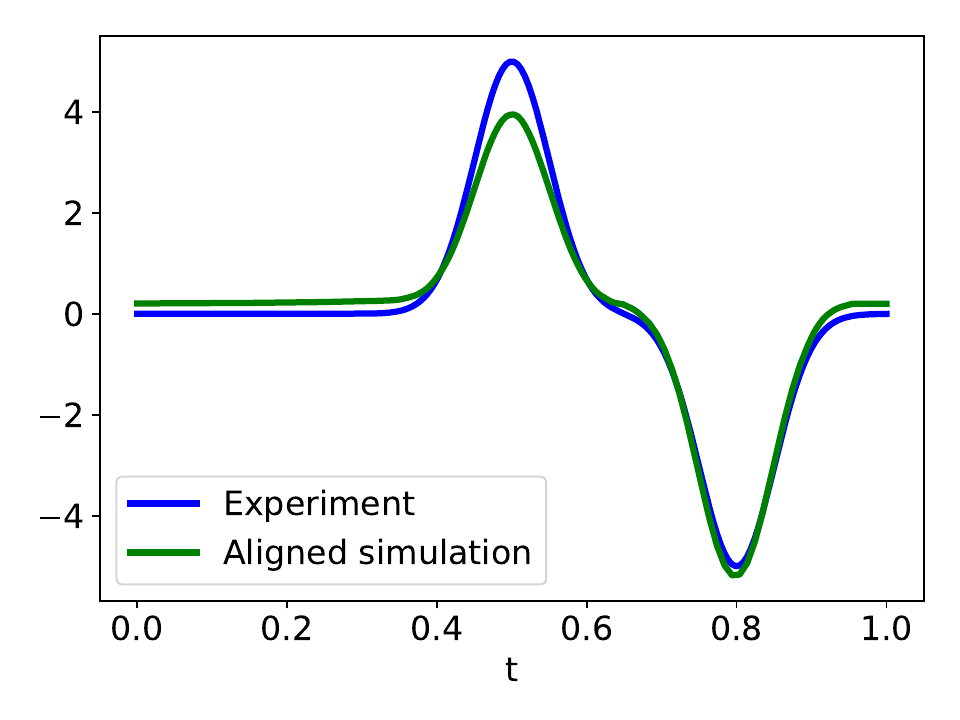}
        \caption{Amplitude}
        \label{fig:amplitudeMAP}
    \end{subfigure}
    \hfill
    \begin{subfigure}{0.49\textwidth}
        \centering
        \includegraphics[width=0.7\textwidth]{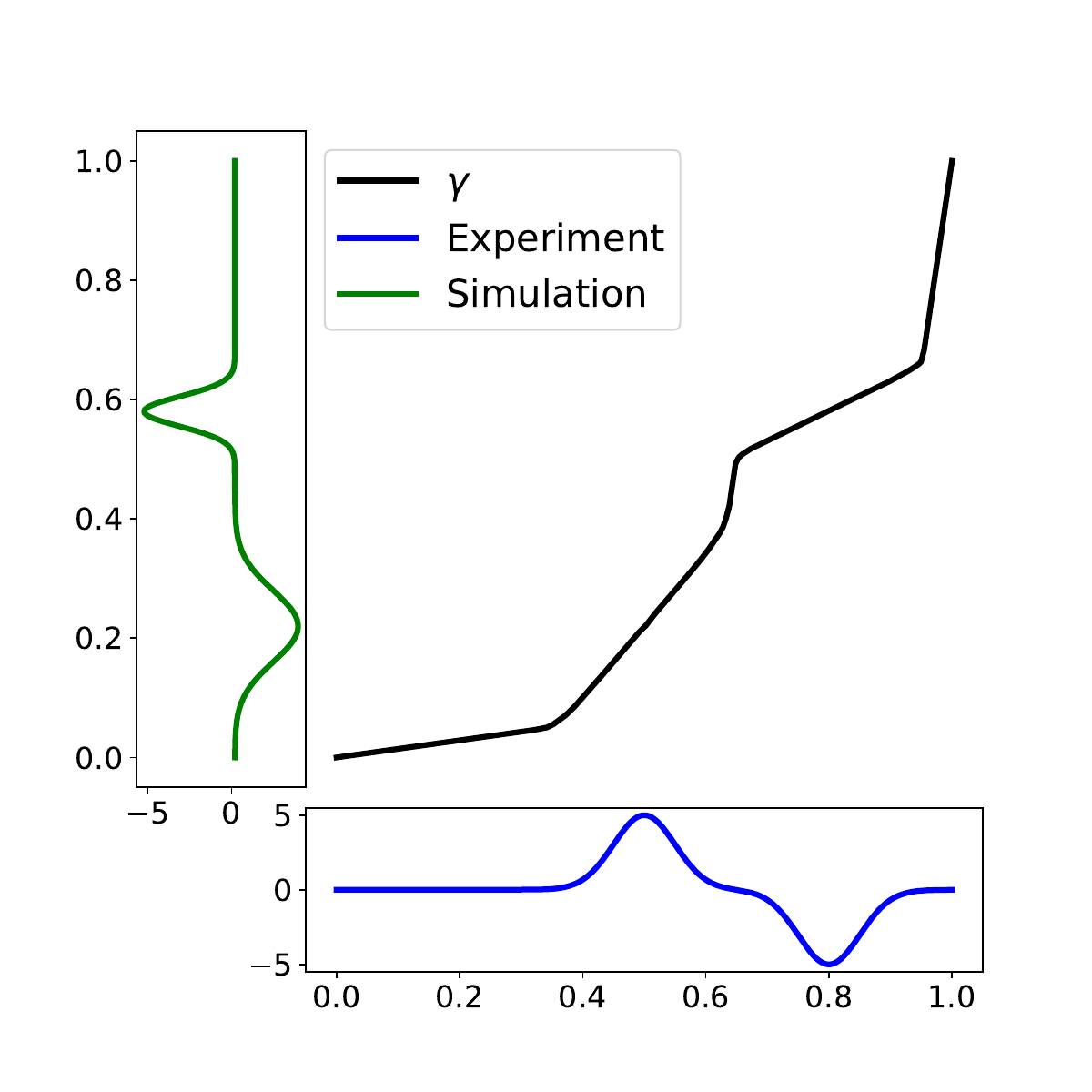}
        \caption{Phase}
        \label{fig:phaseMAP}
    \end{subfigure}
    \caption{Alignment of a simulation on an experiment.}
    \label{fig:distancesMAP}
\end{figure}

With the optimization problem \eqref{opti_alignMAP}, we can define the shooting vector $v_{\mathrm{sim},i}(\beta,\bm{t})$, its associated phase $\gamma_{\mathrm{sim},i}(\beta,\bm{t})$,
and its amplitude $z_{\mathrm{sim},i}(\beta,\bm{t}) = y_{\mathrm{sim}}(x_i,\beta,\gamma_{\mathrm{sim},i}(\beta,\bm{t}))$.
Examples of an amplitude and a phase are shown in Figures~\ref{fig:amplitudeMAP} and \ref{fig:phaseMAP}.
If we consider the alignment of an experimental output on itself, we can define the associated phase $\gamma_{\mathrm{exp},i}(\bm{t})=\bm{t}$,
shooting vector $v_{\mathrm{exp},i}(\bm{t}) = \mathbf{0}$, and amplitude $z_{\mathrm{exp},i}(\bm{t}) = y_{\mathrm{exp},i}(\bm{t})$.

Assumption~\eqref{approx_alignMAP} is a particular case of the one introduced by Kahol \textit{et al.}~\cite{kahol2025efficient}
to avoid sampling hyperparameters. Following similar calculations, we obtain
\begin{align*}
p(\beta|(y_{\mathrm{exp},i}&(\bm{t}))_i)=
\frac{p((v_{\mathrm{sim},i}(\beta,\bm{t}))_i,\beta|(y_{\mathrm{exp},i}(\bm{t}))_i)}{p((v_{\mathrm{sim},i}(\beta,\bm{t}))_i|\beta,(y_{\mathrm{exp},i}(\bm{t}))_i)}\\
&=\frac{p((v_{\mathrm{sim},i}(\beta,\bm{t}))_i,\beta|(y_{\mathrm{exp},i}(\bm{t}))_i)}{\max_{(v_i)_i \in V} p((v_i(\bm{t}))_i|\beta,(y_{\mathrm{exp},i}(\bm{t}))_i)}\\
&= \frac{p((v_{\mathrm{sim},i}(\beta,\bm{t}))_i,\beta|(y_{\mathrm{exp},i}(\bm{t}))_i)}{\prod_i \max_{v_i \in V} p(v_i(\bm{t})|\beta,y_{\mathrm{exp},i}(\bm{t}))}\\
&\propto p((v_{\mathrm{sim},i}(\beta,\bm{t}))_i,\beta|(y_{\mathrm{exp},i}(\bm{t}))_i)\prod_i \mathrm{det} (S_{\beta,y_{\mathrm{exp},i}})^{1/2}\\
&\propto p((y_{\mathrm{exp},i}(\bm{t}))_i|(v_{\mathrm{sim},i}(\beta,\bm{t}))_i,\beta)p_{\mathrm{prior}}^{\beta}(\beta) p_{\mathrm{prior}}^{v}((v_{\mathrm{sim},i}(\beta,\bm{t}))_i) \prod_i \mathrm{det}  (S_{\beta,y_{\mathrm{exp},i}})^{1/2}\\
& \propto p_{\mathrm{prior}}^{\beta}(\beta) \prod_i p(y_{\mathrm{exp},i}(\bm{t})|v_{\mathrm{sim},i}(\beta,\bm{t}),\beta)p_{\mathrm{prior}}^{v}(v_{\mathrm{sim},i}(\beta,\bm{t})) \mathrm{det}  (S_{\beta,y_{\mathrm{exp},i}})^{1/2}\\
\end{align*}

The prior $p_{\mathrm{prior}}^{v}(v_{\mathrm{sim},i}(\beta,\bm{t}))=\mathcal{N}\bigl(v_{\mathrm{sim},i}(\beta,\bm{t}) \mid \mathbf{0}, \Sigma_v\bigr)$ 
can equivalently be written as the likelihood $\mathcal{N}\bigl(v_{\mathrm{exp},i}(\bm{t})  \mid v_{\mathrm{sim},i}(\beta,\bm{t}), \Sigma_v\bigr)$, 
in which $v_{\mathrm{exp},i}(\bm{t}) = \mathbf{0}$ acts as a pseudo-observation.
We can now write:
\begin{equation}
\begin{split}
    p(\beta|(y_{\mathrm{exp},i}(\bm{t}))_i)\propto p_{\mathrm{prior}}^{\beta}(\beta)& \prod_i \mathcal{N}\bigl(z_{\mathrm{exp},i}(\bm{t})\mid z_{\mathrm{sim},i}(\beta,\bm{t}),\Sigma_\delta+\sigma^2 \mathbf{I}_{N_t}\bigr) \\
    &\times \mathcal{N}\bigl(v_{\mathrm{exp},i}(\bm{t}) \mid v_{\mathrm{sim},i}(\beta,\bm{t}), \Sigma_v\bigr)\,
    \mathrm{det}(S_{\beta,y_{\mathrm{exp},i}})^{1/2}
\end{split}
\label{posterior_prod}
\end{equation}

Moreover, assuming that $S_{\beta,y_{\mathrm{exp},i}}$ does not depend on $\beta$,
the posterior distribution~\eqref{posterior_prod} can be rewritten with the same prior on $\beta$ and two independent likelihoods, one for the amplitude and one for the shooting vector:
\begin{align}
z_{\mathrm{exp},i}(\bm{t}) &= z_{\mathrm{sim},i}(\beta,\bm{t}) + \delta(x_i,\bm{t}) + \varepsilon_i(\bm{t}), \label{calibration_amp} \\
v_{\mathrm{exp},i}(\bm{t}) &= v_{\mathrm{sim},i}(\beta,\bm{t}) + \delta_v(x_i,\bm{t}), \label{calibration_pha}
\end{align}
where $\delta_v(x_i,\bm{t}) \sim \mathcal{N}(\mathbf{0},\Sigma_v)$.
In other words, the prior $p_{\mathrm{prior}}^{v}$ has been replaced by the likelihood~\eqref{calibration_pha}.
Under this interpretation, we formally recover the likelihood of Francom \textit{et al.}~\cite{francom}, 
though with a different alignment procedure (see Section~\ref{discussion}). 
Following their terminology, we refer to methods combining a likelihood in the amplitude space with one in the shooting vector space as elastic calibration. 
Both their method and the one introduced in this section fall into this class.

The decomposition of the simulator output into amplitude and shooting vector should in principle be available for any $\beta$.
 Since this is not feasible for computationally expensive codes, we build in practice surrogate models in the amplitude and shooting vector spaces, based on solutions of the optimization problem~\eqref{opti_alignMAP} 
 obtained with the L-BFGS algorithm \cite{liu1989limited} for a dataset of simulator outputs. These surrogate models, denoted $\tilde{z}_{\mathrm{sim},i}(\cdot,\bm{t})$ and $\tilde{v}_{\mathrm{sim},i}(\cdot,\bm{t})$, 
 are constructed as described in Section~\ref{subcalib} and replace the true amplitude $z_{\mathrm{sim},i}(\beta,\bm{t})$ and the true shooting vector $v_{\mathrm{sim},i}(\beta,\bm{t})$ 
 in the Equations~\eqref{calibration_amp} and \eqref{calibration_pha}:
 \begin{equation}
\begin{split}
    p_{\mathrm{elastic}}(\beta|(y_{\mathrm{exp},i}(\bm{t}))_i)&\propto p_{\mathrm{prior}}^{\beta}(\beta) \prod_i 
    \mathcal{N}\bigl(z_{\mathrm{exp},i}(\bm{t})\mid \tilde{\mu}^z_{i}(\beta,\bm{t}),\tilde{\Sigma}^z_{i}(\beta) +\Sigma_\delta +\sigma^2 \mathbf{I}_{N_t}\bigr) \\
    &\times \mathcal{N}\bigl(v_{\mathrm{exp},i}(\bm{t}) \mid \tilde{\mu}^v_{i}(\beta,\bm{t}), \tilde{\Sigma}^v_{i}(\beta) +\Sigma_v\bigr)\,.
\end{split}
\label{posterior_elastic}
\end{equation}

As described in Section~\ref{subcalib}, the posterior distribution of $\beta$
can be sampled once the priors have been specified and the hyperparameters estimated.

\subsection{Discussion}
\label{discussion}
\paragraph{Benefits of the elastic calibration framework}
Elastic calibration methods share three main advantages. 
First, it simplifies surrogate modeling.
Indeed, Xie \textit{et al.}~\cite{xie2024functional} show on several examples that we can capture $99\%$ of the total variance 
of the simulation data with fewer principal components
within an amplitude-phase representation than within the original space for some applications.
Note that surrogate models are used here not only to replace computationally expensive codes, but also to avoid solving problem~\eqref{opti_alignMAP} for each new value of 
$\beta$.

Moreover, this method accounts for both an amplitude discrepancy $\delta(x_i,\bm{t})$ and a phase discrepancy $\delta_v(x_i,\bm{t})$.
The second benefit of this framework is therefore to separate these discrepancies to be able to independently include, 
exclude, or constrain them depending on which sources of modeling error are physically 
plausible for the application.

Finally, in contrast to the posterior~\eqref{posterior_phasediscr}, the separation in amplitude~\eqref{calibration_amp} and shooting vector~\eqref{calibration_pha} implies that the phase
discrepancy no longer needs to be jointly sampled with $\beta$, resulting in a lower-dimensional, and therefore
simpler, sampling problem.

\paragraph{Differences with Francom \textit{et al.}~\cite{francom}}
The main difference between the method introduced in this section and the approach of Francom \textit{et al.}~\cite{francom} is that we view alignment as an approximation of a more 
complete Bayesian model, whereas they rely on an alignment method based on the widely used SRVF framework of Srivastava \textit{et al.}~\cite{srivastava2011}, 
which is not derived from the calibration framework. 
A direct benefit of this Bayesian reinterpretation is that the alignment penalty is no longer a free parameter to be tuned by trial and error, 
as in their work; instead, the penalization weights are determined by the same hyperparameters that govern the calibration errors, 
eliminating an otherwise arbitrary choice. 

Moreover, we align the simulations to the experiment rather than to a common template, since using a template would bias the approximation derived here. 
Because our alignment method does not involve derivatives, unlike theirs, it is less sensitive to noise and therefore does not require introducing a template.
Finally, in contrast to the work of Francom \textit{et al.}~\cite{francom}, our methodology can easily be generalized to
other kinds of non-additive discrepancies.
In particular, in the next section, we will show how our method can naturally be generalized to partial phases.

\section{Bayesian calibration with partial elastic matching}
 \label{section4}

The method of Section~\ref{section3} allows for time deformations within a
fixed time interval, since the warpings $\gamma \in \Gamma$ satisfy
$\gamma(0)=0$ and $\gamma(1)=1$. This implicitly assumes that time
discrepancies between the simulation and the experimental output can only
occur in the interior of the interval, not at its beginning or end. The
condition $\gamma(0)=0$ is generally satisfied in practice, since the time
at which the physical phenomenon begins is usually known and can be set to
$0$. However, the condition $\gamma(1)=1$ is more restrictive: the duration of the
simulated phenomenon may depend on the calibration parameter $\beta$ and may not match
the end time of the experiment, so that a time discrepancy should also be allowed at the end of the
interval, not only in its interior.

We show in this section that the Bayesian approximation adopted in Section~\ref{section3} can be naturally generalized to
relax the constraint $\gamma(1)=1$.
The extension to variable initial times is 
straightforward. 
Throughout this section, the experimental outputs are observed over the interval $[0, 1]$, 
whereas the simulation code is evaluated over a larger interval $[0, t_f]$, with $t_f$ large enough
to make sure that the final time associated with each simulation lies within the interval.

\subsection{Partial phases}

 Bryner and Srivastava~\cite{cut} introduced an elastic framework with
variable end time, based on the SRVF alignment framework (see
Appendix~\ref{partial}). Their set of partial phases allows us to
generalize the calibration method of Section~\ref{section3}.
  We consider here the special case in which only one end time varies, which leads to the set of partial phases
\begin{equation*}
\begin{split}
\Gamma_{\mathrm{p}} = \{\gamma : [0,1] \to [0,t_f] \mid {} & \gamma(0) = 0,\, 0 < \gamma(1) \leq t_f, \\
& \gamma \text{ is differentiable with strictly positive continuous derivative}\}.
\end{split}
\end{equation*}

We observe that the map $\Gamma \times (0,t_f] \to \Gamma_{\mathrm{p}}$,
$(\gamma,s) \mapsto \gamma_{\mathrm{p}} = s\gamma$, is a bijection. Consequently, any partial phase $\gamma_{\mathrm{p}}$ can be uniquely represented as a phase $\gamma \in \Gamma$ paired with a scalar $s = \gamma_{\mathrm{p}}(1) \in (0, t_f]$, which acts as a rescaling factor.

We can now write a calibration relation analogous to Equation~\eqref{calibration_phasediscr} but with partial phases:
\begin{equation}
y_{\mathrm{exp},i}(\bm{t}) = y_{\mathrm{sim}}(x_i,\beta,\gamma_{\mathrm{p}}(x_i,\bm{t})) + \delta(x_i,\bm{t}) + \varepsilon_i(\bm{t}), \quad \varepsilon_i(\bm{t}) \sim \mathcal{N}(\mathbf{0},\sigma^2 \mathbf{I}_{N_t}),
\label{calibration_phasediscr_partial}
\end{equation}
where $\gamma_p(x_i,\cdot) \in \Gamma_p$ is a partial phase that maps part of the simulation onto the whole experimental output.

The Bayesian framework requires a prior on $\gamma_{\mathrm{p}}(x_i, \bm{t})$. Using the 
bijection above, we equivalently specify a prior on $(\gamma,s) \in \Gamma \times (0, t_f]$
and assume prior independence between $\gamma$ and $s$.
For $\gamma$, we retain the previously introduced Gaussian process prior on the associated shooting vector, 
$v(x_i, t) \sim \mathcal{GP}(0, k^v)$.
For the rescaling parameter $s$, we want $\gamma_{\mathrm{p}}$ to remain close to the identity, i.e.,
$s$ close to $1$. We therefore adopt a Gaussian prior centered at $1$ and truncated to the admissible
interval: $s \sim \mathcal{N}(1, \sigma_s^2)$, truncated to $s \in (0, t_f]$. 
The posterior distribution~\eqref{posterior_phasediscr} can then be rewritten with this prior to accommodate partial phases.

\subsection{Partial elastic Bayesian calibration}
\label{partial_elastic}

Following the same approach as in Section~\ref{ela_approx}, we assume that the conditional distribution of
$(v_i(\bm{t}),s_i) \mid \beta, y_{\mathrm{exp},i}(\bm{t})$ is Gaussian:
\begin{equation}
    (v_i(\bm{t}),s_i)|\beta,y_{\mathrm{exp},i}(\bm{t}) \sim \mathcal{N}((v_{\mathrm{sim},i}(\beta,\bm{t}),s_{\mathrm{sim},i}(\beta)),S_{\beta,y_{\mathrm{exp},i}})
    \label{approx_partialalignMAP}
\end{equation}
where
\begin{equation}
    (v_{\mathrm{sim},i}(\beta,\bm{t}),s_{\mathrm{sim},i}(\beta))=\argmax_{\substack{v_i \in V \\ s_i \in (0, t_f]}} p(v_i(\bm{t}),s_i|\beta,y_{\mathrm{exp},i}(\bm{t}))
\label{opti_partialalign}
\end{equation} 
is the mode of the distribution 
and $S_{\beta,y_{\mathrm{exp},i}}$ a covariance matrix.

The optimization problem \eqref{opti_partialalign} defines, for each experiment, a partial 
alignment problem:
\begin{align}
(v_{\mathrm{sim},i}(\beta,\bm{t}),\, &s_{\mathrm{sim},i}(\beta)) = \argmax_{\substack{v_i \in V \\ s_i \in (0, t_f]}} \, p\bigl(y_{\mathrm{exp},i}(\bm{t}) \mid \beta,v_i,s_i\bigr) \, p_{\mathrm{prior}}^{v}(v_i(\bm{t})) \, p_{\mathrm{prior}}^{s}(s_i) \nonumber \\
&= \argmin_{\substack{v_i \in V \\ s_i \in (0, t_f]}} \, \lVert y_{\mathrm{exp},i}(\bm{t}) - y_{\mathrm{sim}}(x_i,\beta,\gamma_{\mathrm{p},i}(\bm{t})) \rVert^2_{\Sigma_{\delta} + \sigma^2 \mathbf{I}} + \lVert v_i(\bm{t}) \rVert^2_{\Sigma_v} + \lVert s_i - 1 \rVert^2_{\sigma_s^2},
\label{opti_partialalignMAP}
\end{align}
where $\gamma_{p,i}$ is the partial phase associated with the shooting vector $v_i$ and the rescaling parameter~$s_i$.

\begin{figure}[h]
    \centering
    \begin{subfigure}{0.49\textwidth}
        \centering
        \includegraphics[width=0.9\textwidth]{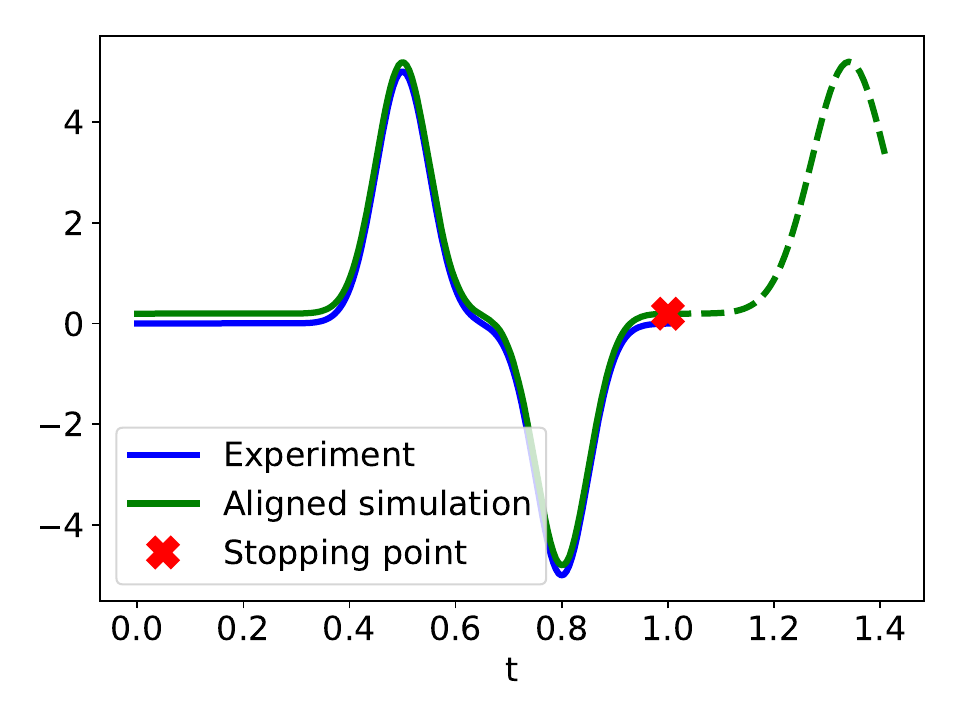}
        \caption{Partial amplitude}
        \label{fig:partialamplitudeMAP}
    \end{subfigure}
    \hfill
    \begin{subfigure}{0.49\textwidth}
        \centering
        \includegraphics[width=0.7\textwidth]{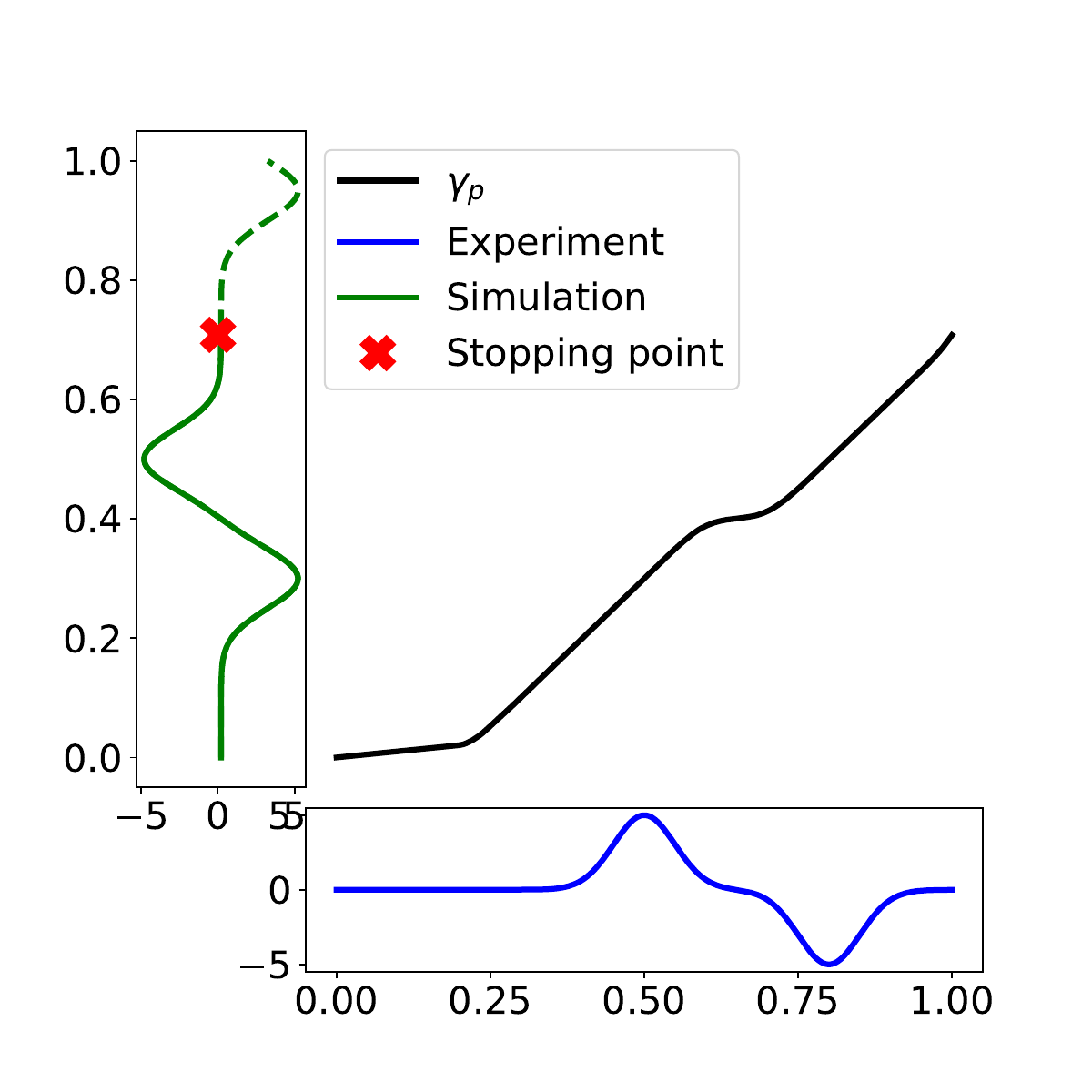}
        \caption{Partial phase}
        \label{fig:partialphaseMAP}
    \end{subfigure}
    \caption{Partial alignment of a simulation on an experiment}
    \label{fig:distancesMAP}
\end{figure}

The solution $(v_{\mathrm{sim},i}(\beta, \bm{t}), s_{\mathrm{sim},i}(\beta))$ defines a 
partial phase $\gamma_{\mathrm{p},i}^{\mathrm{sim}}(\beta, \bm{t})$ and an associated partial 
amplitude $z_{\mathrm{sim},i}(\beta, \bm{t}) = y_{\mathrm{sim}}(x_i, \beta, \gamma_{\mathrm{p},i}^{\mathrm{sim}}(\beta, \bm{t}))$. 
Examples of a partial amplitude and a partial phase are shown in Figures~\ref{fig:partialamplitudeMAP} and~\ref{fig:partialphaseMAP}.

As in Section~\ref{ela_approx}, the prior 
$\mathcal{N}\bigl(s_{\mathrm{sim},i}(\beta) \mid 1, \sigma_s^2\bigr)$ can equivalently be 
written as the likelihood $\mathcal{N}\bigl(s_{\mathrm{exp},i} \mid s_{\mathrm{sim},i}(\beta), \sigma_s^2\bigr)$,
where $s_{\mathrm{exp},i}=1$ acts as a pseudo-observation representing the rescaling 
required to align the experiment with itself. Following the same idea as in Section~\ref{ela_approx},
we obtain:
\begin{equation}
\begin{split}
    p(\beta|(y_{\mathrm{exp},i}(\bm{t}))_i)\propto & p_{\mathrm{prior}}^{\beta}(\beta) \prod_i 
    \mathcal{N}\bigl(z_{\mathrm{exp},i}(\bm{t})\mid z_{\mathrm{sim},i}(\beta,\bm{t}),\Sigma_\delta+\sigma^2 \mathbf{I}_{N_t}\bigr) \\
    &\times \mathcal{N}\bigl(v_{\mathrm{exp},i}(\bm{t}) \mid v_{\mathrm{sim},i}(\beta,\bm{t}), \Sigma_v\bigr)\,
    \mathcal{N}\bigl(s_{\mathrm{exp},i} \mid s_{\mathrm{sim},i}(\beta), \sigma_s^2 \bigr)\,
    \mathrm{det}(S_{\beta,y_{\mathrm{exp},i}})^{1/2}
\end{split}
\label{posterior_prodpartial}
\end{equation}

Moreover, assuming that $S_{\beta,y_{\mathrm{exp},i}}$ does not depend on $\beta$,
if we keep the same prior on $\beta$,
the partial elastic posterior~\eqref{posterior_prodpartial} may be reinterpreted with three independent likelihoods, one in the amplitude space, one in the 
shooting vector space, and one for the scalar rescaling:
\begin{align}
z_{\mathrm{exp},i}(\bm{t}) &= z_{\mathrm{sim},i}(\beta,\bm{t}) + \delta(x_i,\bm{t}) + \varepsilon_i(\bm{t}), & \varepsilon_i(\bm{t}) &\sim \mathcal{N}(\mathbf{0},\sigma^2 \mathbf{I}), \label{calibration_elatf_amp} \\
v_{\mathrm{exp},i}(\bm{t}) &= v_{\mathrm{sim},i}(\beta,\bm{t}) + \delta_v(x_i,\bm{t}), & \delta_v(x_i,\bm{t}) &\sim \mathcal{N}(\mathbf{0},\Sigma_v), \label{calibration_elatf_pha} \\
s_{\mathrm{exp},i}&= s_{\mathrm{sim},i}(\beta) + \delta^s_i, & \delta^s_i &\sim \mathcal{N}(0,\sigma_s^2), \label{calibration_elatf_s}
\end{align}
where $\delta(x_i,\bm{t}) \sim \mathcal{N}(\mathbf{0},\Sigma_\delta)$.

 For computationally expensive codes, in each of the three
spaces, we build a surrogate model,
  denoted $\tilde{z}_{\mathrm{sim},i}(\cdot, \bm{t})$, $\tilde{v}_{\mathrm{sim},i}(\cdot, \bm{t})$
and $\tilde{s}_{\mathrm{sim},i}(\cdot)$, the last being a scalar Gaussian process.
These surrogate models are fitted on a dataset of simulator outputs aligned according to
the solution of Problem~\eqref{opti_partialalignMAP}, which is solved by combining a grid
search over $s_i$ with the L-BFGS algorithm for $v_i$. The surrogate models, with mean $\tilde{\mu}$ and covariance $\tilde{\Sigma}$, replace the true amplitude
$z_{\mathrm{sim},i}(\beta,\bm{t})$, the true shooting vector $v_{\mathrm{sim},i}(\beta,\bm{t})$
and the true rescaling parameter $s_{\mathrm{sim},i}(\beta)$ in Equations~\eqref{calibration_elatf_amp}, \eqref{calibration_elatf_pha} and \eqref{calibration_elatf_s}:
 
\begin{equation}
\begin{split}
    p_{\mathrm{partial}}(\beta|&(y_{\mathrm{exp},i}(\bm{t}))_i) \propto p_{\mathrm{prior}}^{\beta}(\beta) \prod_i 
    \mathcal{N}\bigl(z_{\mathrm{exp},i}(\bm{t})\mid \tilde{\mu}^z_{i}(\beta,\bm{t}),\tilde{\Sigma}^z_{i}(\beta) +\Sigma_\delta +\sigma^2 \mathbf{I}_{N_t}\bigr) \\
    &\times \mathcal{N}\bigl(v_{\mathrm{exp},i}(\bm{t}) \mid \tilde{\mu}^v_{i}(\beta,\bm{t}), \tilde{\Sigma}^v_{i}(\beta) +\Sigma_v\bigr)\,
    \mathcal{N}\bigl(s_{\mathrm{exp},i} \mid \tilde{\mu}^s_{i}(\beta), \tilde{\Sigma}^s_{i}(\beta)+\sigma_s^2 \bigr)\,.
\end{split}
\label{posterior_partial}
\end{equation}

Compared with the calibration relations of Section~\ref{ela_approx}, a third 
relation has been added for the rescaling parameter. Estimating $\sigma_s^2$ is more
 challenging than estimating the other hyperparameters: whereas the amplitude
  and shooting vector relations provide $N_t$ data points per experimental 
  setting, the rescaling relation provides only one. When few experiments are 
  available, $\sigma_s^2$ must therefore be set using expert knowledge or additional assumptions.

\begin{remark}
    \label{rmq_rescaling}
When the variance of the shooting vector prior tends to zero, the partial 
phase reduces to a pure rescaling: $\gamma_{\mathrm{s}}(t) = s\, t$ with $s \in (0, t_f]$. The set of admissible warpings becomes
$\Gamma_{\mathrm{s}} = \{\gamma_{\mathrm{s}} : t \mapsto s\, t \mid 0 < s \leq t_f\},$
and Equation~\eqref{calibration_elatf_pha} is dropped from the system of calibration relations. 
This special case is useful when only a time rescaling is physically meaningful. We refer to this method as calibration with rescaling.
\end{remark}

In this article, we have introduced three calibration methods with time warpings (elastic calibration, partial elastic calibration and calibration with rescaling).
The performance of elastic calibration relies on the simplification provided by the alignment transform.
 In practice, the choice among these three methods should be 
guided both by expert knowledge of how the phase is expected to vary 
and by visual comparison of the resulting amplitude/phase 
decompositions.

\subsection{Hyperparameter optimization}
\label{HP_opti}

We discuss here the optimization of the hyperparameters $(\sigma^2,\theta_{\delta},\theta_v,\sigma_s^2)$.
The same considerations carry over to the elastic calibration of Section~\ref{ela_approx}. In practice, physical knowledge is often available for some of them.
The error variance of the measurement instruments, for instance, may be known. The remaining ones are chosen so as to maximize a criterion, here the joint posterior.

Maximizing the joint posterior raises two difficulties. 
First, the likelihoods~\eqref{calibration_elatf_amp}--\eqref{calibration_elatf_s} depend on the solution of the optimization problem~\eqref{opti_partialalignMAP}, while
 both the likelihoods and that problem depend on the hyperparameters, leading to a bi-level optimization problem. 
Second, surrogate models are built in the amplitude and shooting vector spaces, so the calibration hyperparameters are needed to fit the surrogates, and the surrogates are needed to optimize the hyperparameters.
These difficulties do not arise in Francom \textit{et al.}~\cite{francom}, since the alignment problem is not derived from the calibration relations.

To address these issues, we propose a two-step procedure. In the first step, the alignment is performed using 
arbitrarily chosen hyperparameters, which allows the surrogate models to be fitted and the posterior to be 
maximized to obtain hyperparameter estimates.
 In the second step, we apply the full procedure, starting again from the alignment step, using the estimated hyperparameters.
The method is summarized in Algorithm~\ref{alg:calibration}. 
Note that the first step could be iterated multiple times, each iteration using the hyperparameter estimates obtained from the previous one.
In our experiments, however, additional iterations bring no clear improvement that would justify the extra computational cost. We therefore present only the two-step approach in Section~\ref{section5} and recommend it in practice.

\begin{algorithm}
\caption{Hyperparameter optimization and sampling for partial elastic calibration.}
\label{alg:calibration}
\begin{algorithmic}[1]
\REQUIRE Experimental data $(y_{\mathrm{exp},i}(\bm{t}))_i$, simulations $(y_{\mathrm{sim}}(x_i,\beta_j, \bm{t}))_{i,j}$, initial hyperparameters $\theta_0$
\STATE \textbf{Alignment.} For each $(x_i, \beta_j)$, compute the shooting vector $v_{\mathrm{sim},i}(\beta_j, \bm{t})$ and the rescaling parameter $s_{\mathrm{sim},i}(\beta_j)$ by solving~\eqref{opti_partialalignMAP} with hyperparameters $\theta_0$.
\STATE \textbf{Surrogate.} Fit the surrogate models $\tilde{z}_{\mathrm{sim},i}(\cdot, \bm{t})$, $\tilde{v}_{\mathrm{sim},i}(\cdot, \bm{t})$, and $\tilde{s}_{\mathrm{sim},i}(\cdot)$.
\STATE \textbf{Hyperparameter optimization.} Compute, based on Equation~\eqref{posterior_partial},
\[
(\hat{\beta}, \hat{\theta}) = \arg\max_{\beta,\theta} p_{\mathrm{partial}}(\beta,\theta|(y_{\mathrm{exp},i}(\bm{t}))_i)
\]
\STATE Repeat the \textbf{Alignment} and \textbf{Surrogate} steps with the updated hyperparameters $\hat{\theta}$.
\STATE \textbf{Sampling.} Draw samples from the posterior $p_{\text{partial}}\bigl(\beta \mid (y_{\mathrm{exp},i}(\bm{t}))_i, \hat{\theta}\bigr)$.

\end{algorithmic}
\end{algorithm}

\section{Applications}
\label{section5}

In this section, we illustrate the differences between the previously introduced calibration methods through two examples:
the first one compares the methods on synthetic data generated from an anharmonic 
pendulum model, while the second one applies the methodology of Section~\ref{section4} to calibrate an equation of state from experimental data.
Implementation details are provided in Appendix~\ref{imp_details}.

\subsection{Anharmonic pendulum}
\label{app_pendulum} 
We consider a pendulum that obeys the second-order ordinary differential equation:
\begin{equation}
\frac{\mathrm{d}^2\phi}{\mathrm{d}t^2} + \alpha_1 \frac{\mathrm{d}\phi}{\mathrm{d}t} + \alpha_2 (\phi - \alpha_0 \phi^3) = 0,
\label{eq_pendulum}
\end{equation}
where $\phi$ is the pendulum angle. 

Our goal is to calibrate the three parameters $\alpha_0 \in [0.5,0.98], \alpha_1 \in [0,0.3]s^{-1}$ and $\alpha_2 \in [2,4]s^{-2}$.
For notational simplicity, we introduce normalized parameters  $\beta \in [0,1]^3$ such that $\alpha_0 = 0.48\beta_0+0.5,\alpha_1 = 0.3\beta_1,\alpha_2=2\beta_2+2$.

We generate a single synthetic experiment, with $N_t=100$, corrupted by additive Gaussian white noise $\eta(\bm{t})$ of variance $\sigma^2 = 0.05^2$ and a phase discrepancy:
\begin{equation}
y_{\mathrm{exp}}(\bm{t}) = \phi^*(\bm{t} + 0.05 \sin(0.1 \pi \bm{t}))+\eta(\bm{t}),
\label{exp_pendulum_discr}
\end{equation}
where $\phi^*$ is the solution of Equation~\eqref{eq_pendulum} with $\beta^*=(0.5,0.3,0.8)$ and known experimental settings $\phi_0 = 1 \,\mathrm{rad}$ and $\dot{\phi}_0 = 0 \,\mathrm{rad\cdot s^{-1}}$.

\begin{figure}[h]
    \centering
\includegraphics[width=0.5\textwidth]{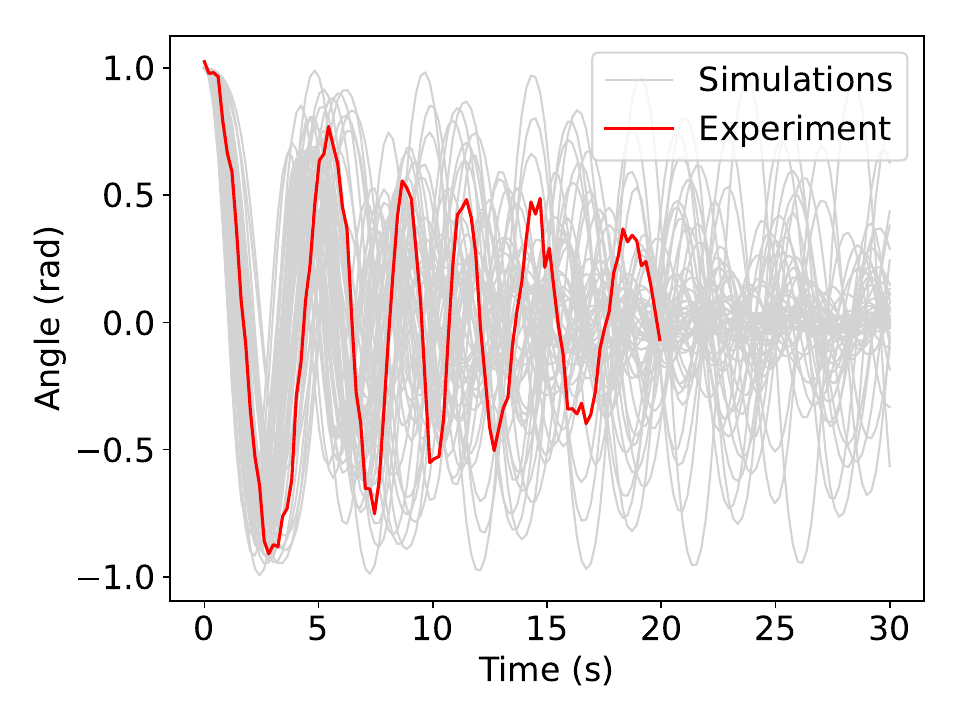}
\caption{Dataset for the anharmonic pendulum with the experiment generated by Equation~\eqref{exp_pendulum_discr}.}
\label{pendulediscr}
\end{figure}

In addition, $100$ simulations are generated using a Latin hypercube sampling (LHS) design in $[0,1]^3$ to train the surrogate models. 
The experiment is recorded over $20$~s and the simulations over $30$~s. 
For calibration methods with fixed end time, 
the simulation outputs are truncated at $20$~s. The simulations and the experimental data are shown in Figure~\ref{pendulediscr}.

Throughout this section, all alignments are computed using the 
approximation of the calibration problem
defined in Equation~\eqref{opti_alignMAP} for elastic alignment and in Equation~\eqref{opti_partialalignMAP} for partial elastic alignment.
A comparison with the
SRVF-based alignment method is provided in the Supplementary Material. 
 We compare the four calibration methods introduced previously, which we refer to as: 
 \begin{itemize}
\item  \emph{non-elastic}: the additive model of Equation~\eqref{calibration_f} based on \cite{kennedy2001bayesian},
\item  \emph{elastic}: the model defined in Equation~\eqref{posterior_elastic} without additive discrepancy (i.e. $\delta=0$),
\item  \emph{partial elastic}: the model defined in Equation~\eqref{posterior_partial} without additive discrepancy (i.e. $\delta=0$)
and where we set $\sigma^2_s = (\Delta t)^2$ based on the time resolution $\Delta t$,
\item \emph{rescaling}: the special case of the partial elastic calibration explained in Remark~\ref{rmq_rescaling} with additive discrepancy, where we set $\sigma^2_s = (\Delta t)^2$.
\end{itemize}
The hyperparameters of each method are estimated using the method explained in Section~\ref{HP_opti}. A uniform prior is assigned to $\beta$.

\begin{figure}[htbp]
\centering
\begin{subfigure}{\textwidth}
\centering
\begin{minipage}{0.32\textwidth}\centering Elastic\\
    \includegraphics[width=\textwidth]{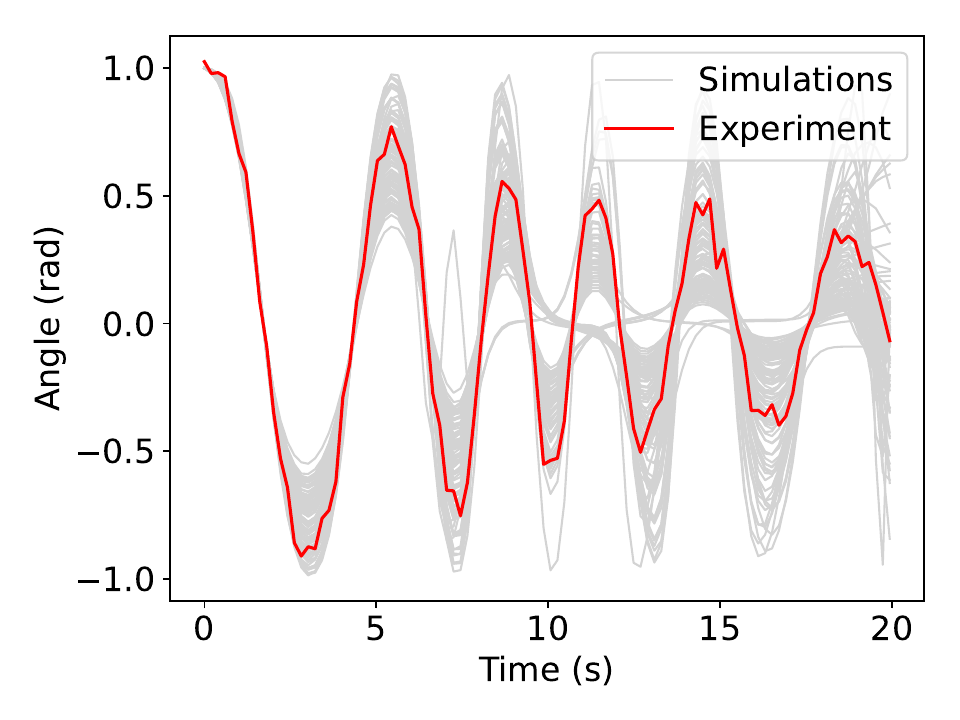}\end{minipage}\hfill
\begin{minipage}{0.32\textwidth}\centering Rescaling\\
    \includegraphics[width=\textwidth]{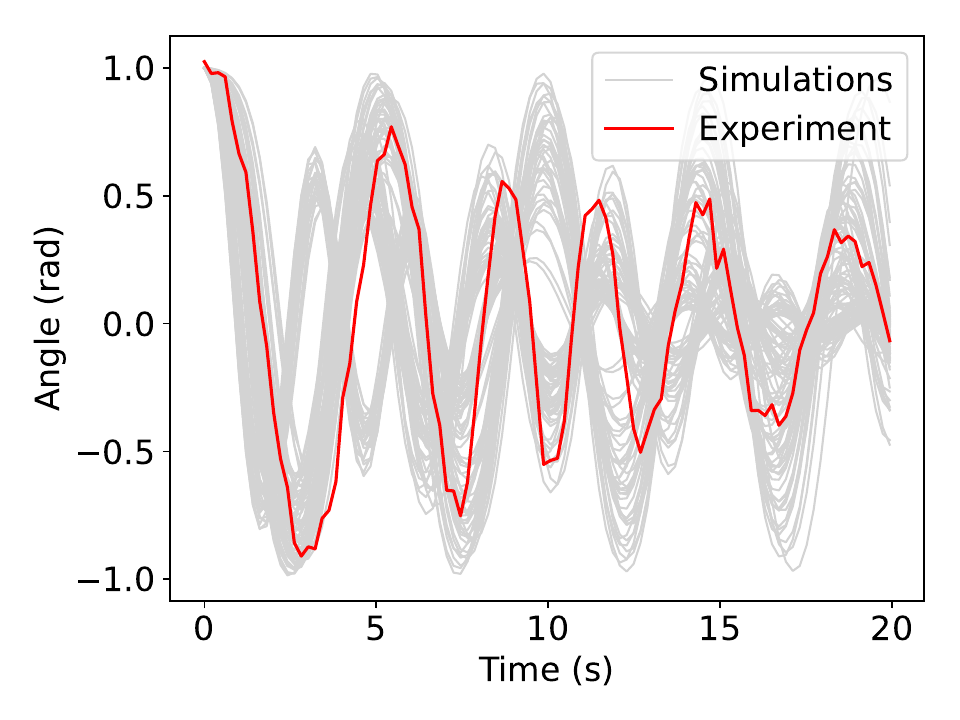}\end{minipage}\hfill
\begin{minipage}{0.32\textwidth}\centering Partial elastic\\
    \includegraphics[width=\textwidth]{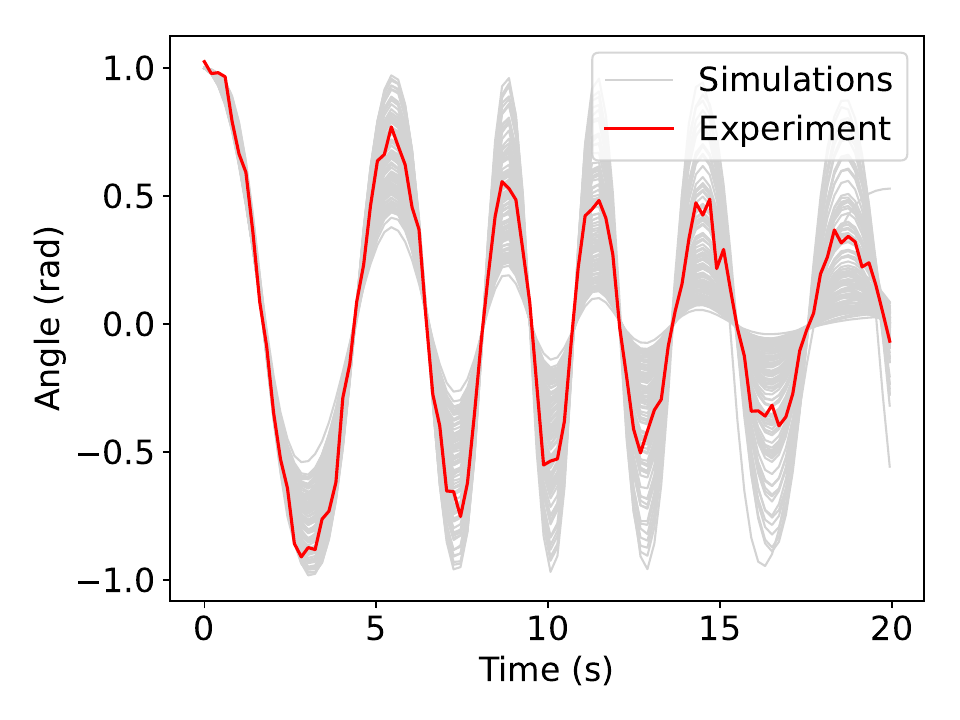}\end{minipage}

\end{subfigure}

\vspace{1em}

\begin{subfigure}{\textwidth}
\centering
\begin{minipage}{0.32\textwidth}\centering
    \includegraphics[width=\textwidth]{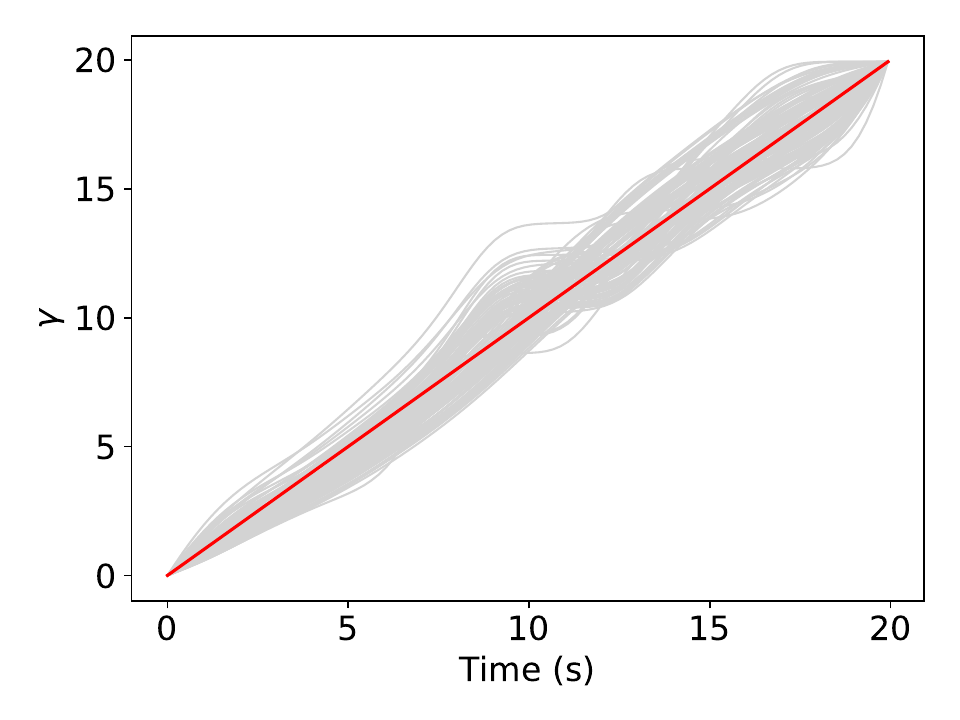}\end{minipage}\hfill
\begin{minipage}{0.32\textwidth}\centering
    \includegraphics[width=\textwidth]{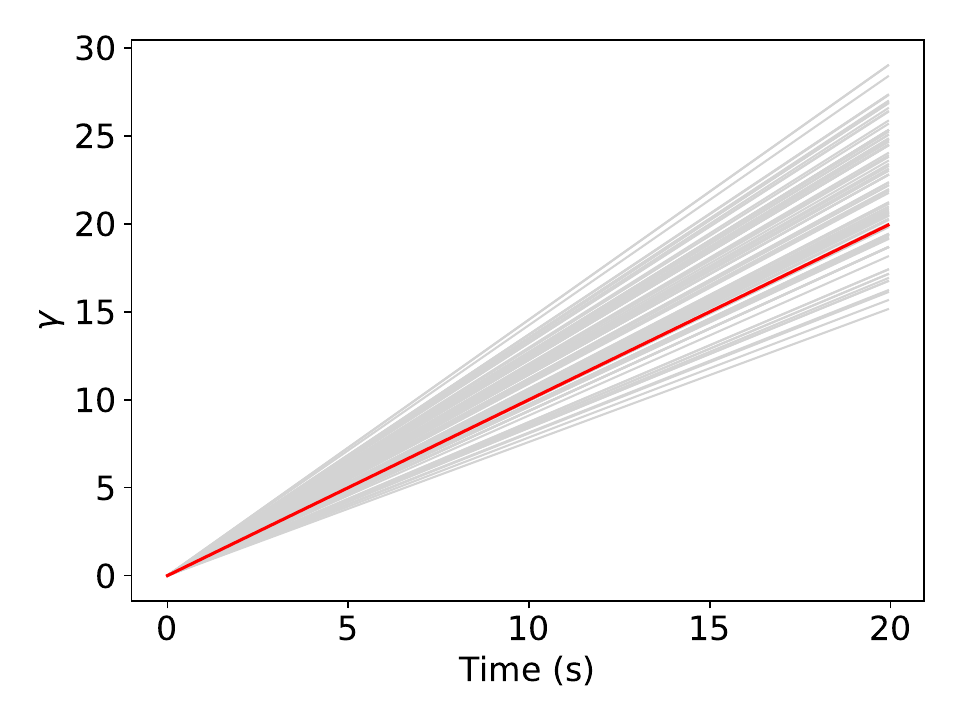}\end{minipage}\hfill
\begin{minipage}{0.32\textwidth}\centering
    \includegraphics[width=\textwidth]{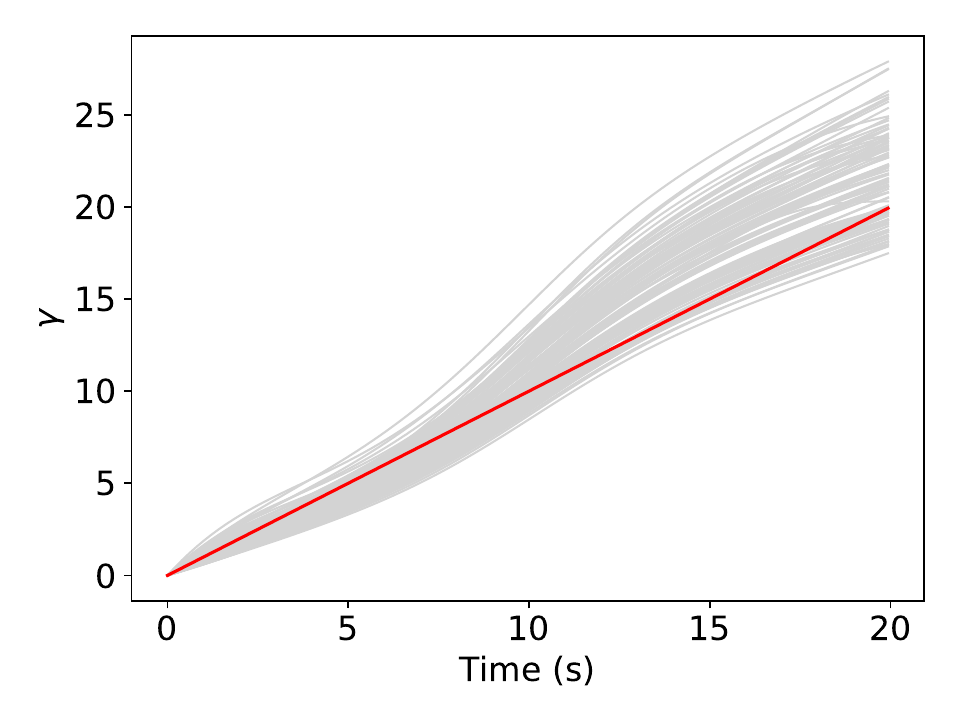}\end{minipage}

\end{subfigure}

\caption{Amplitudes (top) and time warpings (bottom) of the anharmonic pendulum outputs for the different alignment frameworks with the experiment generated by Equation~\eqref{exp_pendulum_discr}.}
\label{pendulediscr_amplitude_phase}
\end{figure}

We first align the simulations with the experimental output using the three alignment methods. 
The resulting amplitudes and time warpings are shown in Figure~\ref{pendulediscr_amplitude_phase}. 
Since the functions do not all exhibit the same number of oscillations, elastic alignment fails to simplify the data representation. 
Rescaling cannot account for the nonlinear phase discrepancy, as it is restricted to linear time warpings.
The partial elastic approach is the only method that leads to a simpler representation of the data.

We then fit the surrogate models. 
The surrogate models are compared on a test dataset $(y_i^{\mathrm{test}})_{i=1,\dots,{N_\mathrm{test}}}$ of $N_{\mathrm{test}}=100$ functions, generated from the inputs $(\beta_i^{\mathrm{test}})_{i=1,\dots,{N_\mathrm{test}}}$, using four criteria
 (see Marrel and Iooss~\cite{marrel2024probabilistic}\footnote{Our definition of the $\mathrm{IAE}$ differs slightly from their definition because the reconstructed surrogate models do not output Gaussian distributions.}):

\begin{itemize}
    \item $\displaystyle Q^2 = 1 - \frac{\sum_{i=1}^{N_{\mathrm{test}}} \lVert y_i^{\mathrm{test}} - \hat{y}(\beta_i^{\mathrm{test}}) \rVert_2^2}{\sum_{i=1}^{N_{\mathrm{test}}} \lVert y_i^{\mathrm{test}} - \bar{y} \rVert_2^2}$,
    where $\hat{y}$ is the posterior mean and $\bar{y}$ is the empirical mean. The closer this quantity is to $1$, the better the mean prediction.
    \item $\displaystyle \mathrm{IAE} = \frac{1}{N_t}\sum_{k=1}^{N_t}\int_0^1 \lvert \hat{\Delta}(\alpha,t_k) - \alpha \rvert \,\mathrm{d}\alpha$, where $\hat{\Delta}(\alpha,t_k)$ denotes the proportion of test observations
     below the predictive quantile of order $\alpha$ at time $t_k$. The closer this quantity is to $0$, the more reliable the credible intervals.
    \item $\displaystyle \mathrm{CRPS} = \frac{1}{N_t}\sum_{k=1}^{N_t}\frac{1}{N_{\mathrm{test}}} \sum_{i=1}^{N_{\mathrm{test}}} \int_{\mathbb{R}} \bigl( F(\beta_i^{\mathrm{test}}, y,t_k) - \mathbf{1}_{y_i^{\mathrm{test}}(t_k) \leq y} \bigr)^2 \,\mathrm{d}y$,
     where $F(\beta_i^{\mathrm{test}}, \cdot,t_k)$ is the predictive cumulative distribution function for the input $\beta_i^{\mathrm{test}}$ at time $t_k$. The closer this quantity is to $0$, the better the predictive accuracy and uncertainty quantification.
    \item $N_{\mathrm{PCA}}$: the number of PCA components used for surrogate modeling. For methods involving multiple surrogate models, we report the number of components in the following order: amplitude, shooting vector, and rescaling factor.
\end{itemize}

\begin{table}[h]
\centering
\begin{tabular}{|c|c|c|c|c|}
\hline
& $Q^2$ & $\mathrm{IAE}$ & $\mathrm{CRPS}$& $N_{\mathrm{PCA}}$ \\
\hline
Non-elastic & 0.78 & 0.10 &0.06& 7 \\
\hline
Elastic & 0.42 & 0.03 &0.12& 16 (9+7+0) \\
 \hline
Rescaling & 0.47 & 0.02 & 0.12 & 8 (7+0+1) \\
\hline
Partial elastic & 0.97 & 0.06 & 0.02 & 7 (3+3+1)\\
\hline
\end{tabular}
\caption{Comparison of the surrogate models for criteria computed on a test dataset for the anharmonic pendulum with the experiment generated by Equation~\eqref{exp_pendulum_discr}.}
\label{tab:pendulediscr_meta}
\end{table} 

To ensure a fair comparison between the different methods and to fully 
account for the rescaling procedures, we compute the metrics on the first half of the test functions (up to $t = 15$~s). 
We compare the reconstructed functions obtained after uncertainty propagation through the surrogate models (in the amplitude, shooting vector, and rescaling spaces) 
using the posterior mean as the predictor.
The criteria are summarized in Table~\ref{tab:pendulediscr_meta}. Surrogate model outputs for four different inputs are shown in Figure~\ref{surrogatediscr_outputs}.

\newcommand{\panel}[2]{%
  \makebox[1.5em][l]{\raisebox{-0.5\height}{\textbf{(#1)}}}%
  \raisebox{-0.5\height}{\includegraphics[width=0.95\textwidth]{#2}}%
}

\begin{figure}[h]
\centering
\panel{a}{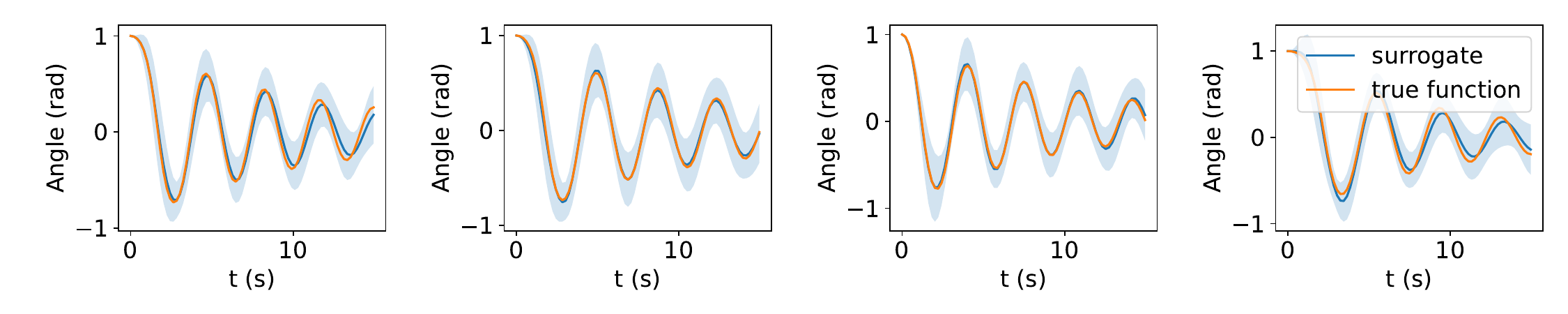}\par\vspace{0.6em}
\panel{b}{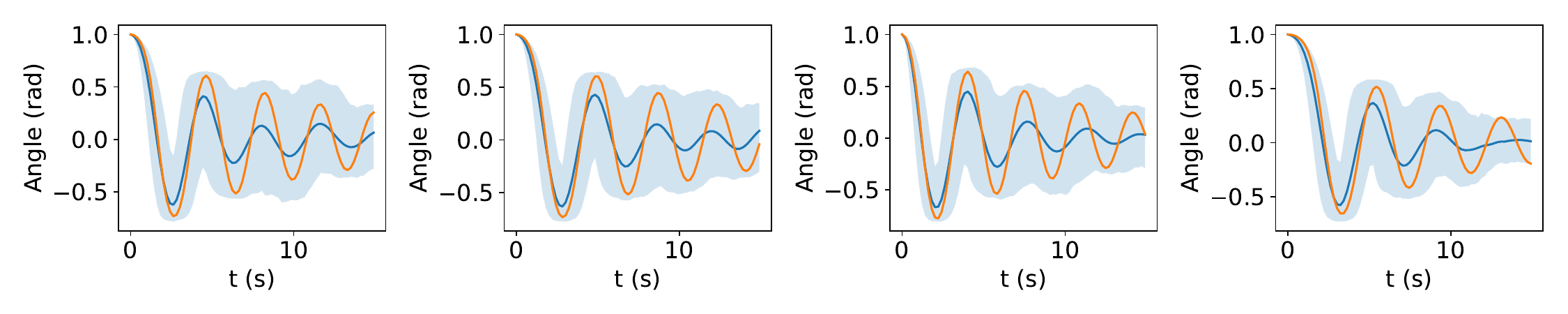}\par\vspace{0.6em}
\panel{c}{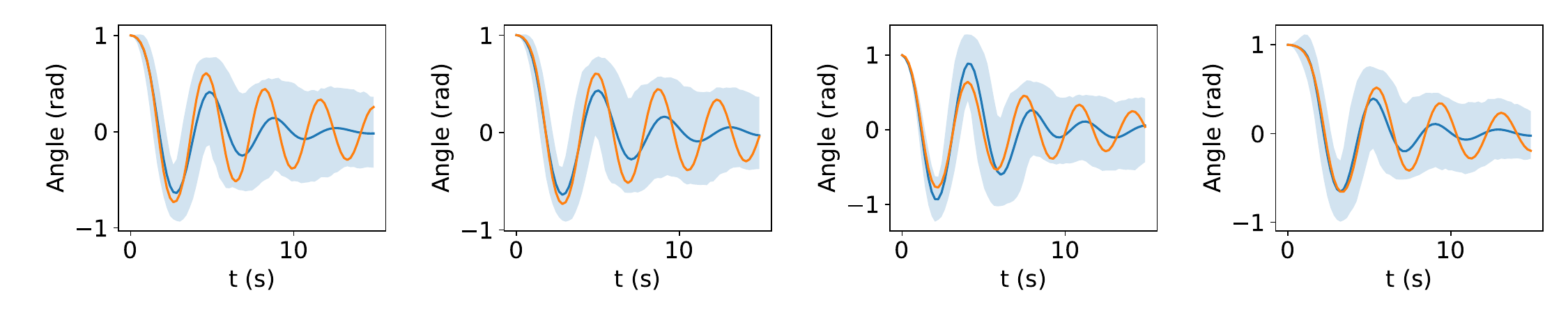}\par\vspace{0.6em}
\panel{d}{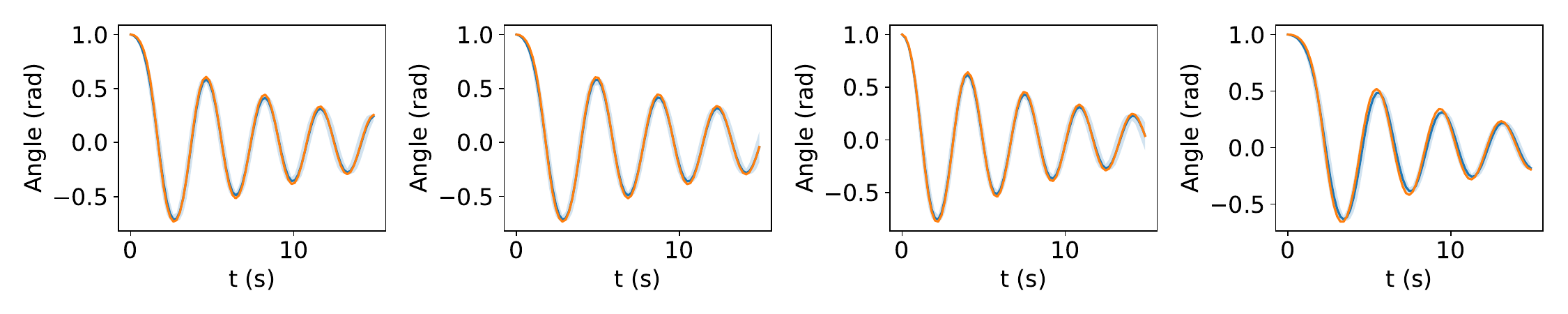}
\caption{Reconstructed anharmonic pendulum outputs for four different test inputs for (a) the non-elastic surrogate, (b) the elastic surrogate, (c) the surrogate with rescaling, and (d) the partial elastic surrogate.}
    \label{surrogatediscr_outputs}
\end{figure}

The best surrogate model is obtained with the partial elastic method. 
For the elastic and rescaling methods, using unadapted alignment methods leads to surrogate models with a low $Q^2$ and more PCA components. For all methods, the $\mathrm{IAE}$ is low indicating good uncertainty quantification.

\begin{figure}[h]
\includegraphics[width=0.98\textwidth]{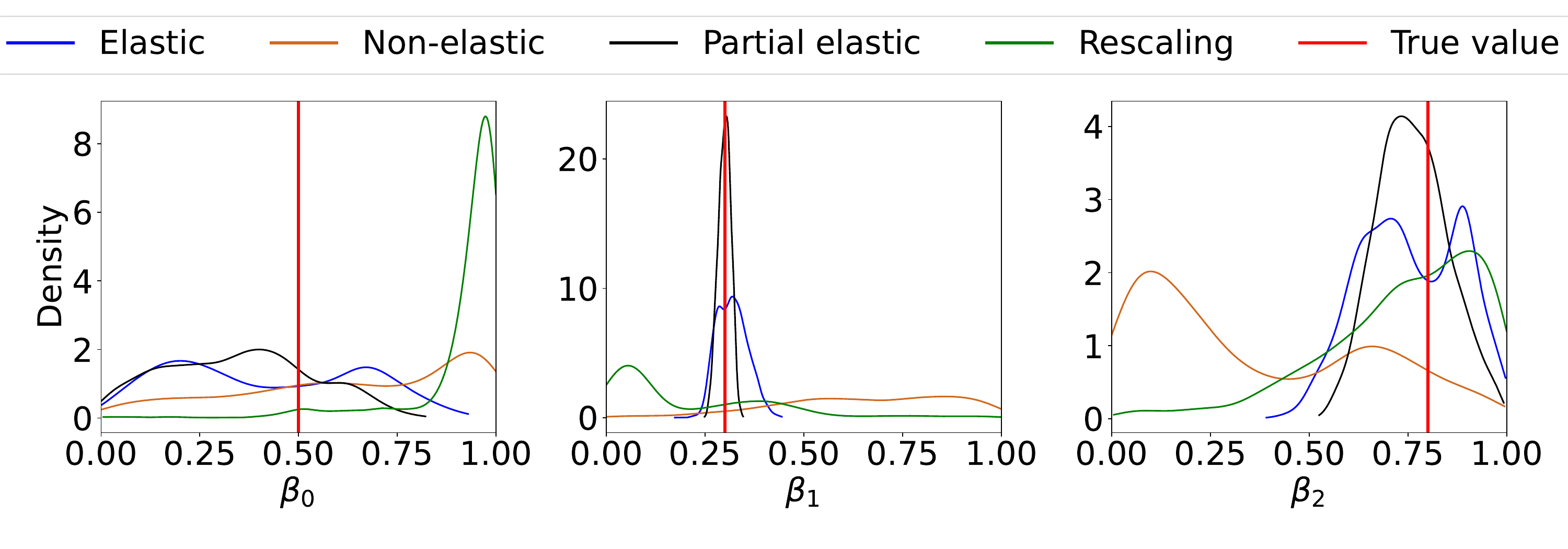}
\caption{Marginals of the posterior for the different calibration methods for the anharmonic pendulum with the experiment generated by Equation~\eqref{exp_pendulum_discr}.}
\label{pendulediscr_sampling}
\end{figure}

The posterior distributions are sampled using sequential Monte Carlo (see~\cite{speich2021sequential}) and shown in Figure~\ref{pendulediscr_sampling}. 
While the non-elastic and rescaling approaches rely on an amplitude discrepancy model, the elastic and partial elastic methods allow 
the discrepancy to be modeled in the shooting vector space and thus correctly account for the phase discrepancy. 
They are therefore the only methods that provide posterior distributions concentrated around the true parameter values.
Moreover, since its surrogate models are more accurate, the partial elastic approach returns a more concentrated posterior than the elastic one.

\begin{table}[h]
\centering
\begin{tabular}{|c|c|c|c|c|}
\hline
& $Q^2$ & $\mathrm{IAE}$ & $\mathrm{CRPS}$& $N_{\mathrm{PCA}}$ \\
\hline
Non-elastic & 0.78 & 0.10 &0.06& 7 \\
\hline
Elastic & 0.65 & 0.08 &0.09& 13 (9+4+0) \\
 \hline
Rescaling & 0.96 & 0.05 & 0.02 & 5 (4+0+1) \\
\hline
Partial elastic & 0.97 & 0.03 & 0.02 & 6 (3+2+1)\\
\hline
\end{tabular}
\caption{Comparison of the  surrogate models for criteria computed on a test dataset for the anharmonic pendulum with the experiment generated by Equation~\eqref{exp_pendulum}.}
\label{tab:pendule_meta}
\end{table} 

\begin{figure}[h]
\includegraphics[width=0.98\textwidth]{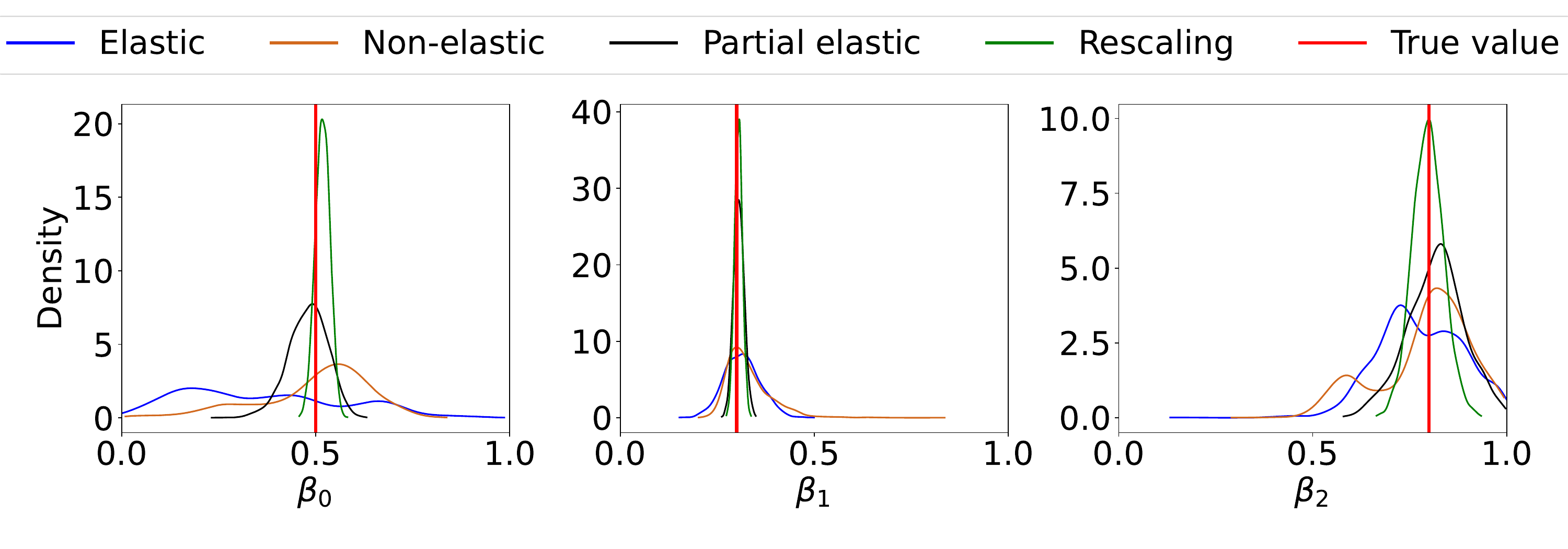}
\caption{Marginals of the posterior for the different calibration methods for the anharmonic pendulum with the experiment generated by Equation~\eqref{exp_pendulum}.}
\label{pendule_sampling}
\end{figure}

Since the simulations themselves exhibit varying phase across the parameter range, the proposed methodology remains beneficial even when no artificial phase discrepancy is introduced in the experimental data, i.e.
\begin{equation}
y_{\mathrm{exp}}(\bm{t}) = \phi^*(\bm{t})+\eta(\bm{t}).
\label{exp_pendulum}
\end{equation}
Table~\ref{tab:pendule_meta} shows that alignment improves the surrogate models, particularly for the rescaling and partial elastic approaches, 
which in turn give more concentrated posterior distributions (Figure~\ref{pendule_sampling}). Further details on this case are given in the Supplementary Material.

\subsection{Equation of state}
In materials physics, to study how a solid responds to a shock, one must know a relation called the equation of state, which links the internal energy, pressure, temperature, and mass density of the material.
Our goal here is to calibrate the parameters of the following Mie-Grüneisen equation of state \cite{menikoff2007empirical}:
\begin{equation}
P(\mu, E) =P_H(\mu,\beta)+\gamma_0 \rho_0 (E-E_H),
\end{equation}
where 
$\rho$ denotes the mass density, $\rho_0$ the initial mass density, $\mu=\frac{\rho}{\rho_0}-1$,
$E$ the internal energy, $P$ the pressure, $E_H$ a reference internal energy and
$\gamma_0$ the Grüneisen coefficient.
The vector $\beta \in [0,1]^4$ gathers, in normalized form, the parameters to be calibrated:
the Grüneisen coefficient $\gamma_0$ and additional parameters entering the Hugoniot pressure $P_H$ whose interpretation 
depends on the chosen model for the $P_H$ function (see, for example, \cite{lukyanov2008equation}). 

\begin{figure}[htbp]
\centering
\begin{minipage}[t]{0.48\textwidth}
\centering
\includegraphics[trim=0cm 2cm 1.5cm 1cm, clip, width=\textwidth]{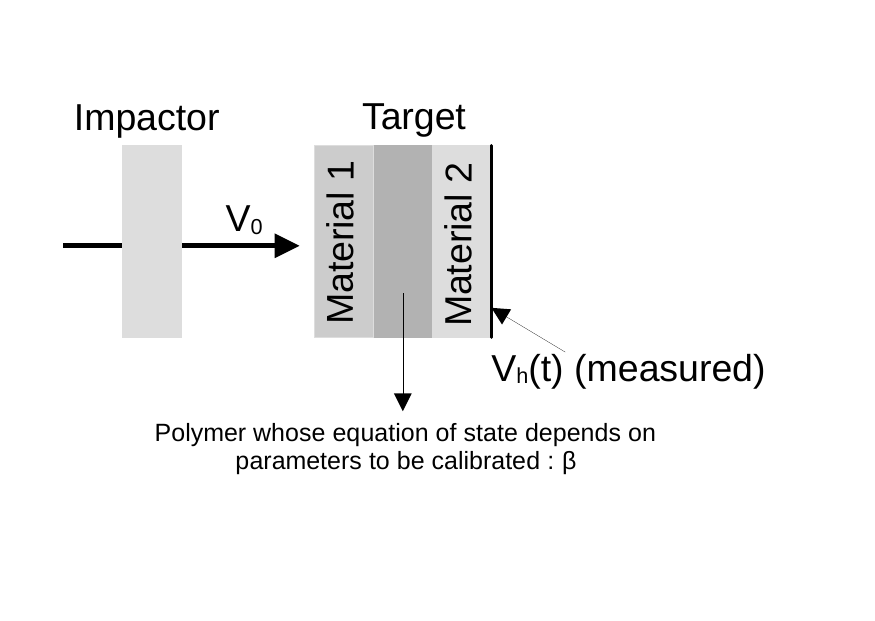}
\caption{Experimental setup for the calibration of the equation of state.}
\label{schema_EOS}
\end{minipage}
\hfill
\begin{minipage}[t]{0.48\textwidth}
\centering
\includegraphics[width=\textwidth]{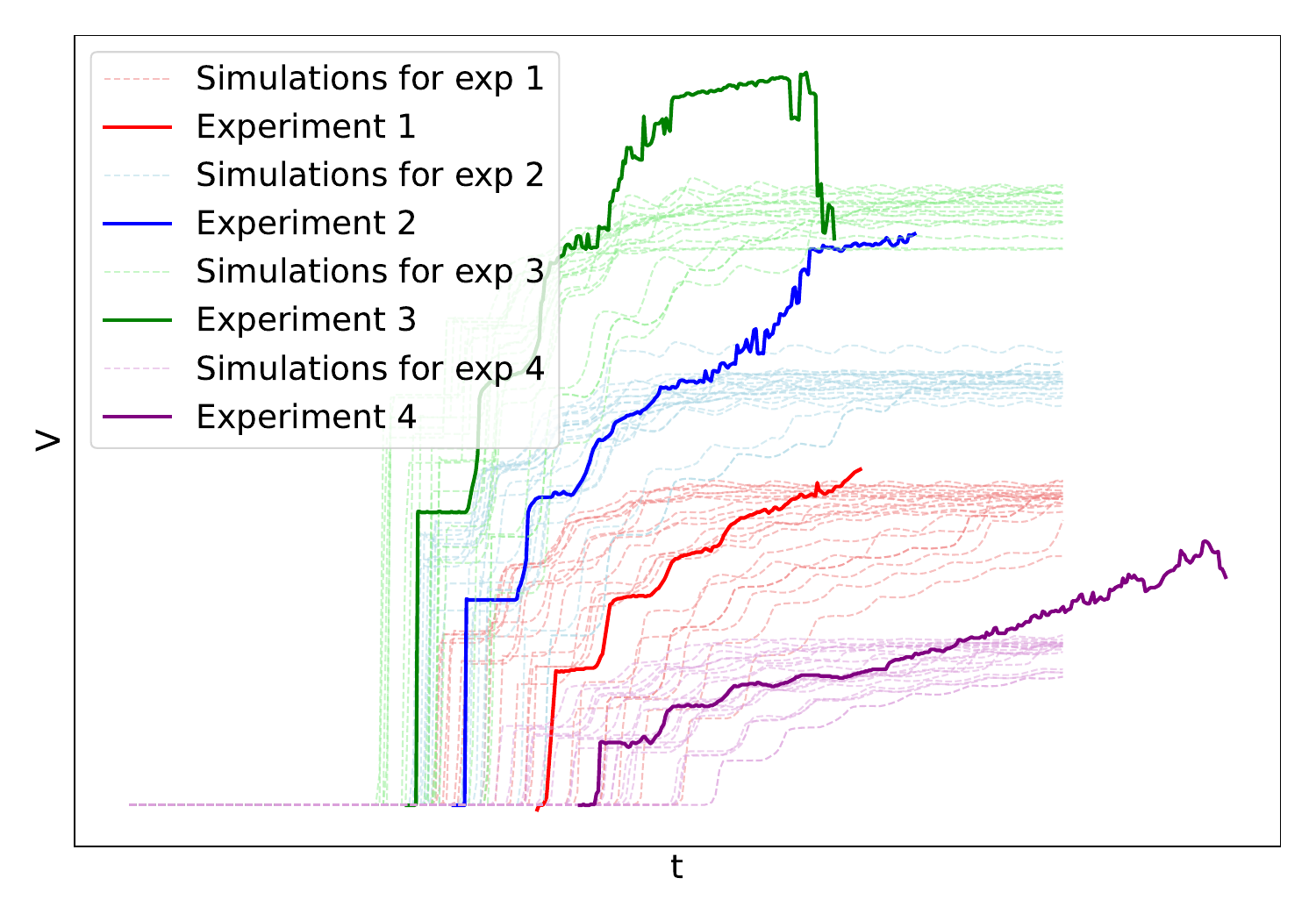}
\caption{Dataset for the calibration of the equation of state.}
\label{dataset_eos}
\end{minipage}
\end{figure}

To estimate these parameters, the experiment depicted in Figure~\ref{schema_EOS} was performed with four different experimental settings. 
An impactor is launched onto a target, generating a shock wave that propagates through the target. 
 The target is made of three layers, the material of interest being placed between two materials
whose equations of state are known. The propagation of the shock wave through the target
therefore depends only on the equation of state of the central material.
The surface velocity of the target is measured \cite{lassig2015analysis}. A hydrodynamic simulation simulates the physical phenomenon.
For each experimental setting, $100$ simulations were performed.
The input parameters were selected using a LHS design in the hypercube $[0,1]^4$.

Figure~\ref{dataset_eos} shows the experimental and simulated outputs. The velocity curves all exhibit similar shock and expansion patterns, 
alternating time intervals where the velocity is nearly constant with intervals where it increases sharply. 
The experimental measurements become highly noisy and unpredictable after several shock/expansion cycles. 
We therefore restrict the calibration procedure to the first three shock/expansion cycles.

\begin{figure}[htbp]
\centering
\includegraphics[width=0.98\textwidth]{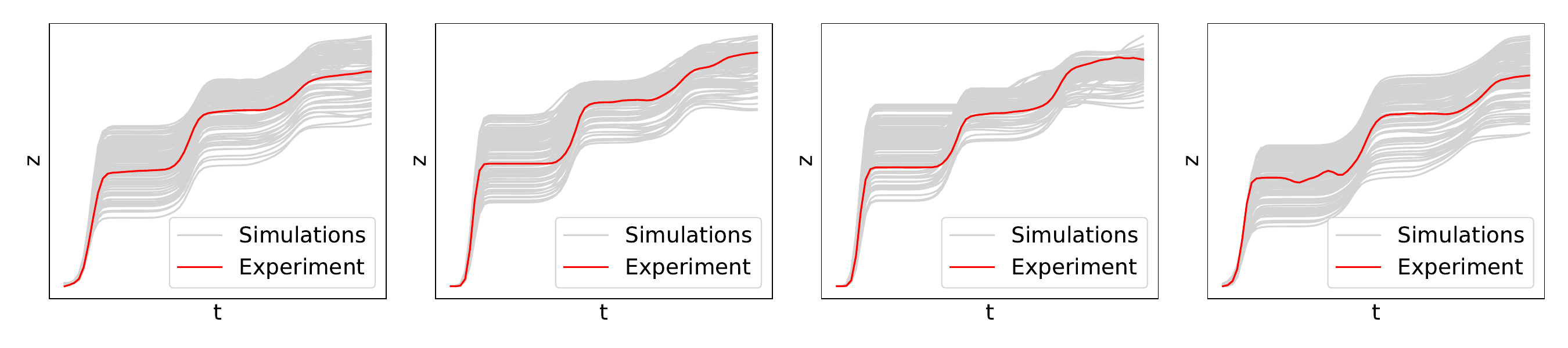}

\vspace{1em}

\includegraphics[width=0.98\textwidth]{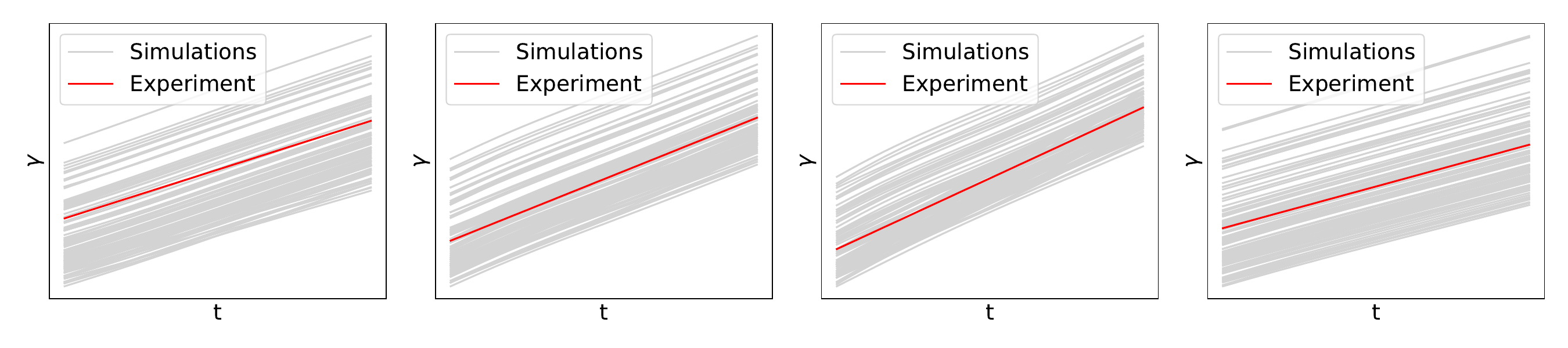}
\caption{Partial amplitudes (top) and partial phases (bottom) of the equation of state for the different experiments}
\label{amplitude_phase_EOS}
\end{figure}

We perform a partial elastic calibration with variable initial and end times. After selecting the time window of interest in the experimental data, the 
simulations are aligned with the experiments using the method described in Section~\ref{partial_elastic}. 
The resulting amplitudes and phases are shown in Figure~\ref{amplitude_phase_EOS}. 
The phase variability is mainly driven by the initial and end times. 
However, using only an affine warping for alignment would prevent experiments and simulations from matching perfectly, 
causing the amplitude discrepancy to exhibit unphysically large fluctuations over short time intervals.
Results obtained with an affine warping are reported in the Supplementary Material.

Regarding the error hyperparameters, for the initial and end times, the error variance is equal to $\frac{1}{N_t^2}$,
except for the first experiment
where based on expert knowledge, we know there is a large measurement error and as a consequence the
standard deviation is chosen $10$ times larger. 
A uniform prior is assigned to the calibration parameters.

\begin{figure}[h]
    \centering
    \includegraphics[width=0.8\textwidth]{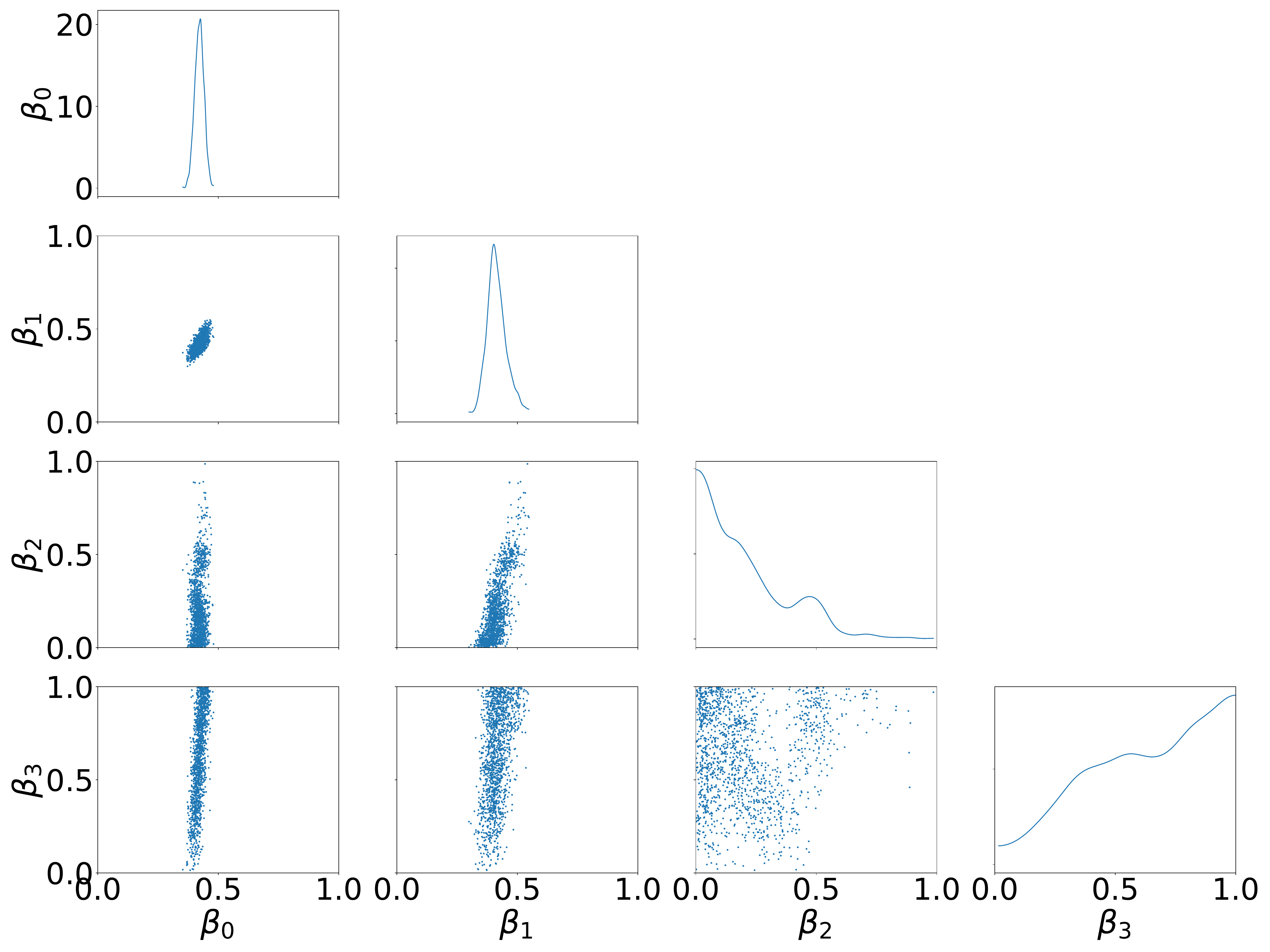}
\caption{Pairplots and marginals of the posterior distribution computed using sequential Monte Carlo for the partial elastic calibration of the equation of state.}
\label{calibration_eos}
\end{figure}

\begin{figure}[h]
    \begin{minipage}{0.48\textwidth}
    \centering
    Experiment 1
    \includegraphics[width=0.9\textwidth]{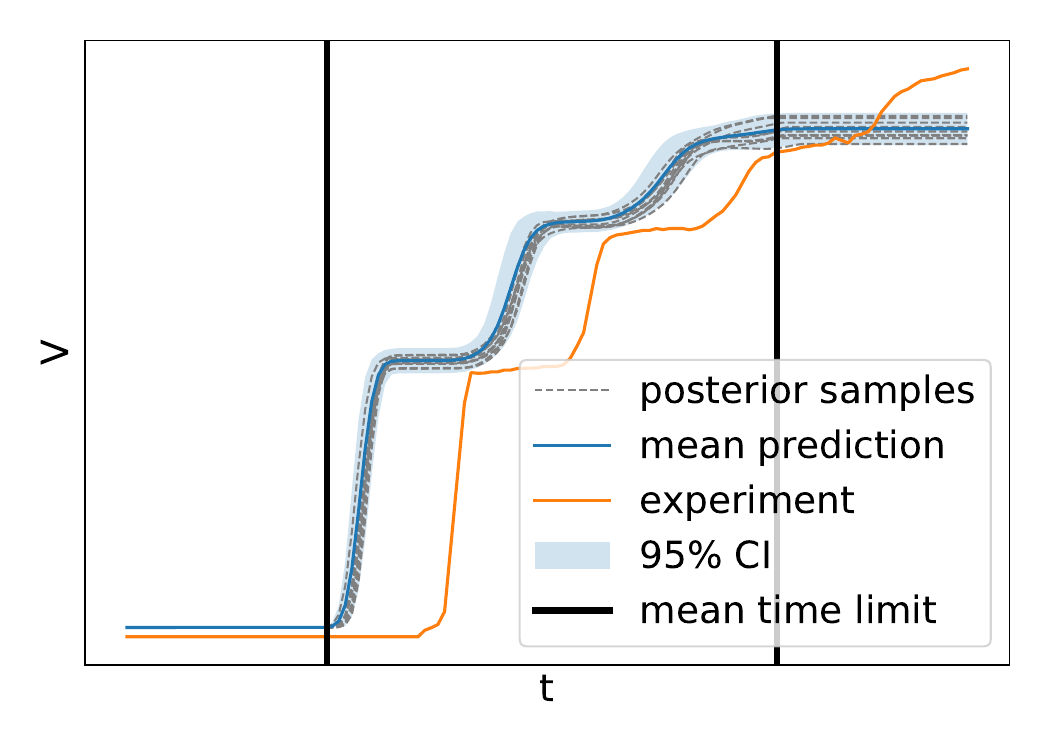}
    \end{minipage}
        \begin{minipage}{0.48\textwidth}
    \centering
    Experiment 2
    \includegraphics[width=0.9\textwidth]{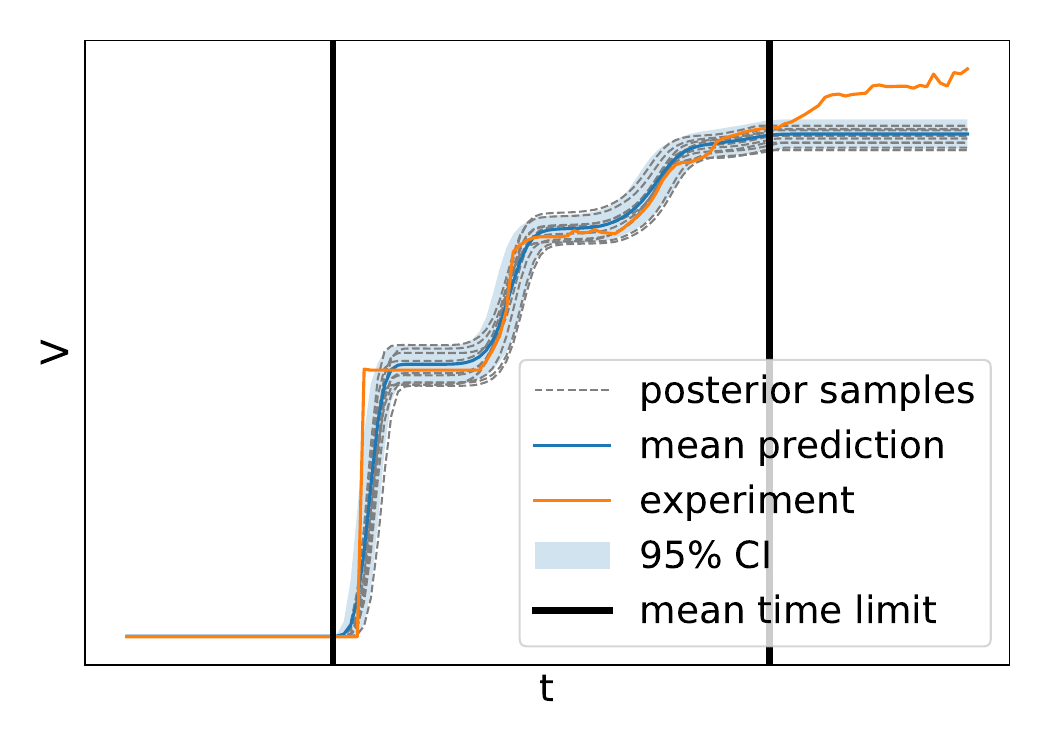}
    \end{minipage}

        \begin{minipage}{0.48\textwidth}
    \centering
    Experiment 3
    \includegraphics[width=0.9\textwidth]{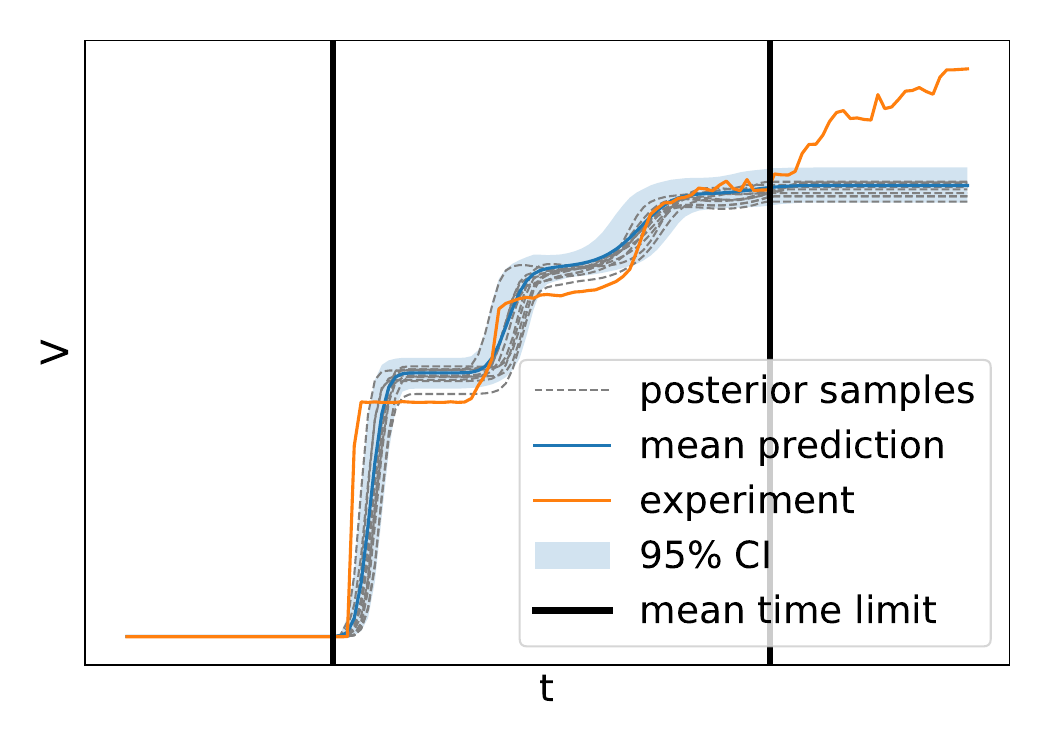}
    \end{minipage}
        \begin{minipage}{0.48\textwidth}
    \centering
    Experiment 4
    \includegraphics[width=0.9\textwidth]{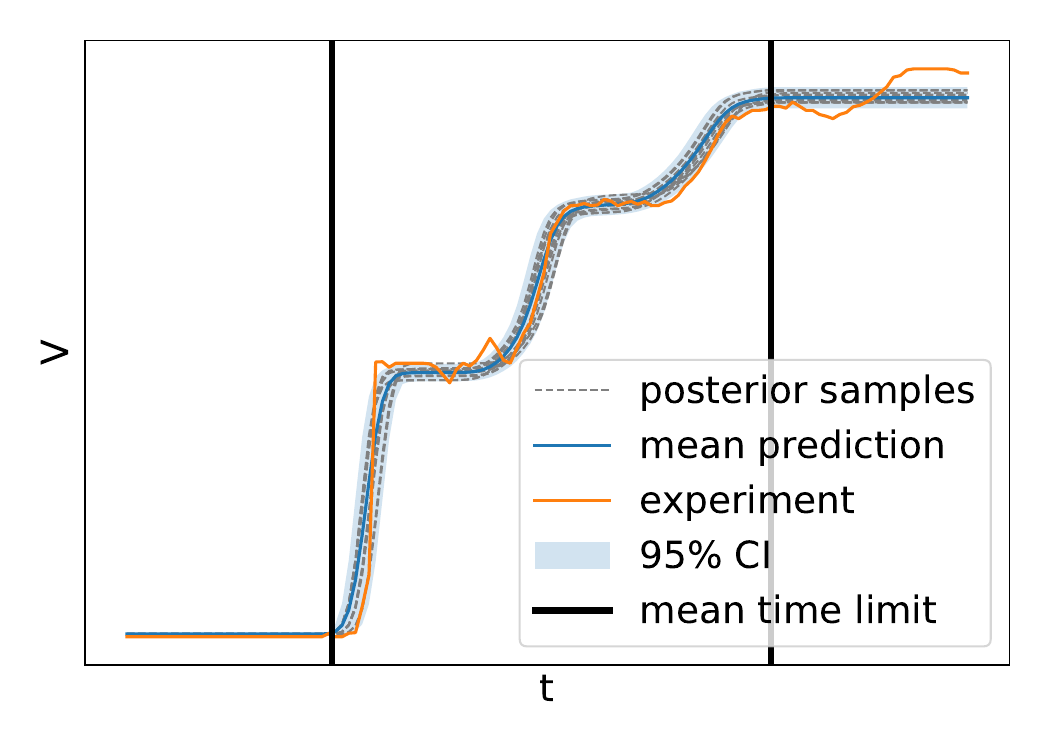}
    \end{minipage}

\caption{Posterior prediction after uncertainty propagation for each experiment. The blue curve 
corresponds to the composition of the mean predictions of the various surrogate models, evaluated at the posterior mean of the calibration parameters. 
The credible intervals, posterior samples and mean time limits are obtained by composing the uncertainties associated with the surrogate models and the calibration parameter.}
\label{posterior_eos}
\end{figure}

Figure~\ref{calibration_eos} shows the sampling results. The posterior distribution is more concentrated than the prior for parameters 
$\beta_0$ and $\beta_1$, whereas little concentration is observed for parameters $\beta_2$ and $\beta_3$. 
Figure~\ref{posterior_eos} shows the mean prediction and uncertainty propagation, accounting for both surrogate and calibration parameter uncertainties, for each experiment.
The decomposition into amplitude and phase 
allows the shock/expansion pattern to be propagated to the samples.
The amplitudes of each experiment are well recovered and 
the decomposition into amplitudes and phases makes it possible to accurately compare the amplitude of the first experiment despite an error in the initial time.

\section{Conclusion}
In this work, we have proposed an interpretation and a generalization of the method of Francom \textit{et al.}~\cite{francom}, based on the amplitude/phase decomposition, to address a broader class of functional calibration problems. 
This generalization relies on an alternative definition of the time warping space. 
The approach was illustrated on a synthetic test case, in which the proposed partial elastic method is shown to effectively simplify the data representation 
whereas standard function alignment methods fail.

We have also observed that simplifying the data representation improves performance in both surrogate modeling and discrepancy modeling. 
When surrogate modeling is based on PCA for dimensionality reduction, simplified representations require fewer principal components, leading to more accurate surrogate models.
Moreover, introducing a discrepancy in the shooting vector space enables sharp variations in the original space to be captured through a smooth phase discrepancy.

Finally, we have identified the assumptions underlying elastic calibration methods. Future work could test these assumptions and develop 
a method for sampling the phase discrepancy directly, which would improve uncertainty quantification at the cost of a substantial increase in computational effort.

\addcontentsline{toc}{section}{References}
\bibliographystyle{siamplain}
\bibliography{biblio}

\newpage 
\appendix

\section{Amplitude/phase decomposition}
\label{align}

Let $f,g:[0,1]\rightarrow \mathbb{R}$ be two absolutely continuous functions. Our goal is to match the function $g$ to $f$ by finding
 the time warping $\gamma$ so that $g\circ \gamma$ is closest to $f$. We denote the time warping group by
\begin{equation*}
\Gamma = \{\gamma : [0,1] \to [0,1] \mid \gamma(0)=0,\, \gamma(1)=1,\, \gamma \text{ is a }\mathcal{C}^2\text{-diffeomorphism}\}.
\end{equation*}

To do so, as explained in \cite{fdabook}, we introduce an optimization problem on $\Gamma$:
 \begin{equation}\gamma_{g\rightarrow f}=\argmin_{\gamma \in \Gamma} E(f,g \circ \gamma)+\lambda\int_0^1 \gamma''(t)^2dt.
\label{align_opti}\end{equation}

$E$ is a criterion that evaluates the quality of the matching and the second term is a penalty controlled by a parameter $\lambda>0$
that favors time warpings close to the identity and smooth time warpings.

To choose the criterion $E$, the book \cite{fdabook} suggests to first apply the SRVF (square root velocity function) transform: $q_f(t)=\text{SRVF}(f)(t)=\text{sgn}(f'(t))\sqrt{|f'(t)|}\in \mathbb{L}^2$.
This transformation has an effect on the time warping: $\text{SRVF}(g \circ \gamma)(t)=\sqrt{\gamma'(t)}\text{SRVF}(g)(\gamma(t))$.
We can now define the criterion:

$$E(f,g \circ \gamma) =||\text{SRVF}(f)-\text{SRVF}(g \circ \gamma)||_2^2$$

To motivate this choice, one may notice that it satisfies the following properties, see \cite{fdabook}:
\begin{itemize}
\item \textbf{Invariance to the same warping:} For $\gamma \in \Gamma$, $E(f \circ \gamma, g\circ \gamma)=E(f,g)$
\item \textbf{Inverse warping:} If $\gamma=\text{arg} \min_{\gamma \in \Gamma} E(f, g \circ \gamma)$, then $\gamma^{-1}=\text{arg} \min_{\gamma \in \Gamma} E(f\circ \gamma, g)$
\item \textbf{Distance in the amplitude space:} Let us define the amplitude space as the quotient space $\mathbb{L}^2/\Gamma$ i.e. an element of this space is an orbit 
$[q]=\{\sqrt{\gamma'}q\circ \gamma:\gamma \in \Gamma\}$.
The criterion $d_a([q_f],[q_g])=\inf _{\gamma\in \Gamma }E(f,g \circ \gamma)$ is a distance in the quotient space. 
This result shows that this distance is not dependent on the element of the orbit.\\
\end{itemize}

We have to distinguish the amplitude of $g$: $[q_g]=\{\sqrt{\gamma'}q_g\circ \gamma:\gamma \in \Gamma\}$ which does not depend on $f$ from the phase relative to $f$: $\gamma_{g\rightarrow f}=\text{arg} \min_{\gamma \in \Gamma} E(f,g \circ \gamma). $
and the amplitude relative to $f$, $g \circ \gamma_{g\rightarrow f}$.
We consider here only amplitudes and phases relative to a template function.
When the template does not need to be explicitly specified, we will simply write amplitude and phase.

Different algorithms have been proposed to solve \eqref{align_opti}. 
Tucker \textit{et al.}~\cite{tucker2013generative} suggest to use a dynamic programming algorithm, Huang \textit{et al.}~\cite{huang2016} argue for a Riemannian BFGS algorithm and 
\cite{cheng2016,lu2017} propose Bayesian algorithms that account for uncertainties of the transform. 

To implement calibration into the amplitude/phase space, all the time series have to be split into an amplitude and 
a phase relative to a template. This template can either be the experiment or a simulation close to the experiment.

\section{Time warpings and shooting vectors}
\label{shooting}
The time warping space $\Gamma$ is a nonlinear manifold which does not have the basic properties of a vector space. Francom \textit{et al.}\cite{francom}
proposed to simplify this space to work in a Hilbert space. The article first applies the SRVF transform to phases
$$\psi=\text{SRVF}(\gamma)=\sqrt{\gamma'} $$

Since $\gamma(0)=0$ and $\gamma(1)=1$, $||\psi||_2^2=\int_0^1 \psi(t)^2dt= 1$ and as a result $\Psi=\{\text{SRVF}(\gamma):\gamma \in \Gamma\}$ is
an open manifold contained within the positive orthant of the unit $\mathcal{L}^2$-sphere. This geometry is simpler but still nonlinear and thus Francom \textit{et al.}~\cite{francom} consider the following tangent space at the point $\psi_0\in\Psi$: 
$T_{\psi_0}(\Psi) = \left\{ v \in L^2 \ \bigg| \ \int_{0}^{1} v(t)\,\psi_0(t)\,dt = 0 \right\}$ and the inverse-exponential map:
$$
\begin{aligned}
\exp^{-1}_{\psi_0} : \Psi &\longrightarrow  T_{\psi_0}(\Psi)\\
\psi &\longmapsto \frac{\kappa}{\sin(\kappa)}\bigl(\psi -\cos(\kappa)\psi_0 \bigr) \quad\text{ where }\kappa =\cos^{-1}(\int_0^1\psi\psi_0)
\end{aligned}
$$

The point $\psi_0$ corresponds to the origin of the tangent space. We choose $\gamma_0(t)=t$ and thus $\psi_0(t)=1$.
For a phase $\gamma \in \Gamma$, we call $v=\exp^{-1}_{\psi_0}(\text{SRVF}(\gamma))$ its shooting vector. When we compare
shooting vectors, we are comparing a transformation of the derivatives of $\gamma$ rather than the values of $\gamma$.
Depending on our data and the optimization method used to compute the phases, comparing phases using $\gamma$ or $v$ may lead to different results.

From a shooting vector $v \in T_{\psi_0}(\Psi)$, we can compute the associated phase with the exponential map:
    $$\psi=
\exp_{\psi_0}(v)
    = \cos\!\left(\|v\|\right)\,\psi_0
    + \sin\!\left(\|v\|\right)\,\frac{v}{\|v\|}
 $$

 We then have to check that for all $ t \in [0,1], \psi(t)>0$ to ensure that the shooting vector is associated with a valid phase and then, we can compute 
$\gamma(t)=\int_0^t \psi(s)^2ds$.

\section{Partial elastic matching}
\label{partial}

Bryner and Srivastava\cite{cut} generalize the framework of Srivastava and Klassen~\cite{fdabook} to allow for partial matching. We summarize their method here.

We can extend the previously defined functions $f$ and $g$ to the interval $[0,\infty)$ by setting for all $ t>1, f(t)=f(1), g(t)=g(1)$.
Note that for $t>1, q_f(t)=q_g(t)=0$.
With this new setting, the optimization criterion can be rewritten:
$$\tilde{E}(f,g)=\int_0^{\infty} (q_f(t)-q_g(t))^2dt.$$

To define partial matching, we introduce the partial time warping group: 
$$\Gamma_p=\{\gamma_p:[0,\infty)\rightarrow[0,\infty): \gamma_p(0)=0,\gamma_p \text{ is a strictly increasing diffeomorphism}\}$$

With these new definitions, the partial matching problem is
 \begin{equation}\gamma_{p} = \text{arg} \min_{\gamma_p \in \Gamma_p} \tilde{E}(f,g \circ \gamma_p).
\label{alignpartial_opti}\end{equation}

Depending on the value of $\gamma_p(1)$, we can rewrite the criterion $\tilde{E}(f,g\circ \gamma_p)$:

\begin{itemize}
\item If $\gamma_p(1)<1$, we can set $s=\gamma_p(1)$ and $\gamma=\frac{\gamma_p}{s} \in \Gamma$ then 
$$\tilde{E}(f,g\circ \gamma_p)=\int_0^1 (q_f(t)-q_g(s\gamma(t))\sqrt{s\gamma'(t)})^2 dt+\int_{s}^1 q_g(t)^2dt.$$

\item If $\gamma_p(1)>1$, we can set $s=\gamma_p^{-1}(1)<1$ and $\gamma=\gamma_p(t\,s)\in \Gamma$, then 
$$\tilde{E}(f,g\circ \gamma_p)=\int_0^1(q_f(s\gamma^{-1}(t))\sqrt{s(\gamma^{-1})'(t)}-q_g(t))^2dt+\int_{s}^1 q_f(t)^2dt.$$
\end{itemize}

In both cases, the problem \eqref{alignpartial_opti} can be rewritten as an optimization problem over $(\gamma ,s) \in \Gamma \times \mathbb{R}^{+}$.
The first integral corresponds to a standard elastic matching after rescaling the function $f$ or $g$ by $s$ and the second integral is a penalty term that encourages $s$ to be close to $1$ i.e 
not to censor the function too much. 
To control this penalization, Bryner and Srivastava\cite{cut} suggest to multiply the second term by a constant $\kappa>0$.

In Figure~\ref{fig:partialamplitudeMAP}, the first integral is associated with the solid lines, the second integral with the dashed lines and the parameter $s$ with the red cross.\\

The first approach consists in defining a grid for possible values of \(s\), and for each value optimizing \(\gamma\) using standard
elastic matching algorithms. Alternatively, Bryner and Srivastava\cite{cut} present a gradient-based method to jointly optimize \((\gamma, s)\).

\section{Implementation details}
\label{imp_details}
This section gathers the practical choices made in the numerical
experiments of Section~\ref{section5}. 

\paragraph{Surrogate modeling}
The Gaussian processes are taken with a zero mean function and a
Mat\'ern $5/2$ covariance function. The number of principal components
retained is the smallest one for which the explained variance reaches
$99\%$. The models are implemented with the GPy library~(v1.13.2), and
the kernel hyperparameters are estimated by maximum likelihood.

\paragraph{Initialization}
The discrepancy terms are modeled as zero-mean Gaussian processes with a
Mat\'ern $5/2$ kernel parameterized by a lengthscale $l$ and a standard
deviation $\sigma_\delta$. When the discrepancy is not fixed to $0$, these
hyperparameters must be given starting values in
Algorithm~\ref{alg:calibration}; the values used for each test case are
reported in Table~\ref{tab:init_hyperparams}. For the amplitude
discrepancy of the equation of state, the initial standard deviation is
set relative to the magnitude of the output rather than to a fixed value:
$a_1$ denotes the value of the first plateau of the experimental curve
being calibrated, so that the initialization is $\sigma_\delta = 0.05\,a_1$ for each
experiment.

\begin{table}[h]
\centering
\begin{tabular}{|l|c|c|c|}
\hline
Test case & Discrepancy & $l$ & $\sigma_\delta$ \\
\hline
\multirow{2}{*}{Anharmonic pendulum} & Amplitude & $0.1$ & $0.1$ \\
\cline{2-4}
& Phase & $0.3$ & $0.1$ \\
\hline
\multirow{2}{*}{Equation of state} & Amplitude & $0.1$ & $0.05\,a_1$ \\
\cline{2-4}
& Phase & $0.2$ & $0.05$ \\
\hline
\end{tabular}
\caption{Initial values of the discrepancy hyperparameters used in
Algorithm~\ref{alg:calibration}. For the amplitude discrepancy of the
equation of state, $a_1$ denotes the value of the first plateau of each experiment.}
\label{tab:init_hyperparams}
\end{table}

\paragraph{Sampling method}
The posterior distributions are sampled with the sequential Monte Carlo
sampler implemented in PyMC~(v5.12). The algorithm propagates a
population of weighted particles through a sequence of tempered
distributions proportional to $\pi(\beta)\,L(\beta)^{\alpha}$, where
$\pi$ is the prior, $L$ the likelihood, and $\alpha$ increases from $0$ (prior) to $1$ (posterior). The number of Markov chain Monte Carlo
steps applied at each stage is selected automatically, so that the
correlation with the samples of the previous stage falls below a given
threshold. We use $500$ particles
with $5$ independent runs of the sampler, which yields $2500$ posterior
samples.

\newcounter{main_figure}
\setcounter{main_figure}{\value{figure}}
\renewcommand\thefigure   {SM\fpeval{\value{figure}-\value{main_figure}}}

\newcounter{main_equation}
\setcounter{main_equation}{\value{equation}}
\renewcommand\theequation   {SM\fpeval{\value{equation}-\value{main_equation}}}

\newcounter{main_table}
\setcounter{main_table}{\value{table}}
\renewcommand\thetable   {SM\fpeval{\value{table}-\value{main_table}}}


\setcounter{section}{0}
\setcounter{subsection}{0}
\let\section\SMorigsection

\renewcommand{\thesection}{SM\arabic{section}}
\renewcommand{\thesubsection}{SM\arabic{section}.\arabic{subsection}}



\vspace*{1cm} 

\section*{\centering SUPPLEMENTARY MATERIAL}

\addtocontents{toc}{\protect\vspace{20pt}}
\addcontentsline{toc}{section}{SUPPLEMENTARY MATERIAL}

\input{pebmc_suppmat.tex}

\end{document}

%% file: ex_shared.tex
\usepackage{lipsum}
\usepackage{amsfonts}
\usepackage{graphicx}
\usepackage{epstopdf}
\usepackage{algorithm,algorithmic}

\usepackage[english]{babel}

\usepackage{amsmath}
\usepackage{multirow}
\usepackage{bm}
\usepackage{amssymb}
\usepackage{graphicx}
\usepackage{bbm}
\usepackage{tikz}
\usepackage{hyperref}
\usepackage[nocompress]{cite}
\usepackage{subcaption}
\DeclareMathOperator*{\argmin}{arg\,min}
\DeclareMathOperator*{\argmax}{arg\,max}

\ifpdf
  \DeclareGraphicsExtensions{.eps,.pdf,.png,.jpg}
\else
  \DeclareGraphicsExtensions{.eps}
\fi

\usepackage{enumitem}
\setlist[enumerate]{leftmargin=.5in}
\setlist[itemize]{leftmargin=.5in}

\newsiamremark{remark}{Remark}
\newsiamremark{hypothesis}{Hypothesis}
\crefname{hypothesis}{Hypothesis}{Hypotheses}
\newsiamthm{claim}{Claim}
\newsiamremark{fact}{Fact}
\crefname{fact}{Fact}{Facts}

\headers{Bayesian Calibration with Functional Outputs Using
Elastic Partial Matching}{P. Casteras, J. Bect, J. Garnier, G. Salin}

\title{Bayesian Calibration with Functional Outputs Using
Elastic Partial Matching}

\author{%
  Paul Castéras\footnotemark[2] \footnotemark[3] \thanks{%
    CEA, DAM, DIF, F-91297, Arpajon, France %
    (\email{paul.casteras@cea.fr}, \email{gwenael.salin@cea.fr}).} %
  \and Julien Bect\thanks{%
    Université Paris-Saclay, CNRS, CentraleSupélec, %
    Laboratoire des signaux et systèmes, 91190 Gif-sur-Yvette, France. %
    \email{julien.bect@centralesupelec.fr}.} %
  \and Josselin Garnier\thanks{%
    Centre de Mathématiques Appliquées, Ecole polytechnique, %
    Institut Polytechnique de Paris, 91120 Palaiseau, France. %
    \email{josselin.garnier@polytechnique.edu}}
  \and Gwenaël Salin\footnotemark[1]%
}

\usepackage{amsopn}

\makeatletter
\AtBeginDocument{\let\SMorigsection\section}
\makeatother

\makeatletter
\renewcommand{\tableofcontents}{%
  \begin{center}
    \mbox{\color{header1}\footnotesize\HLtext CONTENTS}
  \end{center}
  \addvspace{.15in}\nopagebreak
  \@starttoc{toc}%
  \addvspace{.25in}}

\newcommand{\l@section}[2]{\def\@linkcolor{black}\addvspace{1em} \@dottedtocline{1}{0em}{3.2em}{\bfseries  #1}{#2}}
\newcommand{\l@subsection}[2]{\def\@linkcolor{black}\@dottedtocline{2}{3.2em}{4.5em}{#1}{#2}}
\newcommand{\l@subsubsection}{\def\@linkcolor{black}\@dottedtocline{3}{7.7em}{6em}}
\newcommand{\l@paragraph}{\@dottedtocline{4}{10em}{6em}}
\makeatother

%% file: pebmc_suppmat.tex
\section{Details about the anharmonic pendulum example with the experiment generated by Equation~\eqref{exp_pendulum}}

\begin{figure}[h]
    \centering
\includegraphics[width=0.5\textwidth]{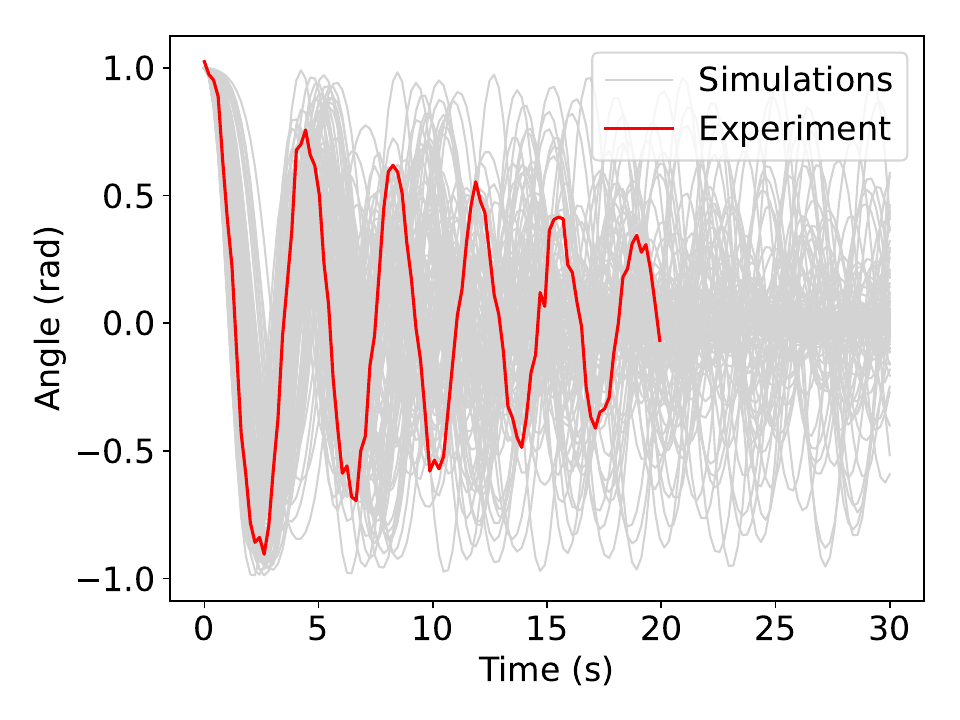}
\caption{Dataset for the anharmonic pendulum with the experiment generated by Equation~\eqref{exp_pendulum}.}
\label{pendule}
\end{figure}

As mentioned at the end of Section~\ref{app_pendulum}, the proposed method remains applicable when no artificial phase discrepancy is introduced in the experimental data,
and can still improve surrogate modeling. This section provides further details on that setting.
The simulations and the experimental data are shown in Figure~\ref{pendule}.

\begin{figure}[h]
\centering
\begin{minipage}[b]{0.32\textwidth}
\centering
    Elastic
\end{minipage}
\hfill
\begin{minipage}[b]{0.32\textwidth}
\centering
    Rescaling
\end{minipage}
\hfill
\begin{minipage}[b]{0.32\textwidth}
\centering
    Partial elastic
\end{minipage}

\begin{minipage}[b]{0.32\textwidth}
\centering
\includegraphics[width=\textwidth]{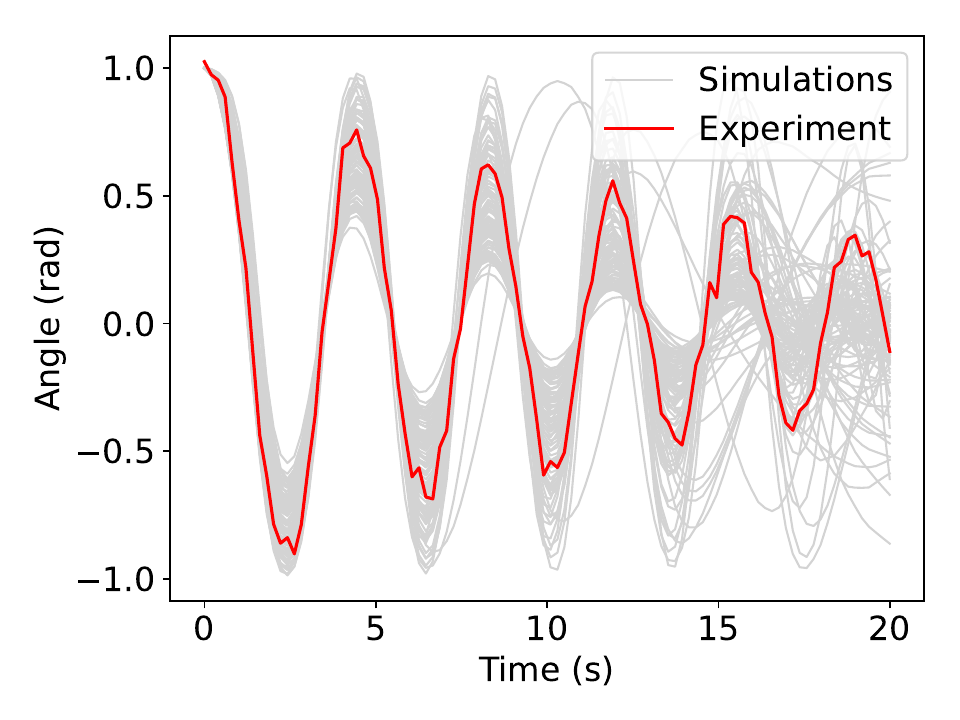}
\end{minipage}
\hfill
\begin{minipage}[b]{0.32\textwidth}
\centering
\includegraphics[width=\textwidth]{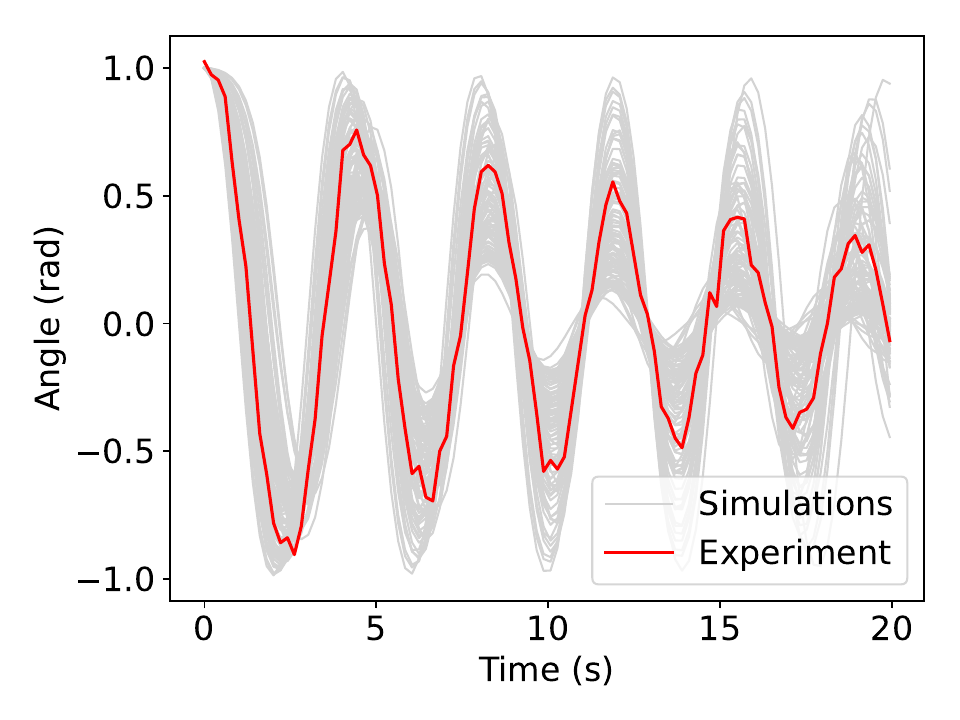}
\end{minipage}
\hfill
\begin{minipage}[b]{0.32\textwidth}
\centering
\includegraphics[width=\textwidth]{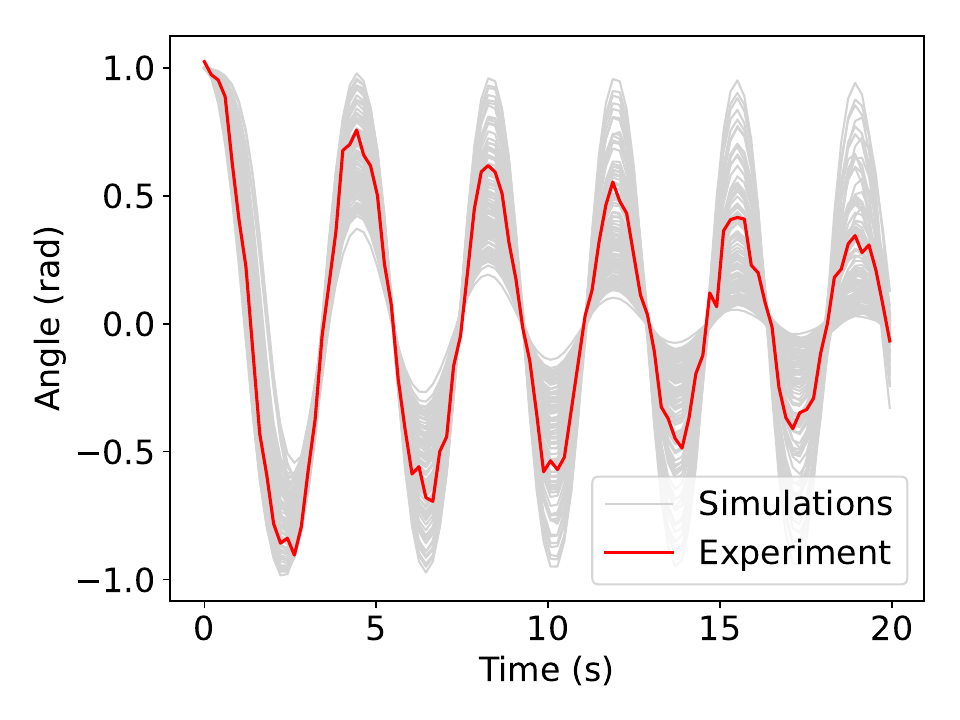}
\end{minipage}

\vspace{1em}

\begin{minipage}[b]{0.32\textwidth}
\centering
\includegraphics[width=\textwidth]{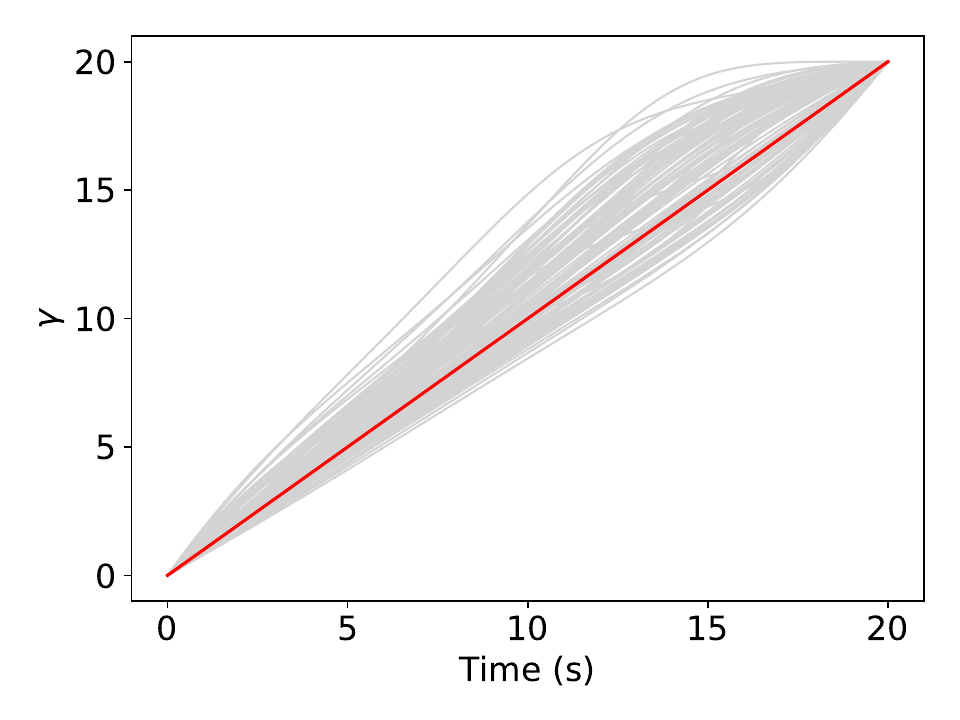}
\end{minipage}
\hfill
\begin{minipage}[b]{0.32\textwidth}
\centering
\includegraphics[width=\textwidth]{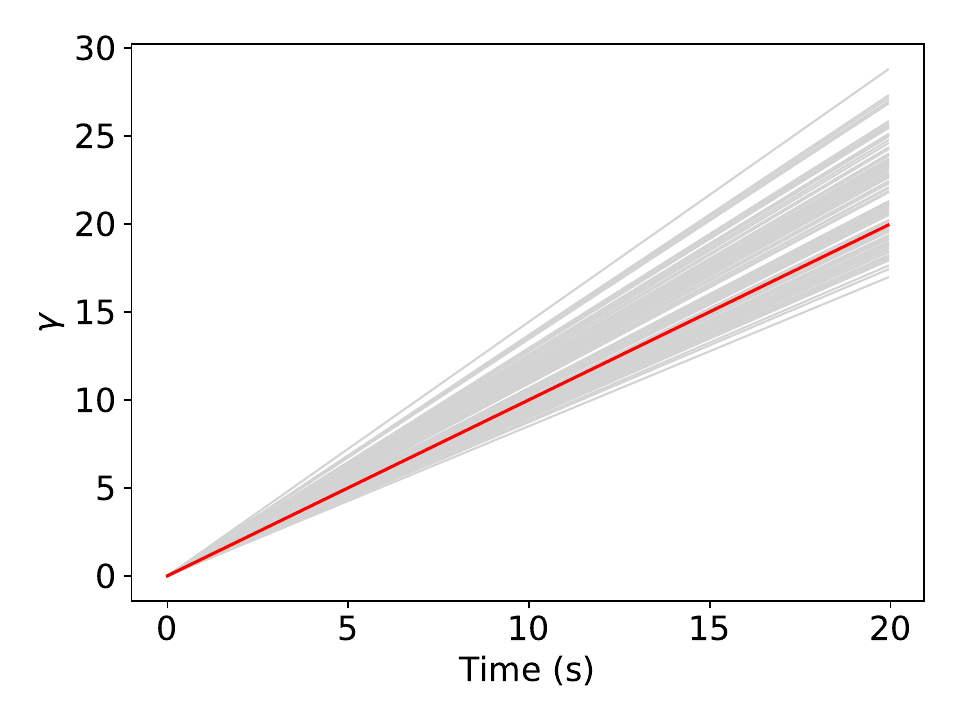}
\end{minipage}
\hfill
\begin{minipage}[b]{0.32\textwidth}
\centering
\includegraphics[width=\textwidth]{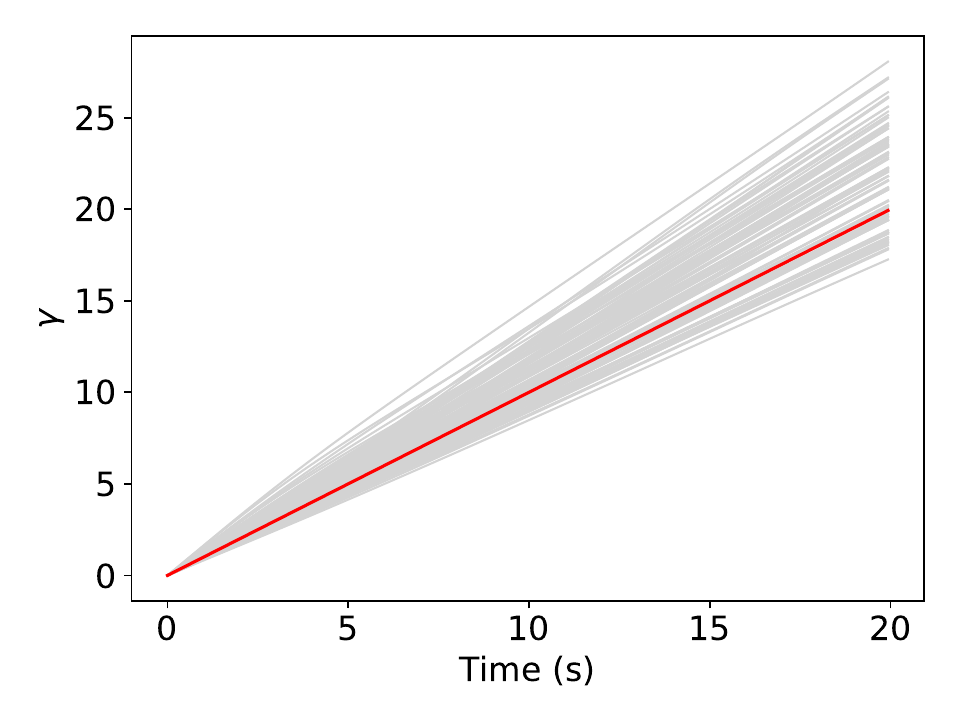}
\end{minipage}

\caption{Amplitudes (top) and time warpings (bottom) of the anharmonic pendulum with the experiment generated by Equation~\eqref{exp_pendulum} for the different alignment frameworks.}
\label{pendule_amplitude_phase}
\end{figure}

We first align the simulations with the experiment using the three different alignment methods. 
The resulting amplitudes and time warpings are shown in Figure~\ref{pendule_amplitude_phase}. 
Since the functions do not all exhibit the same number of oscillations, the elastic alignment fails to simplify the data representation. 
Both the rescaling and the partial elastic approaches lead to a simpler representation of the data.
However, the partial elastic method aligns the simulations to the experiment more accurately.

\begin{figure}[h]
\centering
\panel{a}{Images/pendule_meta_notela.pdf}\par\vspace{0.6em}
\panel{b}{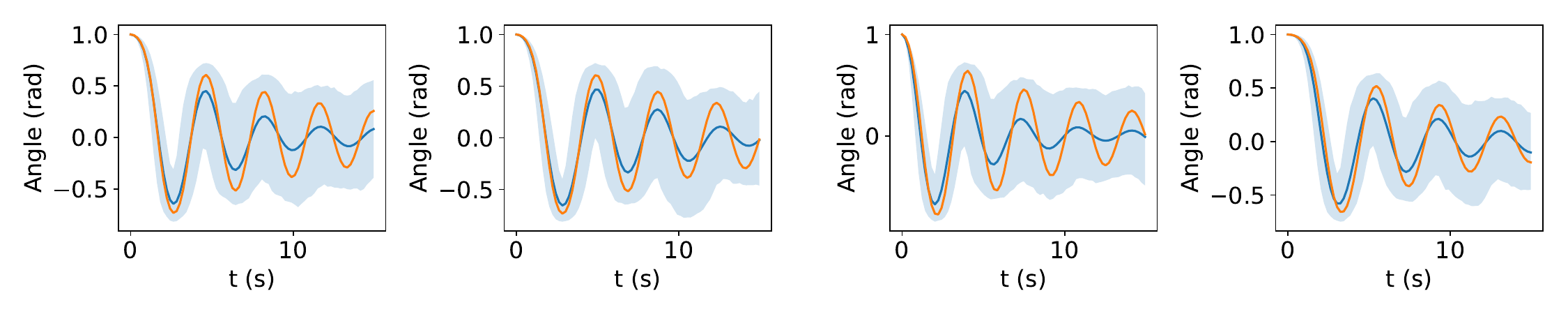}\par\vspace{0.6em}
\panel{c}{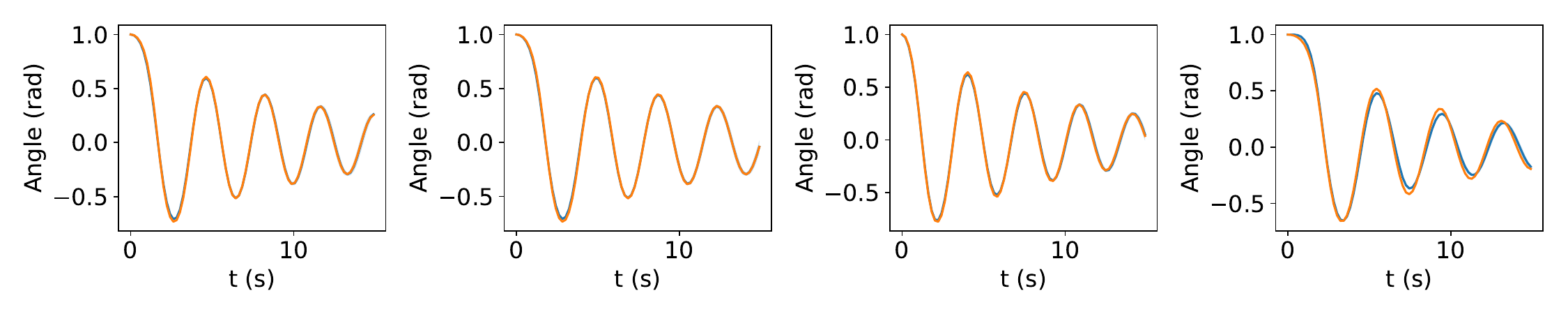}\par\vspace{0.6em}
\panel{d}{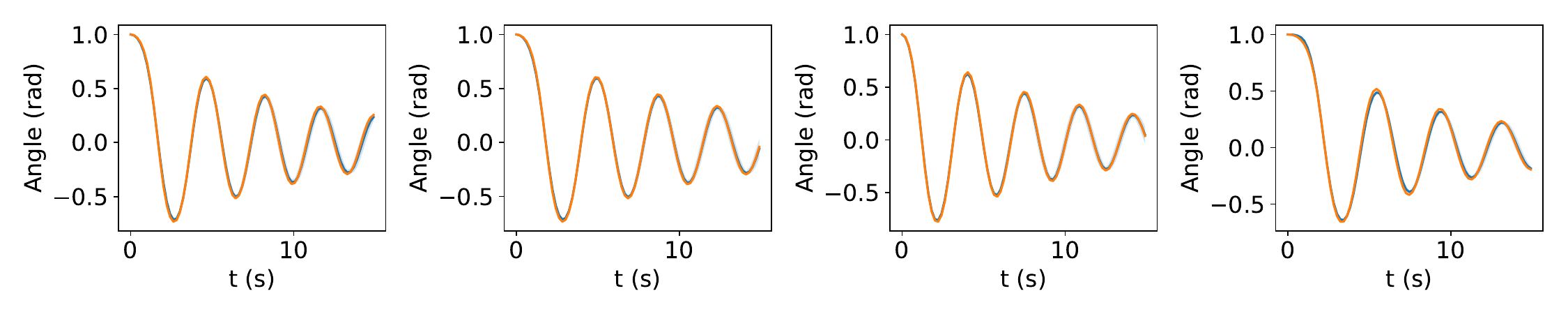}
\caption{Reconstructed anharmonic pendulum outputs with the experiment generated by Equation~\eqref{exp_pendulum} for four different test inputs for (a) the non-elastic surrogate, (b) the elastic surrogate, (c) the surrogate with rescaling, and (d) the partial elastic surrogate.}
    \label{surrogate_outputs}
\end{figure}

The surrogate model outputs for four different inputs are shown in Figure~\ref{surrogate_outputs}.

\begin{figure}[h]
\includegraphics[width=0.8\textwidth]{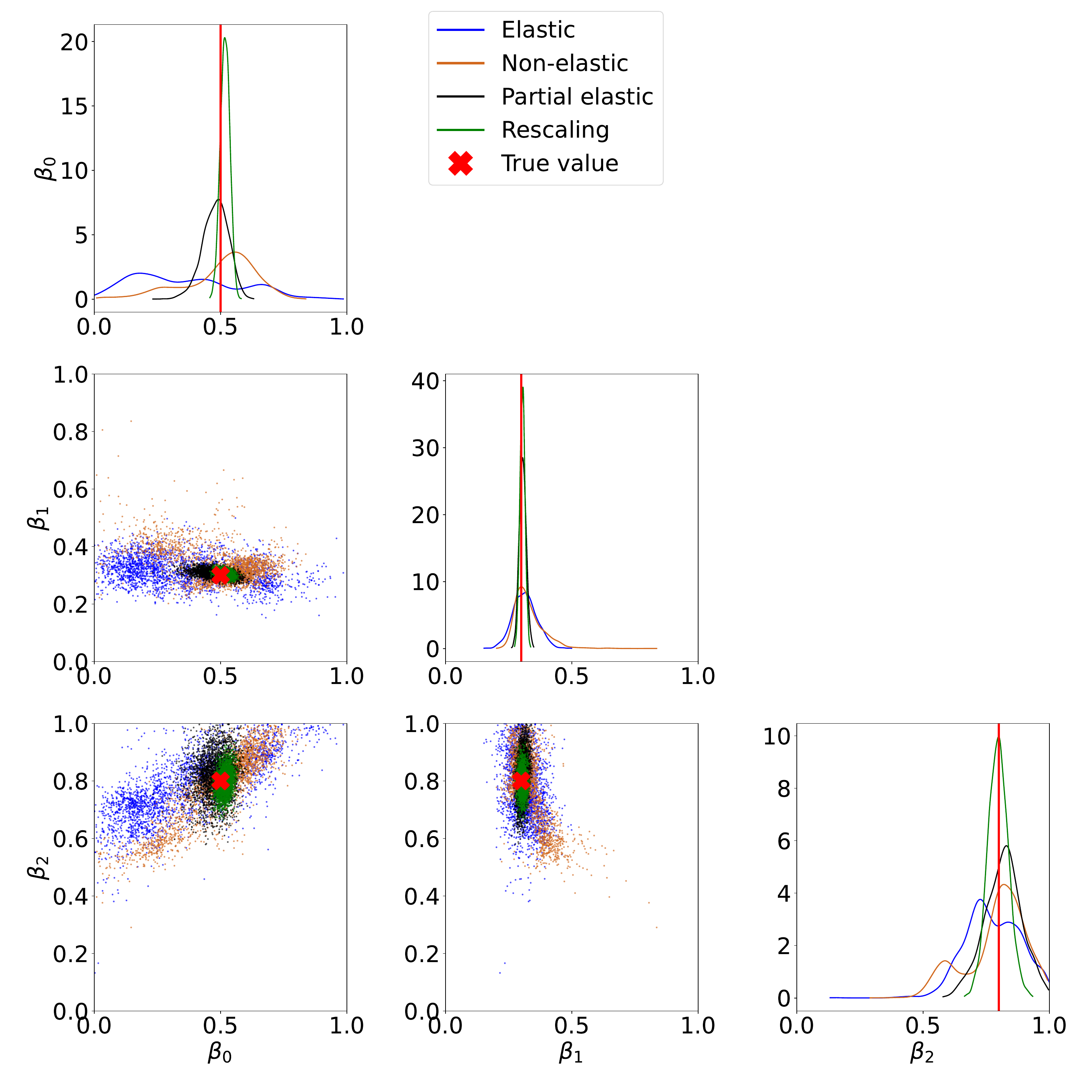}
\caption{ Sampling results (pairplots and marginals) for the different calibration methods with the experiment generated by Equation~\eqref{exp_pendulum} for the anharmonic pendulum.}
\label{pendule_sampling}
\end{figure}

The posterior distributions are shown in Figure~\ref{pendule_sampling}. 
All the  posterior distributions are similar for the  parameter $\beta_2$.
The distributions obtained with the non-elastic and elastic methods are more diffuse for the other two parameters, which can be explained by the lower accuracy of their surrogate models.
While the rescaling and partial elastic methods lead to similar posterior distributions for the parameter $\beta_1$, 
the distribution of $\beta_0$ is more concentrated under the rescaling approach. This parameter governs the initial behavior of the simulations. 
In the rescaling framework, this effect is captured through the amplitude representation, whereas in the partial elastic method, it is encoded in the shooting vector.

\section{Calibration of the anharmonic pendulum with SRVF alignment}
\label{partial_srvf}

\subsection{With the experiment generated by Equation~\eqref{exp_pendulum_discr}}

In Section~\ref{app_pendulum}, the calibration results were obtained 
using an alignment method based on the approximation of a calibration 
problem. As an alternative, Francom \textit{et al.}~\cite{francom} rely on an SRVF-based 
alignment method, which is the current state of the art for function 
alignment. We therefore reproduce the calibration study with this 
method to assess the sensitivity of our results to the choice of 
alignment procedure. The SRVF-based method (see 
Appendices~\ref{align} and~\ref{partial}) involves two hyperparameters 
that penalize deviations of the time warping from the identity, which 
we set to $\lambda = 0.05$ and $\kappa = 0$.

\begin{figure}[h]
\centering
\begin{minipage}[b]{0.32\textwidth}
\centering
        Elastic
\includegraphics[width=\textwidth]{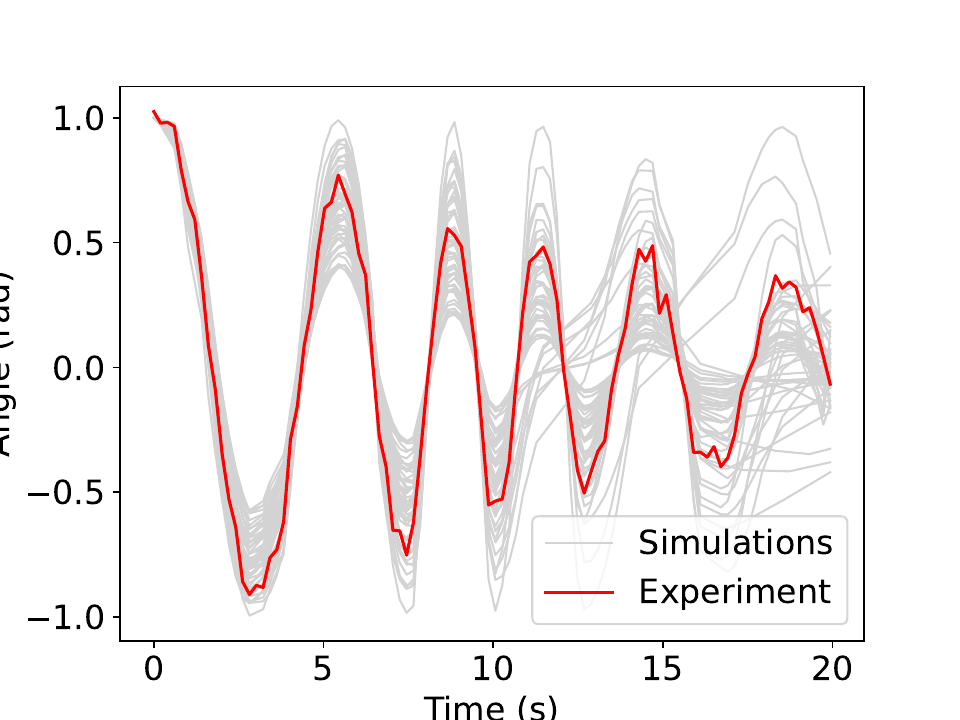}
\end{minipage}
\hfill
\begin{minipage}[b]{0.32\textwidth}
\centering
        Rescaling
\includegraphics[width=\textwidth]{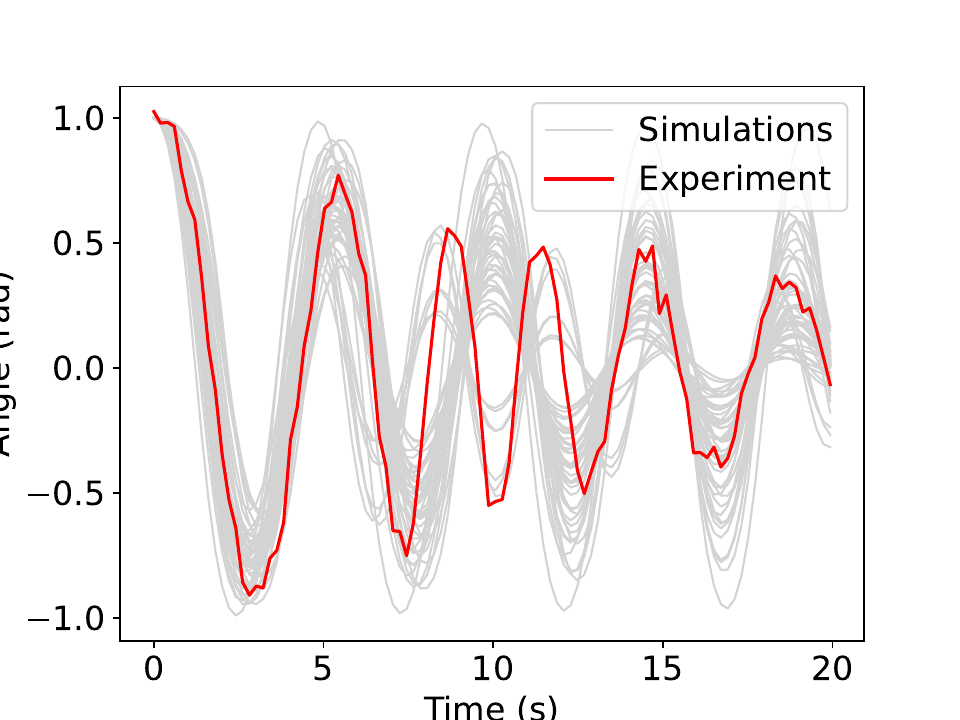}
\end{minipage}
\hfill
\begin{minipage}[b]{0.32\textwidth}
\centering
        Partial elastic
\includegraphics[width=\textwidth]{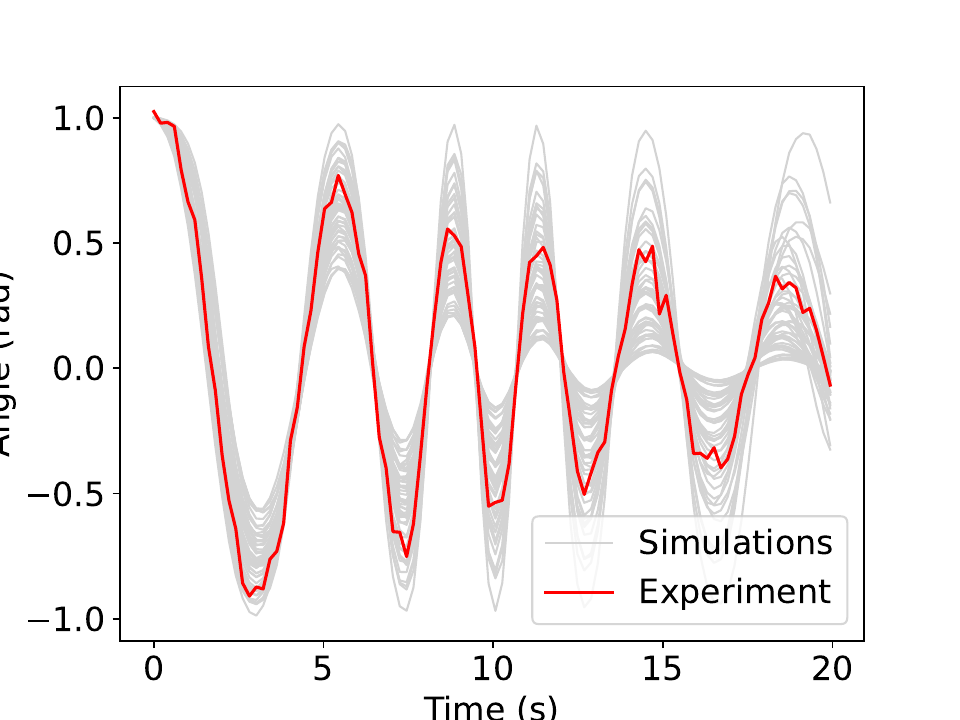}
\end{minipage}
\caption{Amplitudes of the anharmonic pendulum outputs with the experiment generated by Equation~\eqref{exp_pendulum_discr} for the different
SRVF-based alignment frameworks.}
\label{pendule_amplitude_SRVFdiscr}
\end{figure}

The amplitudes obtained after alignment are displayed in
Figure~\ref{pendule_amplitude_SRVFdiscr}.

\begin{figure}[h]
\includegraphics[width=0.98\textwidth]{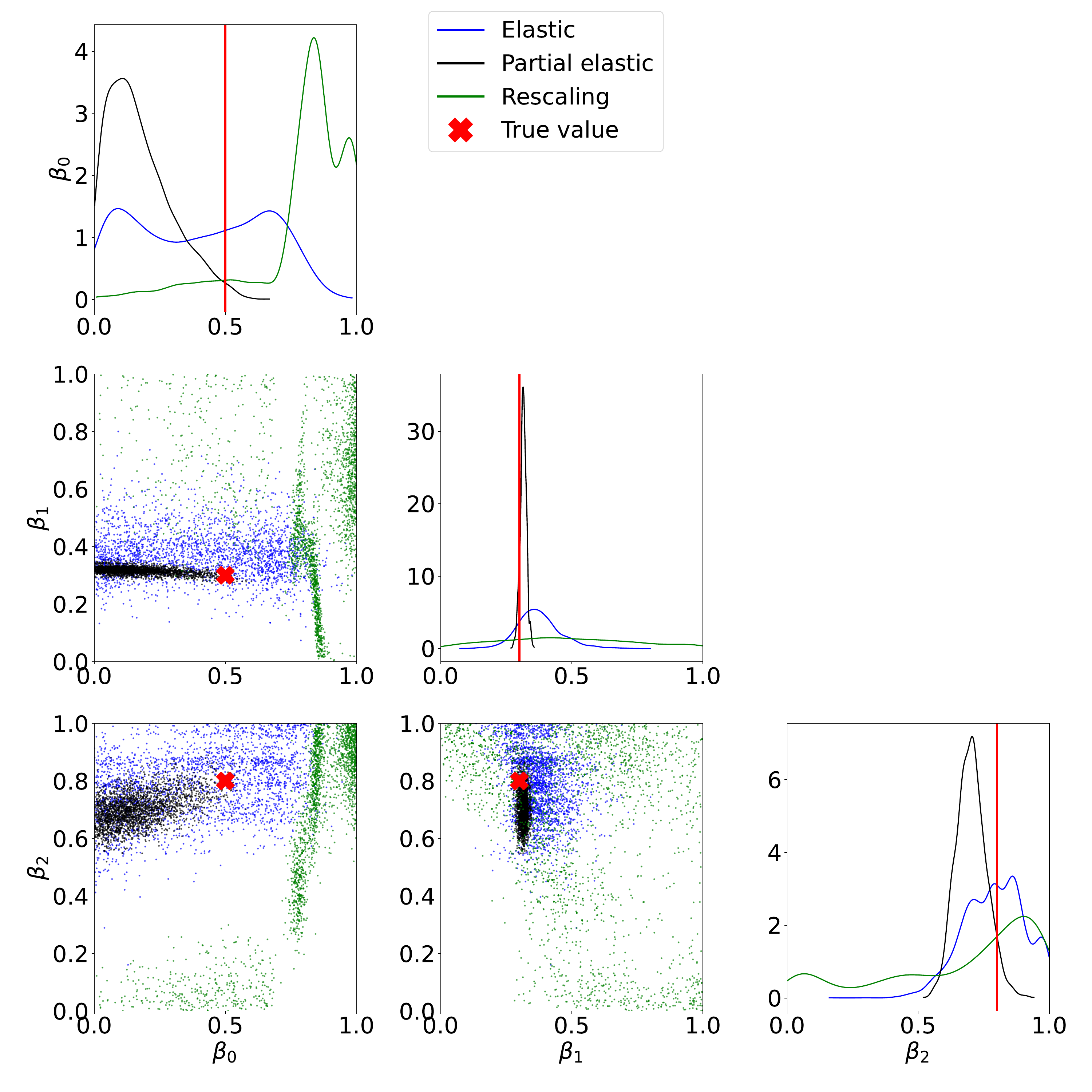}
\caption{Sampling results (pairplots and marginals) for the different
methods based on SRVF alignments with the experiment generated by Equation~\eqref{exp_pendulum_discr}.}
\label{pendule_sampling_SRVF_discr}
\end{figure}

\begin{figure}[h]
\includegraphics[width=0.8\textwidth]{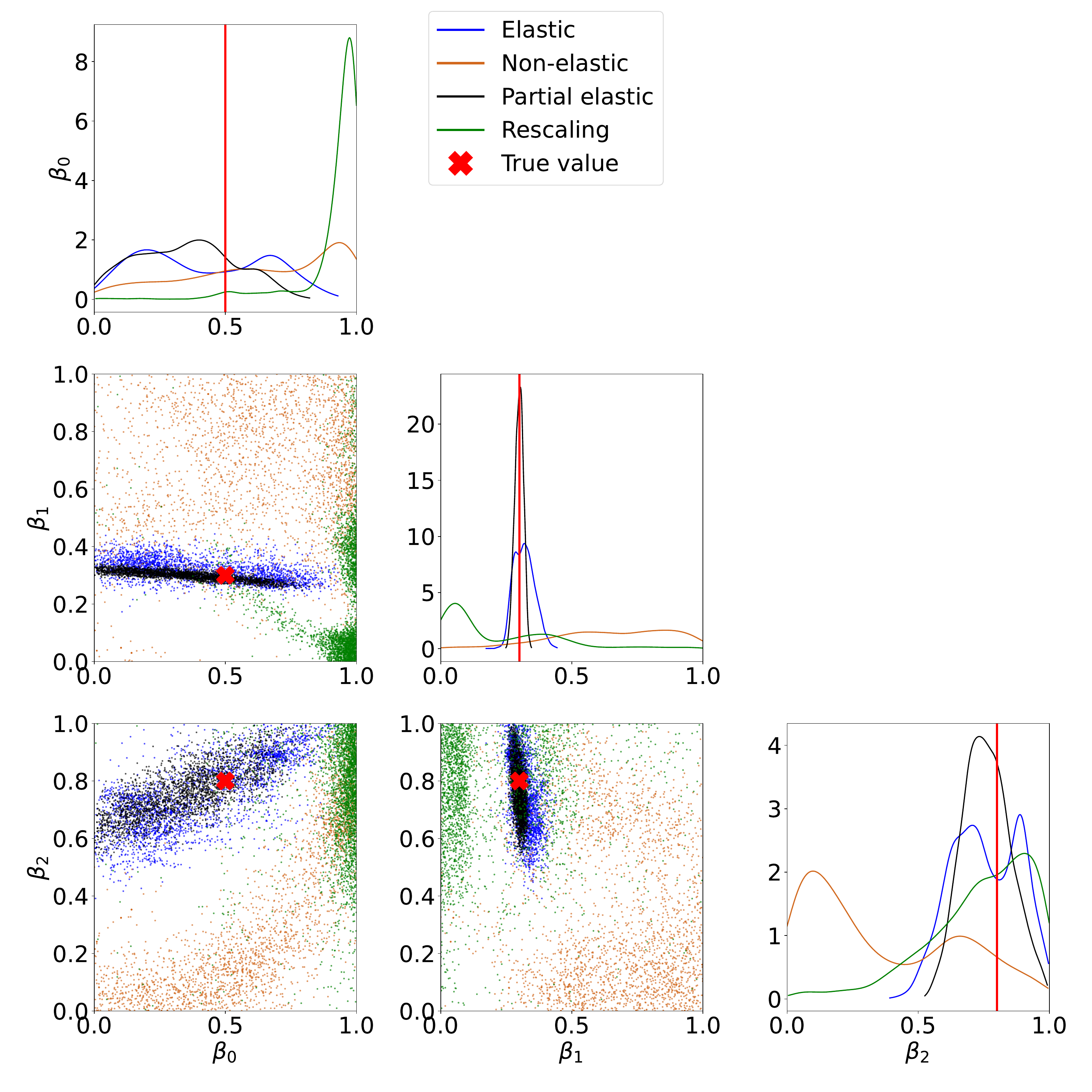}
\caption{Sampling results for the different methods for the anharmonic pendulum with the experiment generated by Equation~\eqref{exp_pendulum_discr}.}
\label{pendule_samplingdiscr}
\end{figure}

The corresponding sampling results are reported in
Figure~\ref{pendule_sampling_SRVF_discr} and can be compared directly
with those obtained using the approximate calibration-based alignment,
reproduced in Figure~\ref{pendule_samplingdiscr}. 
In both cases, the most concentrated posterior is the one using partial elastic alignment.

\subsection{With the experiment generated by Equation~\eqref{exp_pendulum}}

We now repeat the same comparison in the setting where the experiment is generated by Equation~\eqref{exp_pendulum}.
As before, the alignment step relies on the SRVF-based method of
\cite{francom} (see Appendices~\ref{align} and~\ref{partial}), with the
same hyperparameter values $\lambda = 0.05$ and $\kappa = 0$.

\begin{figure}[h]
    \centering
    \begin{minipage}[b]{0.32\textwidth}
        \centering
        Elastic
        \includegraphics[width=\textwidth]{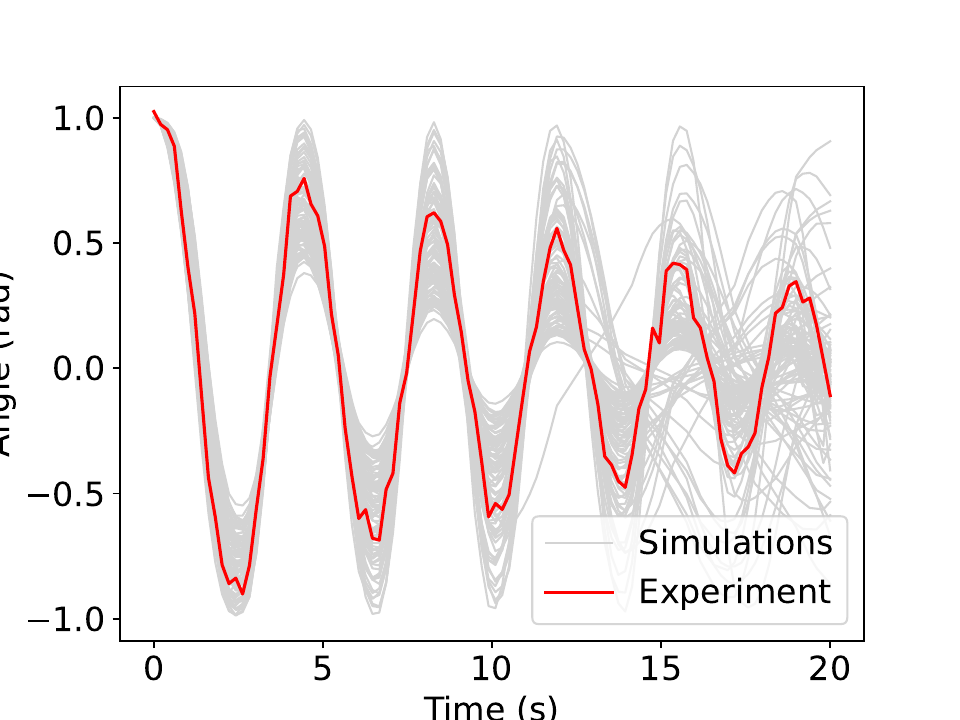}
    \end{minipage}
    \hfill
    \begin{minipage}[b]{0.32\textwidth}
        \centering
        Rescaling
        \includegraphics[width=\textwidth]{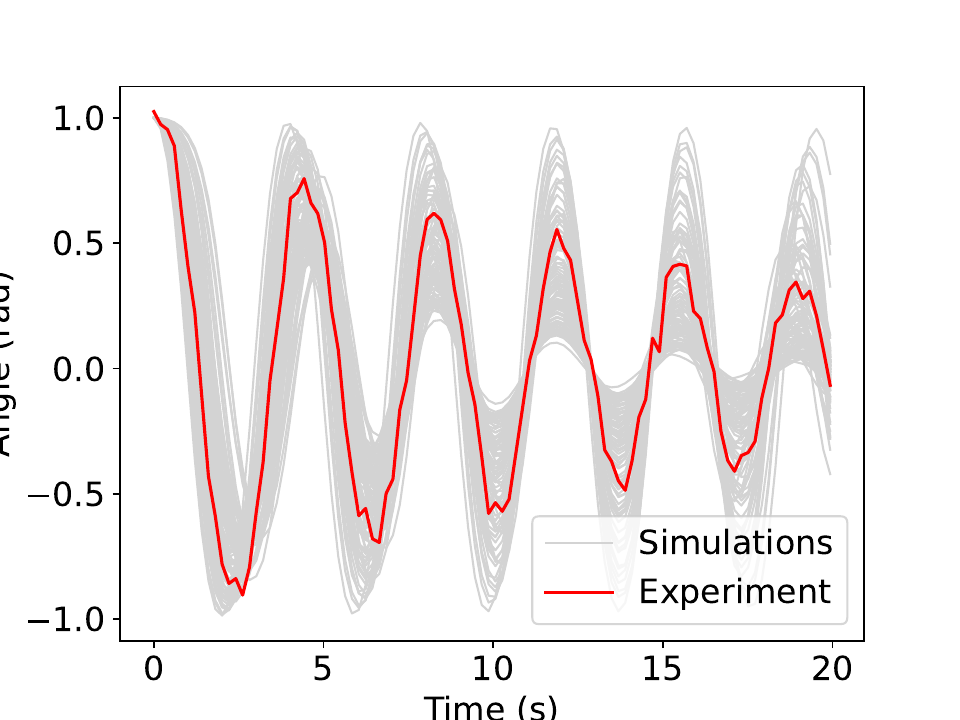}
    \end{minipage}
    \hfill
    \begin{minipage}[b]{0.32\textwidth}
        \centering
        Partial elastic
        \includegraphics[width=\textwidth]{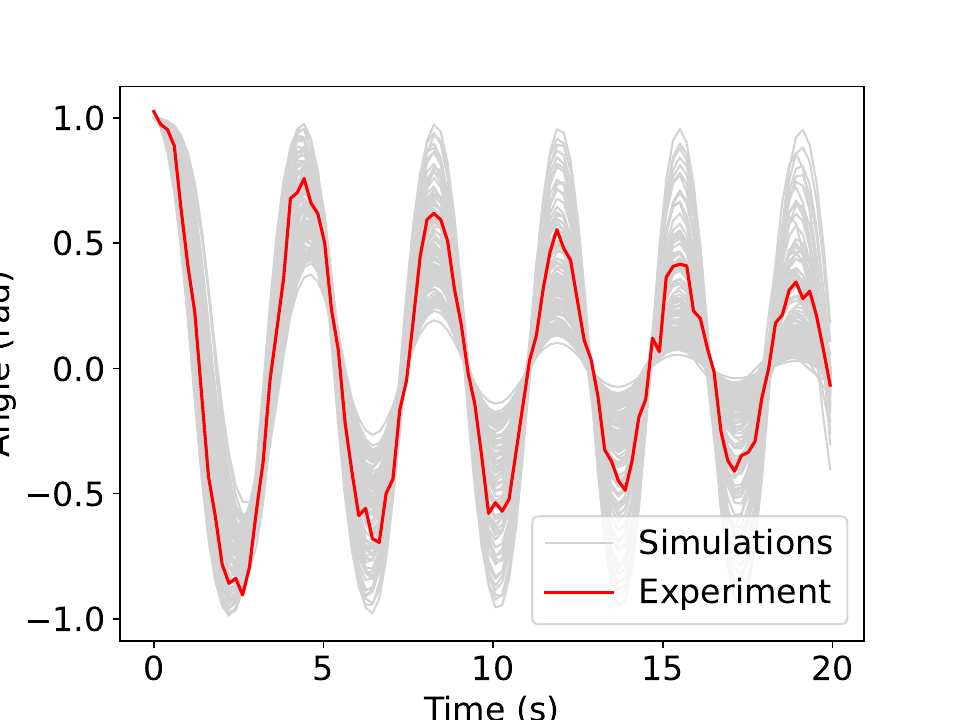}
    \end{minipage}
\caption{Amplitudes of the anharmonic pendulum outputs for the different SRVF-based alignment frameworks with the experiment generated by Equation~\eqref{exp_pendulum}.}
\label{pendule_amplitude_SRVF}
\end{figure}

\begin{figure}[h]
\includegraphics[width=0.98\textwidth]{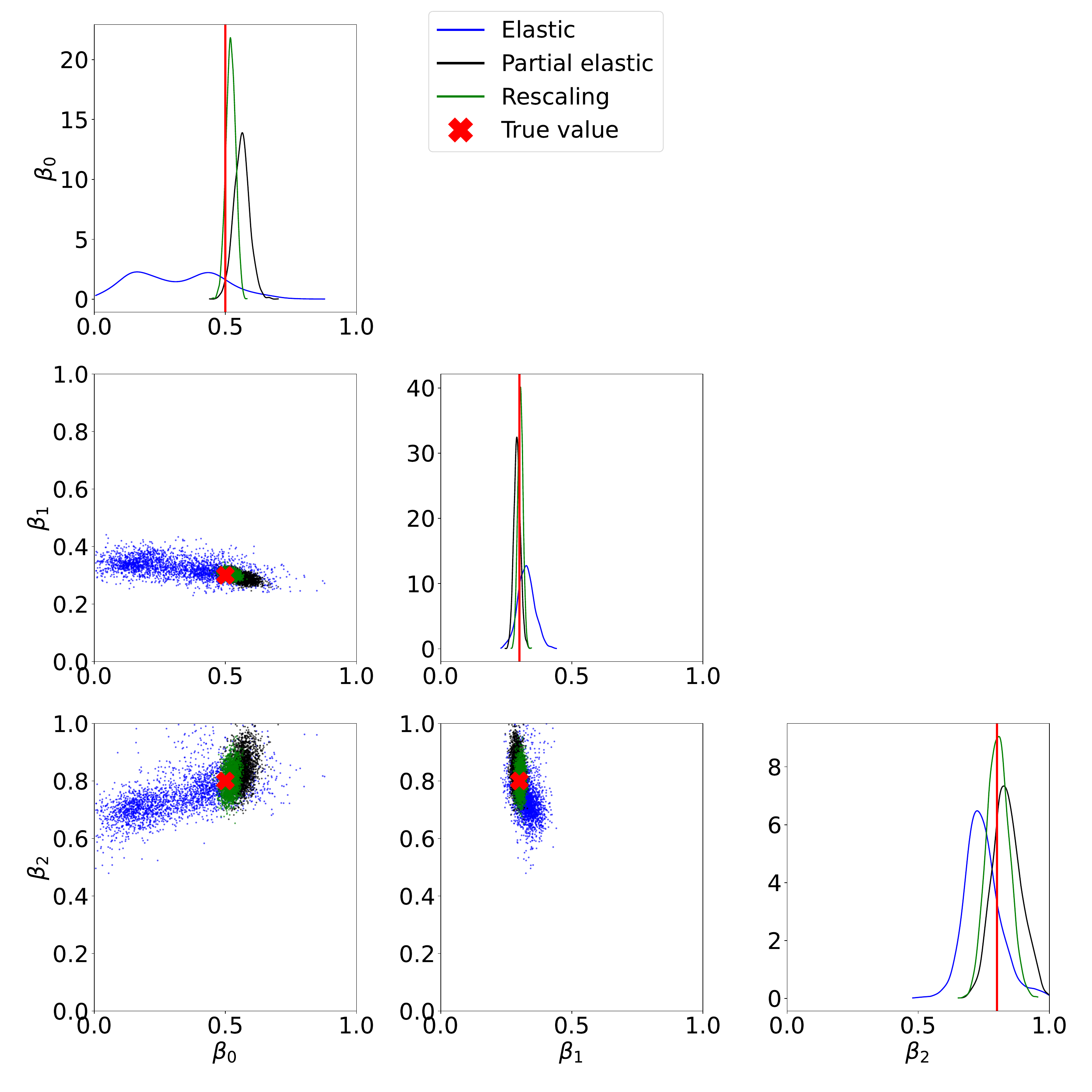}
\caption{ Sampling results (pairplots and marginals) for the different methods based on SRVF alignments with the experiment generated by Equation~\eqref{exp_pendulum}.}
\label{pendule_sampling_SRVF}
\end{figure}

The alignment results are shown in Figure~\ref{pendule_amplitude_SRVF}: 
the amplitudes obtained with both methods are very close. 
Table~\ref{tab:pendule_meta_SRVF} reports a slight difference in the 
number of PCA components selected by each method, which translates 
into small differences in the other metrics. Finally, the sampling 
results in Figure~\ref{pendule_sampling_SRVF} lead to the same 
conclusions as before: partial elastic calibration and calibration 
with rescaling yield more concentrated posterior distributions than 
the other two methods. Overall, the two alignment procedures produce 
similar results.

\begin{table}[h]
\centering
\begin{tabular}{|c|c|c|c|c|}
\hline
& $Q^2$ & $\mathrm{IAE}$ & $\mathrm{CRPS}$& $N_{\mathrm{PCA}}$ \\
\hline

Elastic & 0.57 & 0.04 &0.10& 15 (10+5+0) \\
 \hline
Rescaling & 0.95 & 0.05 & 0.02 & 4 (3+0+1) \\
\hline
Partial elastic & 0.96 & 0.04 & 0.02 & 8 (4+3+1)\\
\hline
\end{tabular}
\caption{Comparison of the different surrogate models for different metrics computed on a test dataset for the anharmonic pendulum for SRVF-based alignment methods
with the experiment generated by Equation~\eqref{exp_pendulum}.}
\label{tab:pendule_meta_SRVF}
\end{table}

\section{Calibration of the equation of state with affine rescaling}

\begin{figure}[h]
    \centering
    \includegraphics[width=0.98\textwidth]{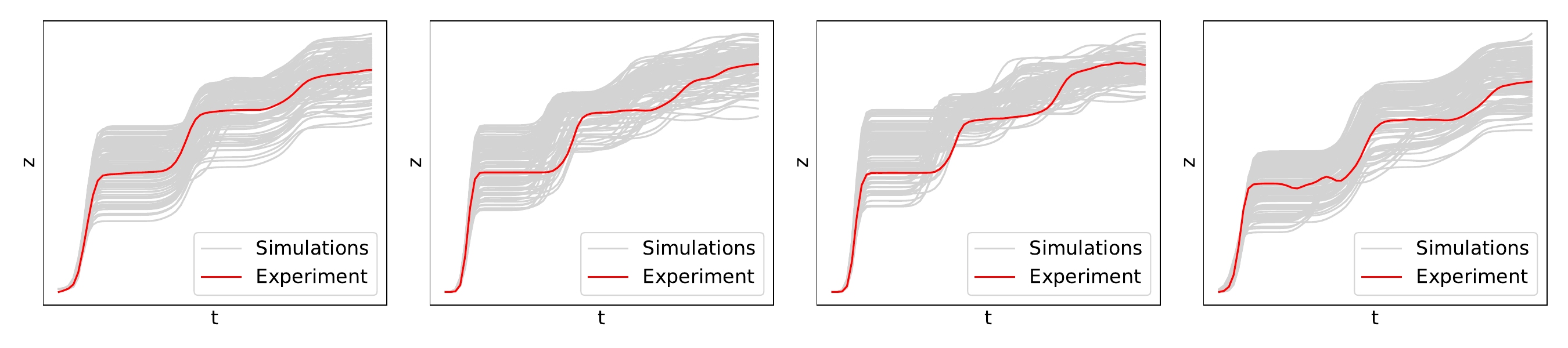}
\caption{Partial amplitudes for the affine time warping of the equation of state for the different experiments.}
\label{amplitude_EOS_affine}
\end{figure}

\begin{figure}[h]
    \centering
    \includegraphics[width=0.98\textwidth]{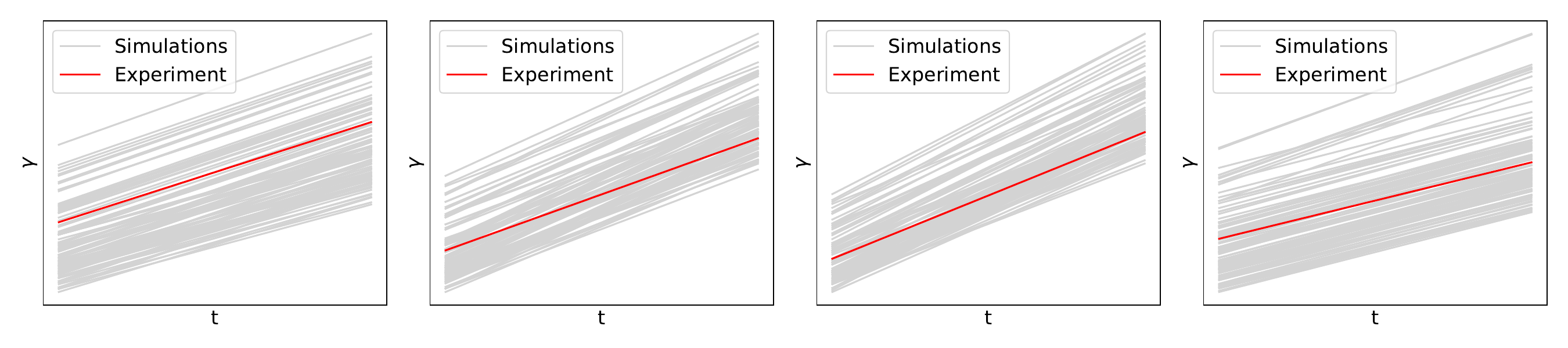}
\caption{Partial phases for the affine time warping of the equation of state for the different experiments.}
\label{phase_EOS_affine}
\end{figure}

We compare here the method proposed in the article for the calibration of the equation of state with one based on an affine rescaling. 
The amplitudes and phases obtained with the latter are shown in Figures~\ref{amplitude_EOS_affine} and~\ref{phase_EOS_affine}. 

\begin{figure}[h]
    \centering

    \includegraphics[width=0.98\textwidth]{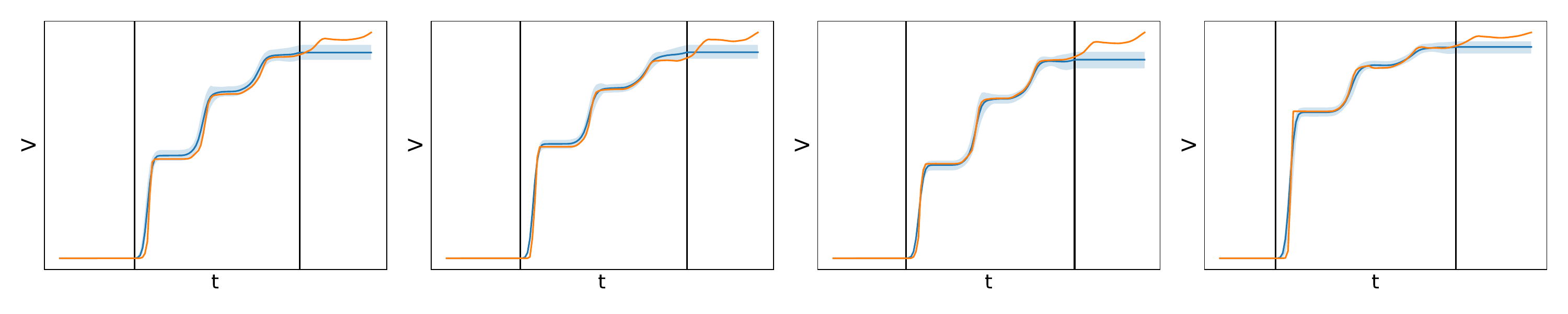}\\
    Experiment 1\par\vspace{0.6em}

    \includegraphics[width=0.98\textwidth]{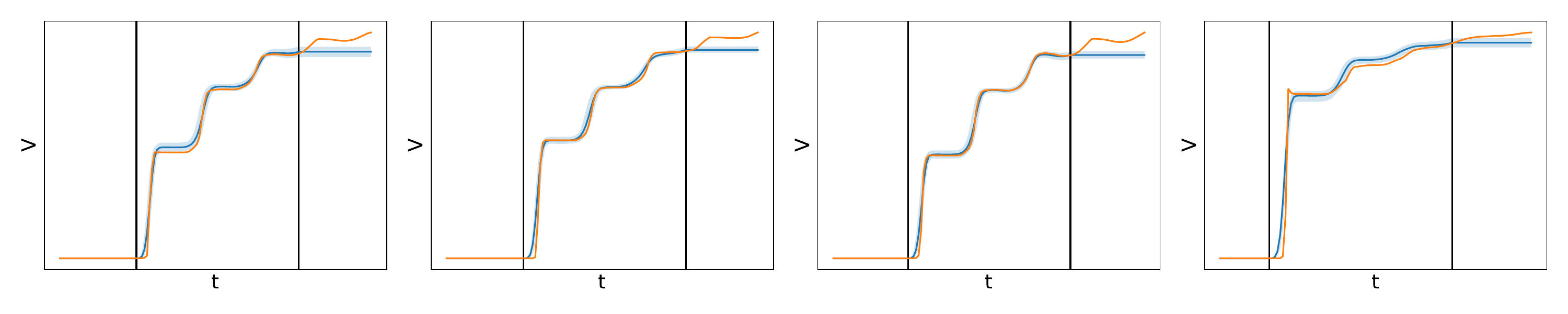}\\
    Experiment 2\par\vspace{0.6em}

    \includegraphics[width=0.98\textwidth]{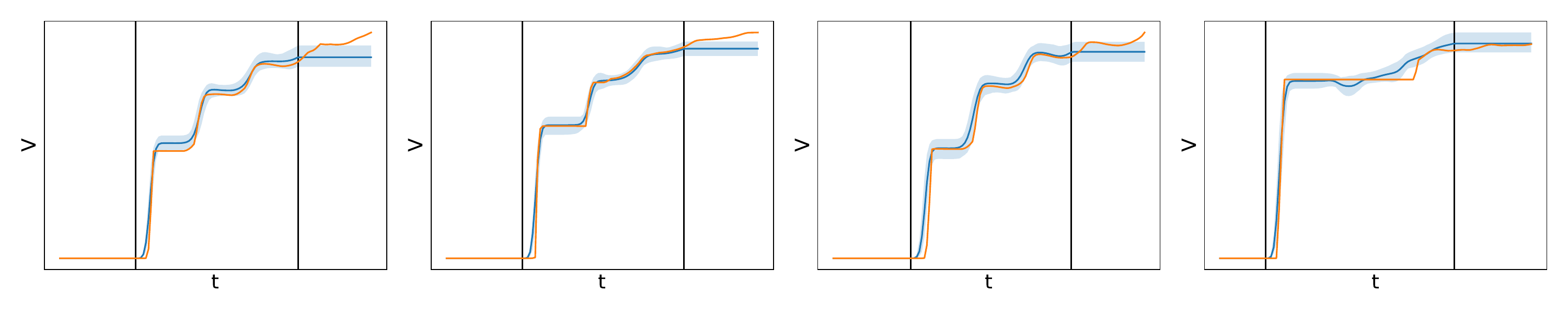}\\
    Experiment 3\par\vspace{0.6em}

    \includegraphics[width=0.98\textwidth]{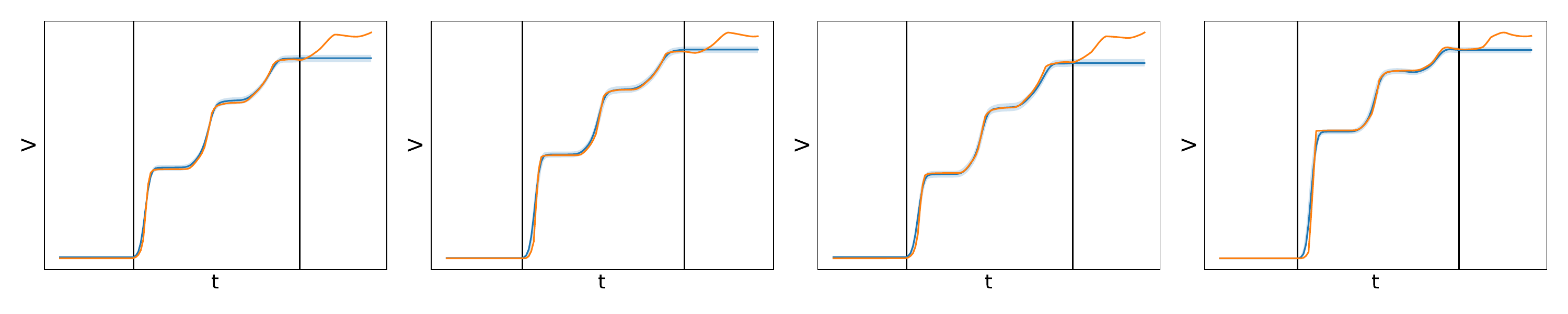}\\
    Experiment 4\par\vspace{0.6em}

    \caption{Reconstructed outputs for the four experiments of the equation of state with partial elastic surrogate.}
    \label{surrogate_partial_EOS}
\end{figure}

\begin{figure}[h]
    \centering

    \includegraphics[width=0.98\textwidth]{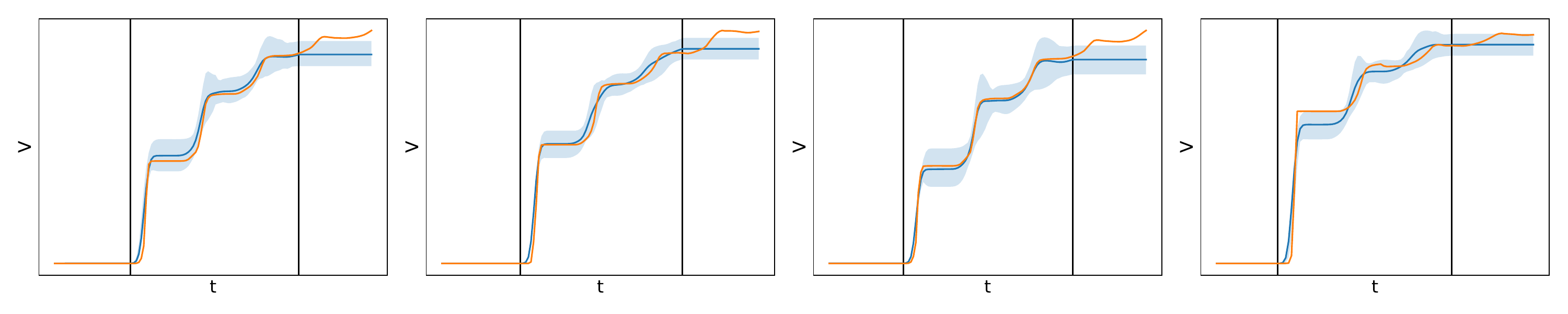}\\
    Experiment 1\par\vspace{0.6em}

    \includegraphics[width=0.98\textwidth]{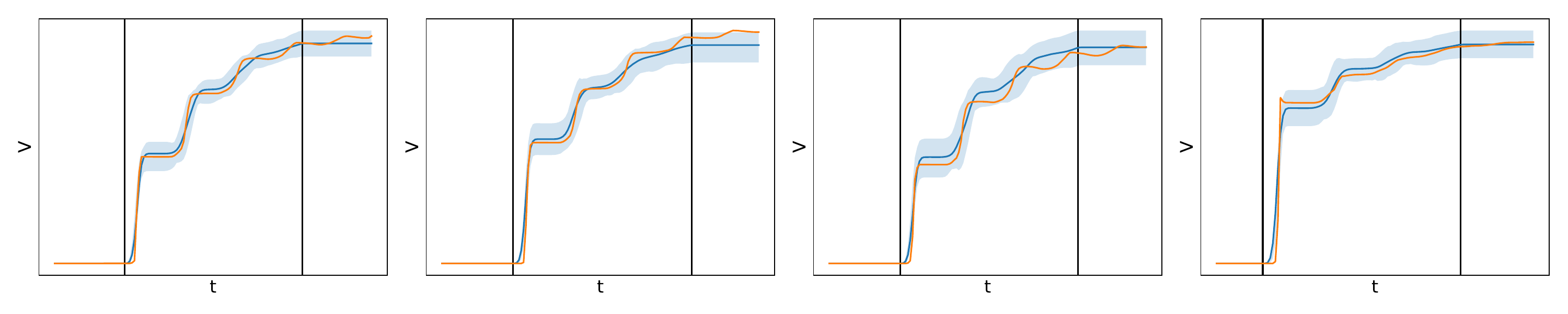}\\
    Experiment 2\par\vspace{0.6em}

    \includegraphics[width=0.98\textwidth]{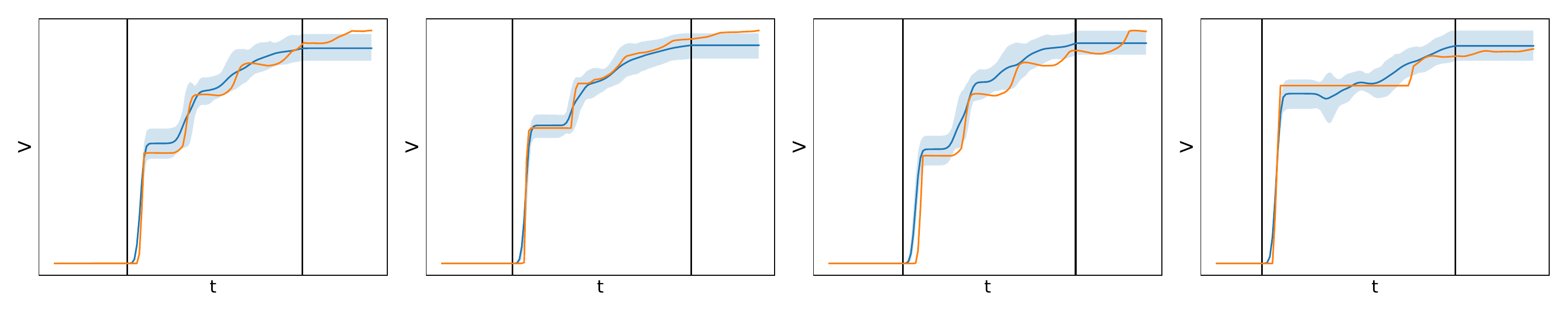}\\
    Experiment 3\par\vspace{0.6em}

    \includegraphics[width=0.98\textwidth]{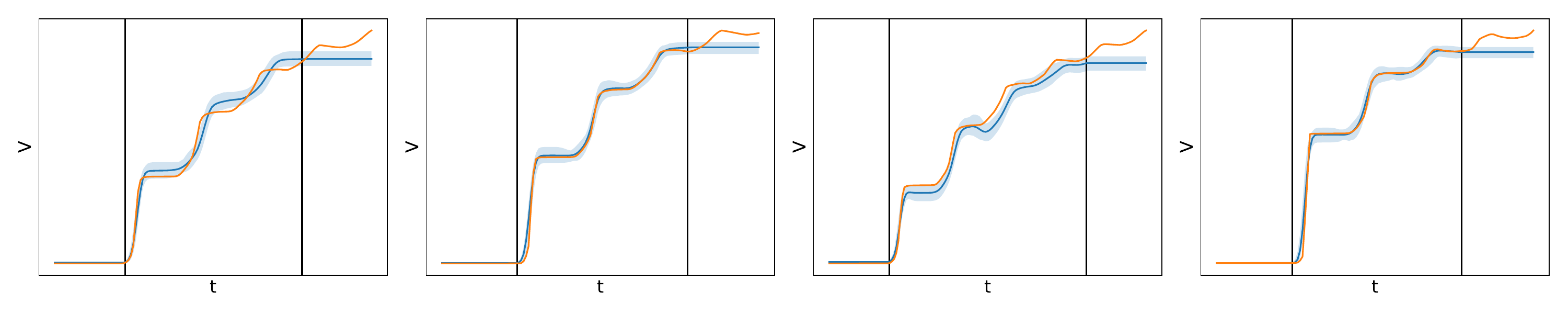}\\
    Experiment 4\par\vspace{0.6em}

    \caption{Reconstructed outputs for the four experiments of the equation of state with affine surrogate.}
    \label{surrogate_affine_EOS}
\end{figure}

We compare the surrogate models using four metrics ($\mathrm{MSE}$, $\mathrm{IAE}$, $\mathrm{CRPS}$, and $N_{\mathrm{PCA}}$). 
The results are reported in Table~\ref{tab:surrogate_EOS_metrics}.
Figures~\ref{surrogate_partial_EOS} and~\ref{surrogate_affine_EOS} show the outputs of the partial elastic and affine surrogates, respectively, for four representative inputs across the different experiments.

\begin{figure}[h]
\includegraphics[width=0.9\textwidth]{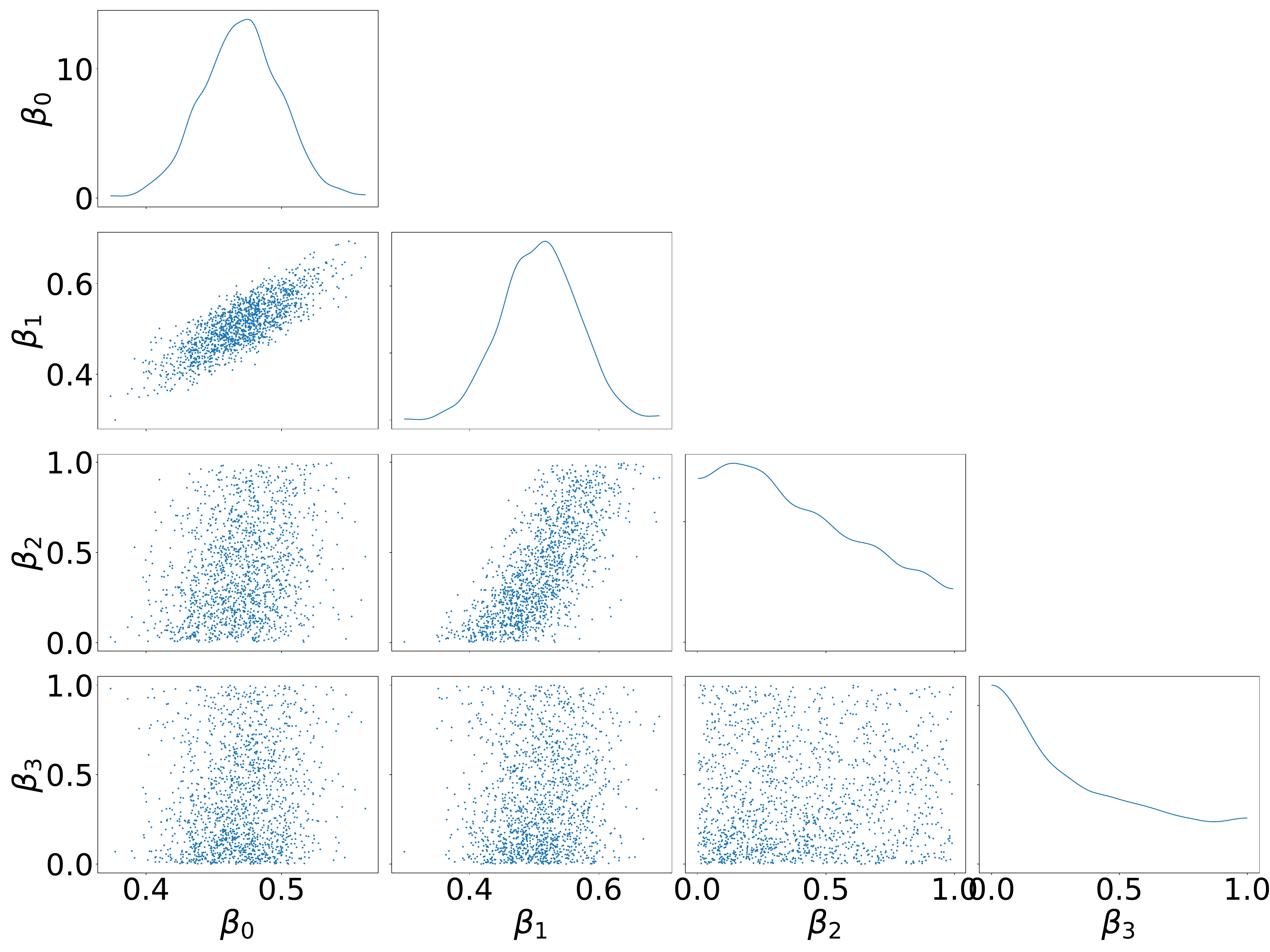}
\caption{Pairplots and marginals of the posterior distribution computed using sequential Monte Carlo for the affine calibration of the equation of state.}
\label{eos_sampling_affine}
\end{figure}

\begin{figure}[h]
    \begin{minipage}{0.48\textwidth}
    \centering
    Experiment 1
    \includegraphics[width=0.98\textwidth]{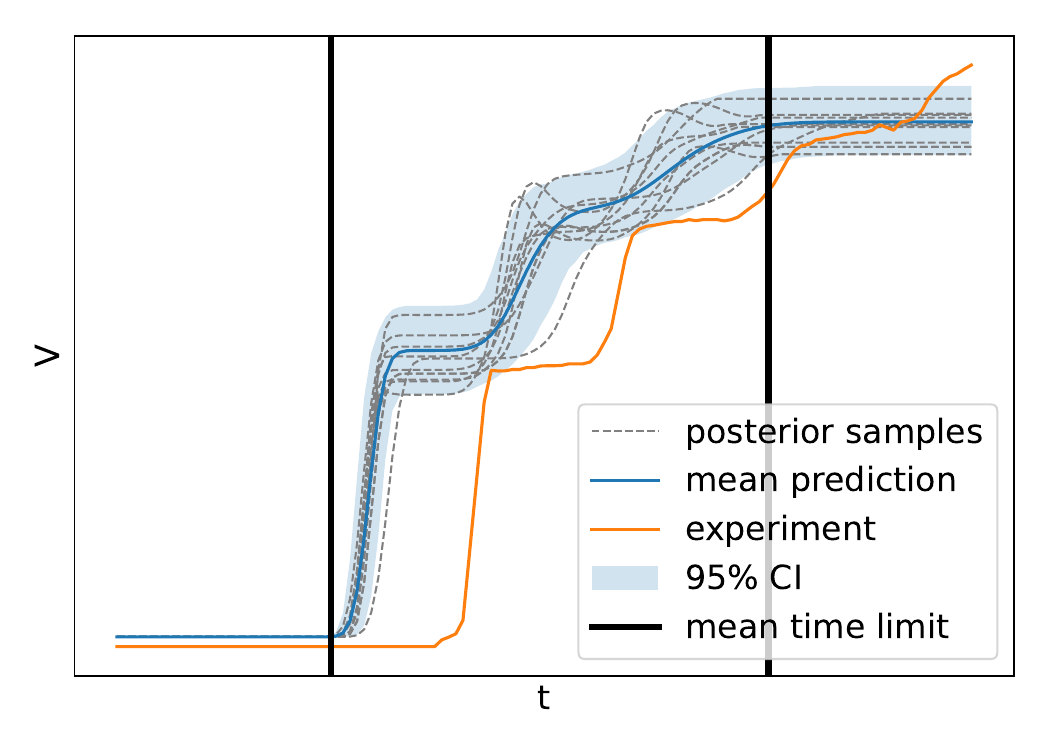}
    \end{minipage}
        \begin{minipage}{0.48\textwidth}
    \centering
    Experiment 2
    \includegraphics[width=0.98\textwidth]{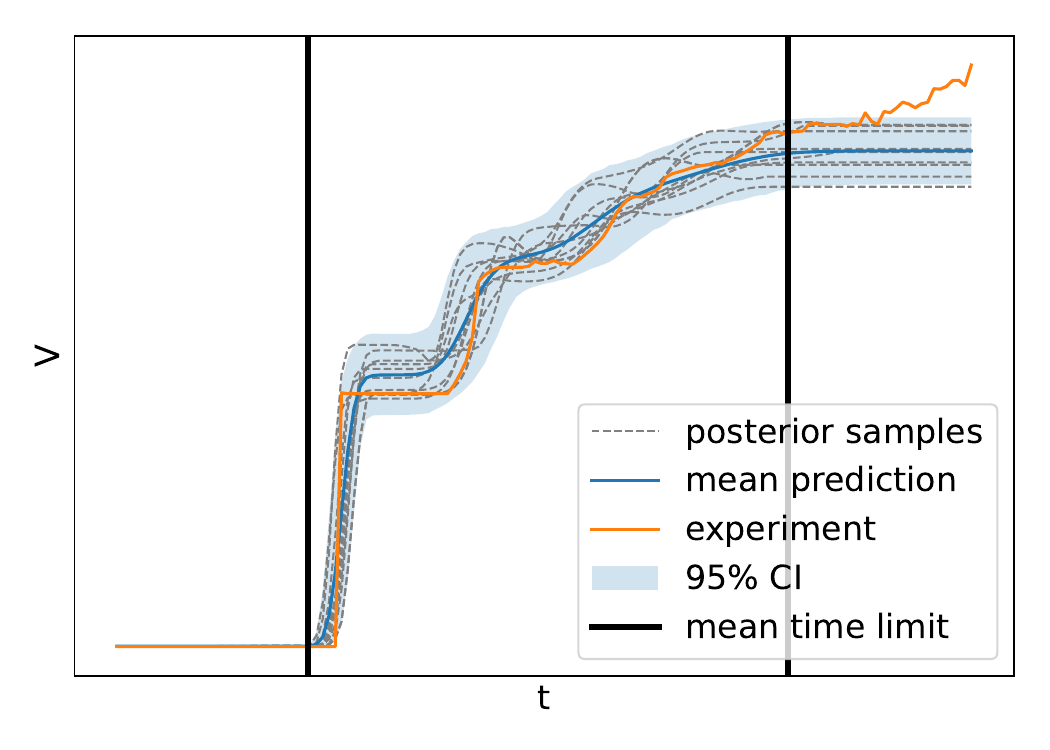}
    \end{minipage}

        \begin{minipage}{0.48\textwidth}
    \centering
    Experiment 3
    \includegraphics[width=0.98\textwidth]{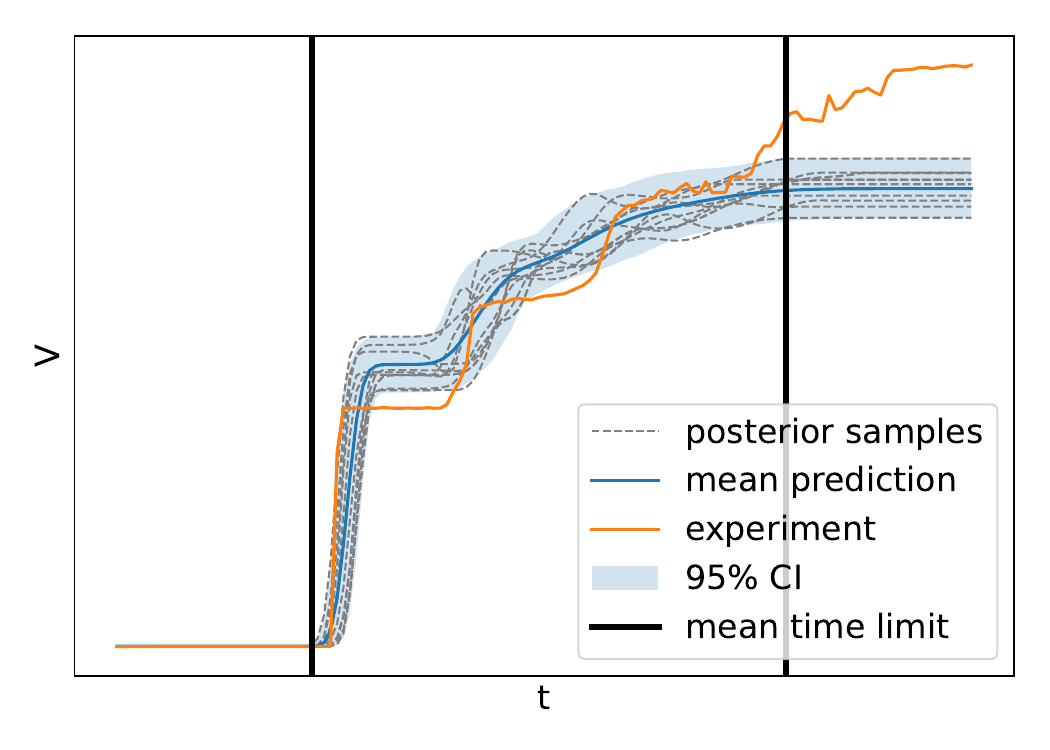}
    \end{minipage}
        \begin{minipage}{0.48\textwidth}
    \centering
    Experiment 4
    \includegraphics[width=0.98\textwidth]{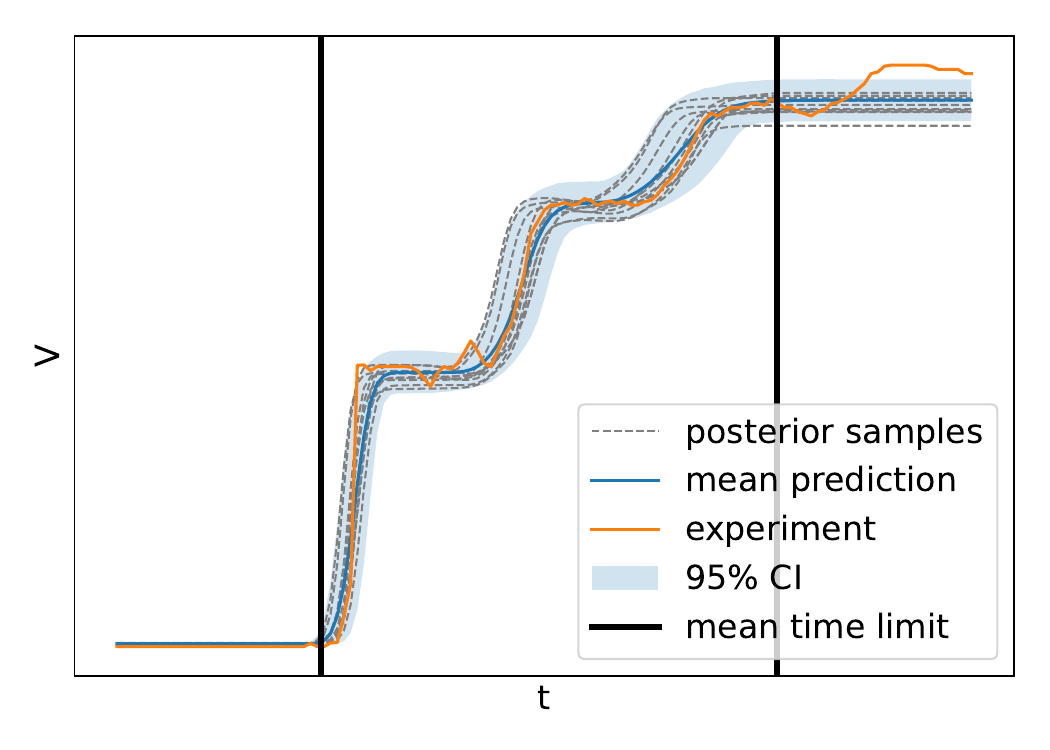}
    \end{minipage}

\caption{Posterior prediction after uncertainty propagation for each experiment. The blue curve 
corresponds to the composition of the mean predictions of the various surrogate models, evaluated at the posterior mean of the calibration parameters. 
The credible intervals and posterior samples are obtained by composing the uncertainties associated with the surrogate models and the calibration parameter.}
\label{posterior_eos_affine}
\end{figure}

We use the same assumptions as in the article to model the various errors. 
Sampling results are shown in Figure~\ref{eos_sampling_affine}, and the associated predictions in Figure~\ref{posterior_eos_affine}. 
With the affine rescaling, the surrogate models are slightly worse, and the distributions of the calibration parameters and predictions are less concentrated than with the partial elastic method.

\begin{table}[h]
\centering
\begin{tabular}{|c|c|c|c|c|}
\hline
& $\mathrm{MSE}$ & $\mathrm{IAE}$ & $\mathrm{CRPS}$ & $N_{\mathrm{PCA}}$ \\
\hline
Affine, exp 1 & 55 & 0.04 & 26 &5(4+0+1) \\
\hline
Partial elastic, exp 1 & 47 & 0.05 & 18 & 6(3+2+1)\\
\hline
Affine, exp 2 & 76 & 0.06 & 35 &6(5+0+1) \\
\hline
Partial elastic, exp 2 & 72 & 0.06 & 28 & 6(3+2+1)\\
\hline
Affine, exp 3 & 131 & 0.04 & 58 &  9(8+0+1)\\
\hline
Partial elastic, exp 3 & 129 & 0.03 & 50 &9(6+2+1)\\
\hline
Affine, exp 4 & 19 & 0.04 & 9 & 5(4+0+1) \\
\hline
Partial elastic, exp 4 & 18 & 0.05 & 8 & 5(2+2+1)\\
\hline
\end{tabular}
\caption{Comparison of the affine and partial elastic surrogate models  for different metrics computed with cross validation for the different experiments.}
\label{tab:surrogate_EOS_metrics}
\end{table}

